\documentclass[]{jfm}

\usepackage{graphicx}
\usepackage{epstopdf,epsfig}
\usepackage{natbib}
\usepackage{hyperref}
\usepackage{mathtools}
\usepackage{amsmath}
\hypersetup{
    colorlinks = false,
    urlcolor   = blue,
    citecolor = black,
    citebordercolor  = {0 1 0},
}

\usepackage{comment}
\usepackage{bm}
\usepackage{float}

\newcommand{\RomanNumeralCaps}[1]
\linenumbers

\title{
       A viscous drop in a planar linear flow - the role of deformation on streamline topology
      }

\author{Sabarish V. Narayanan\aff{1}        
        \and Ganesh Subramanian\aff{2}
        \corresp{\email{sganesh@jncasr.ac.in}}
       }

\affiliation{
            \aff{1}Robert Frederick Smith School of Chemical and Biomolecular Engineering, Cornell University, Ithaca, NY 14853, USA. 
            \aff{2}Engineering Mechanics Unit, JNCASR, Jakkur, Bangalore - 560064, India.
            }

\begin{document}

\maketitle

\begin{abstract}
Planar linear flows are a one-parameter family, with the parameter $\hat{\alpha}\in [-1,1]$ being a measure of the relative magnitudes of extension and vorticity; $\hat{\alpha} = -1$, $0$ and $1$ correspond to solid-body rotation, simple shear flow and planar extension, respectively. For a neutrally buoyant spherical drop in a hyperbolic planar linear flow with $\hat{\alpha}\in(0,1]$, 
the near-field streamlines are closed 
for $\lambda > \lambda_c = 2 \hat{\alpha} / (1 - \hat{\alpha})$, $\lambda$ being the drop-to-medium viscosity ratio; all streamlines are closed for an ambient elliptic linear flow with $\hat{\alpha}\in[-1,0)$.
We use both analytical and numerical tools to show that drop deformation, as characterized by a non-zero capillary number\,($Ca$), destroys the aforementioned closed-streamline topology. While inertia has previously been shown to transform closed Stokesian streamlines into open spiraling ones that run from upstream to downstream infinity, the streamline topology around a deformed drop, for small but finite $Ca$, is more complicated. Only a subset of the original closed streamlines transforms to open spiraling ones, while the remaining ones densely wind around a configuration of nested invariant tori. Our results contradict previous efforts 
 pointing to the persistence of the closed streamline topology exterior to a deformed drop, 
and have important implications for transport and mixing.
\end{abstract}

\begin{keywords}
\end{keywords}


\section{Introduction}\label{sec:intro}
Disperse multiphase systems are a common occurrence, and thence of great relevance, in both natural and industrial settings. Fundamental questions with regard to these systems arise in the context of their non-trivial rheological behavior, and in relation to estimating the rates of transport of mass or heat between the different phases. We consider a scenario where the disperse and continuous phases are Newtonian liquids of the same density, with the ratio of the disperse\,($\hat{\mu}$) to continuous\,($\mu$) phase viscosities denoted by $\lambda$. When the disperse phase (drop) volume fraction is small, interactions among the different disperse phase constituents can be neglected, and the problem reduces to a single Newtonian drop in an ambient Newtonian fluid undergoing a shearing flow. Further, for neutrally buoyant drops with sizes\,($a$) small compared to the characteristic flow scale\,(say $L$), an arbitrary shearing flow appears as an unbounded linear one at leading order in $a/L$. The small drop size also implies that viscous effects are dominant, and that the Reynolds numbers for both the fluid motion inside and outside the drop, defined as $Re = \dot{\gamma} a^2 \nu$ and $\hat{Re} = \dot{\gamma}a^2/\hat{\nu}$, are therefore small; here, $\dot{\gamma}$ is a scale for the ambient shear rate, and $\nu$ and $\hat{\nu}$ are the ambient and drop kinematic viscosities, respectively. In general, for the drop to remain spherical, interfacial tension must dominate ambient viscous stresses, and this requires that the dimensionless parameter  quantifying the relative strengths of the two, the capillary number $Ca = \mu \dot{\gamma} a /\sigma$\,($\sigma$ being the coefficient of interfacial tension), be small. 

In this work, we examine a weakly deformed drop in an ambient hyperbolic planar linear flow in the Stokesian regime, the focus being on the drop-deformation-induced alteration of the exterior streamline topology. The geometry of both interior and exterior streamlines is known to profoundly influence transport characteristics in the convection-dominant limit, corresponding to large Peclet numbers, and the role of the interior streamline topology in this regard has been examined in an independent effort\,(\cite{Pavan_2023}). Drop deformation depends on both $Ca$ and $\lambda$, and the validity of the weak deformation assumption requires either a large interfacial tension ($Ca \ll 1$, $\lambda \sim O(1)$) as already mentioned above, or a large drop viscosity ($\lambda \gg 1$, $\lambda Ca \sim O(1)$) for ambient linear flows with vorticity that allow for a tank-treading motion of the deformed interface at steady state. Our focus is on the former small-$Ca$ limit.

We begin with a detailed overview of the literature pertaining to the streamline pattern around a spherical (undeformed) drop or a particle in ambient planar linear flows. \citet{Cox68} analysed the streamline pattern around a neutrally buoyant rigid sphere in an ambient simple shear flow, for $Re = 0$, finding a region of closed streamlines around the rotating sphere. The significance of closed exterior streamlines became known from the work of \citet{Acrivos71} who showed, for the above scenario, that the non-dimensional scalar transport rate from the sphere\,(the Nusselt number $Nu$) plateaus at a finite value for $Pe \gg 1$; here, $Pe = \dot{\gamma}a^2/D$ is the Peclet number denoting the ratio of the convective to the diffusive transport rate, with $D$ being the heat\,(or mass) diffusivity. The above studies were extended to the one-parameter family of planar linear flows by \citet{Poe76}, who showed that closed streamlines exist around the rotating sphere for all cases except planar extension, in which case the sphere does not rotate; the parameter characterizing planar linear flows is $\hat{\alpha} \in [-1,1]$ with $\hat{\alpha} = -1$, $0$ and $1$ corresponding to solid-body rotation, simple shear and planar extension, respectively. For simple shear flow in particular, the aforementioned closed streamline region is infinite in extent within the flow-vorticity plane. \cite{Poe76} also showed that, for all the cases with a closed streamline region around the rigid sphere, $Nu$ again plateaus to an $\hat{\alpha}$-dependent value for $Pe \gg 1$; as expected, this limiting value diverges for $\hat{\alpha} \rightarrow 1$. Closed streamlines therefore lead to diffusion-limited transport, in contrast to the continued enhancement expected for an open-streamline topology, owing to boundary-layer-driven transport at large $Pe$; in this latter case, $Nu$ increases as $Pe^{\frac{1}{3}}$ for $Pe \rightarrow \infty$ for rigid particles. \citet{Powell83} extended the work of \citet{Cox68} in analyzing both the interior and exterior streamline topologies for a neutrally-buoyant spherical drop ($Ca = 0$) in an ambient planar linear flow. Unlike a solid particle, the author found that the existence of a closed streamline region in the exterior now depends on both $\hat{\alpha}$ and $\lambda$, with such a region arising 
only for $\lambda > \lambda_c(\hat{\alpha}) = 2 \hat{\alpha} / (1 - \hat{\alpha})$; also see \cite{Kris18a}. The threshold viscosity ratio $\lambda_c(\hat{\alpha})$ diverges for $\hat{\alpha} \rightarrow 1$ so that, even for a drop, all streamlines remain open in planar extension regardless of $\lambda$. In contrast, $\lambda_c(0) = 0$, so a closed streamline region exists in simple shear flow for any non-zero $\lambda$, and in fact, remains infinite in extent within the flow-vorticity plane.

Closed streamlines are a consequence of Stokesian reversibility. Reversibility constraints are no longer applicable in presence of inertia, that is, when the Reynolds number, either $Re$ or $\hat{Re}$\,(for a drop), is finite. \citet{Sub06a,Sub06b} analysed the effect of weak inertia\,($Re \ll 1$) on the streamline topology around a rigid sphere in an ambient planar linear flow with $\hat{\alpha} < 1$, and found the originally closed streamlines to be replaced by ones that come in from infinity, tightly spiral around the rotating sphere, and  eventually move off to infinity. The emergence of open spiraling streamlines in the exterior eliminates the diffusion-limited transport that prevails in the Stokesian limit. Thus, for any non-zero $Re$, and for sufficiently large $Pe$ such that $Re Pe \gg 1$, the said authors found $Nu \propto (Re Pe)^{\frac{1}{3}}$, a prediction confirmed in later simulations\citep{Yang_2011}. It is worth mentioning that earlier attempts\citep{Poe75,Morris2004}, that attempted to include inertial effects, had erroneously assumed closed streamlines to persist at finite $Re$, likely motivated by the 2D scenario involving a circular cylinder where closed streamlines cannot open up, despite the onset of inertial effects, on account of incompressibility \citep{Rob70a,Koss74}. The effect of inertia on the exterior streamline topology for a neutrally-buoyant spherical drop, suspended in a planar hyperbolic linear flow, was first analyzed by \citet{Kris18b} who found that, much like a rigid sphere, fluid inertia opens up the closed streamlines for $\lambda > \lambda_c(\hat{\alpha})$. An analysis of scalar transport leads to $Nu \propto (Re Pe)^{\frac{1}{2}}$ for $Re \ll 1, RePe \gg 1$, corresponding to the leading order convection being driven by the finite-$Re$ spiraling streamlines; note that there exists the usual boundary-layer-driven enhancement for $\lambda < \lambda_c(\hat{\alpha})$ with $Nu \propto Pe^{\frac{1}{2}}$ for $Pe \gg 1$\citep{Kris18a}. The geometry of the spiraling streamlines for a drop is more complicated, with each turn of a near-surface spiraling streamline resembling a Jeffery orbit\,(a spherical ellipse) with an $\hat{\alpha},\lambda$-dependent aspect ratio, rather than being circular as for a rigid sphere. The analysis of scalar transport in 
\citet{Kris18b} also 
showed the existence of a second critical, $\lambda = \lambda_{c2}(\hat{\alpha})$, demarcating distinct spiraling-streamline regimes: (a) the `single-wake' regime where all near-surface streamlines spiral in from infinity along the vorticity axis, and spiral out in the vicinity of the flow-gradient plane; and (b) the `bifurcated-wake' regime where near-surface streamlines spiral in both along the vorticity axis, and within the flow-gradient plane, and then spiral out along a pair of intermediate directions symmetrically disposed about the plane of shear.

While inertia is the only source of irreversibility for a solid particle, for a drop, one also has irreversible interfacial forces coming into play when the drop deforms\,(non-zero $Ca$), and that drive the drop towards sphericity regardless of the sense of the ambient shear. There have been several studies on deformed drops in an ambient linear flow and we briefly survey these efforts, before moving on to a discussion of the streamline topology around a deformed drop. GI Taylor first determined the steady state drop deformation in simple shear flow\citep{Taylor32}, and in planar extension\citep{Taylor34}. In the former case, Taylor considered both the limit of a dominant interfacial tension\,($Ca, \lambda Ca \ll 1$), and that of a dominant drop viscosity\,($\lambda, \lambda Ca \gg 1$), although the latter large-$\lambda$ limit was more restrictive than need be. For small $Ca$, the drop deforms into an ellipsoid with its longest axis aligned with the ambient extension, consistent with symmetry arguments that require the 
deformation to be linear in the ambient rate of strain tensor. For $\lambda Ca \gg 1$, the small-$Ca$ linearity arguments no longer hold, and Taylor found the deformed drop to align along the flow direction instead. The implication of these findings is that the inclination of the deformed drop with respect to the flow direction is a function of $\lambda Ca$, decreasing from $\frac{\pi}{4}$ for $\lambda Ca \ll 1$, to $0$ for $\lambda Ca \gg 1$. This was confirmed by \citet{Cox69} who also analyzed the transient dynamics of alignment, showing that the approach towards the steady $\frac{\pi}{4}$-alignment for $\lambda Ca \ll 1$ was monotonic, while that towards near-flow-aligned orientations for $\lambda Ca \gg 1$ had an oscillatory character, with damped shape oscillations, with a frequency of $O(\dot{\gamma})$, preceding eventual alignment. The relaxation time derived in \cite{Cox69}'s analysis was incorrect, as pointed out\,(and corrected) by \citet{Rall80}, although the principal features pertaining to the drop dynamics remain unchanged. An important conclusion from the effort of \cite{Cox69} is the singular nature of the miscible limit\,($Ca = \infty$), in which case almost any initial drop shape leads to persistent undamped oscillations, with there being no steady state for simple shear flow\,(and, more generally, for any of the planar linear flows with vorticity); the only exception being the trivial scenario where the initial configuration is identical to the stationary one - for simple shear, this is the flow-aligned configuration derived by \cite{Taylor32} for $\lambda Ca = \infty$. For planar extension alone, a miscible drop will be stretched out to infinity\,(leading to eventual breakup) regardless of $\lambda$. These features, with regard to miscible drop dynamics in an ambient linear flow, have been confirmed in the later analyses of Tucker and co-workers\citep{WetzelTucker1999,WetzelTucker2001}.

Following \cite{Cox69}, \citet{Frankel70} calculated the drop deformation to second order in $Ca$, which was further extended by \citet{Biesel73} to $O(Ca^3)$. Even though both perturbation expansions were developed for small values of a general deformation parameter $\epsilon$ which could equal either $Ca$ or $1/\lambda$, solutions were found only for the former case since, for $\epsilon = 1/\lambda$, one needs to account for terms in the convected\,(Jaumann) derivative that jump order when $\lambda Ca \sim O(1)$\citep{Rall80}. Later, \citet{Greco02} recognized, for $Ca \ll 1, \lambda Ca \sim O(1)$, that the corrections to the leading order spherical-drop disturbance field involved a characteristic quadratic combination of the rate of strain and vorticity tensors, and used this insight to simplify the calculation of the higher-order terms in $Ca$. Later studies of Vlahovska and co-workers\citep{Vlav05,Vlav09} have extended the small-deformation analysis to surfactant-laden drops where the Marangoni number\,($Ma$) enters as an additional parameter; in the limit of a clean interface\,($Ma \rightarrow 0$), the authors' results improve upon those of \cite{Biesel73}. Finally, following \citet{Rall80}'s suggestion, \citet{Olivera15} calculated the higher order corrections in $1/\lambda$, albeit for asymptotically large rather than order unity values of $\lambda Ca$. While the discussion above has focused on earlier analytical efforts, there have also been numerical investigations of drop deformation and break-up, originally using the boundary integral technique\citep{Rall78,Stone89,Stone90,Kennedy94,Kwak98}, and later using other methods\citep{Li2000,Amani2019}. Experimental investigations began with Taylor himself\citep{Taylor32,Taylor34}, followed by others\citep{Mason61,Bentleyleal86}, all of which have been well reviewed\citep{Rall84,Stone94}. 

The focus of the extensive literature referred to above has always been on drop deformation/break-up and the resulting implications for emulsion microstructure and rheology. Very little attention has been paid to the finite-$Ca$ streamline topology which is expected to play an important role in inter-phase transport processes. The reason for this lack of attention may perhaps be attributed to the experiments of \citet{Torza71} which involved observing the evolution of the dyed region, and paths of tracer particles, around a deformed drop in a counter-rotating Couette device. The observations were along the vorticity direction, and despite sufficient ambiguity in the nature of the tracer trajectories, the authors concluded that closed (asymmetric)\,streamlines persist within the deformed drop even until a Taylor deformation parameter of $0.25$, implying that the streamline topology around a deformed drop was qualitatively identical to that around a spherical one. This conclusion appeared to be reinforced by the boundary integral\,(BEM) computations of \citet{Kennedy94}; see Fig.$6$ therein. Although the emphasis was not on the streamline patterns, the recent effort of \citet{Komrakova14}, based on the Lattice-Boltzmann Method\,(LBM) method, did show several snapshots of the flow within a deformed drop with one or more regions of apparently closed streamlines\,(referred to as `vortices' by the authors). It is worth mentioning that numerically distinguishing between spiraling and closed streamlines requires in general a much higher resolution than that needed for obtaining converged values of the Taylor deformation parameter. The computations of \citet{Singh11} appear to be the only ones that examine the nature of spiraling streamlines around a neutrally buoyant drop, in simple shear flow, in some detail. The authors find a reversal in the direction of spiraling, with streamlines sufficiently far from the flow-gradient plane spiraling out along the vorticity direction, in apparent contradiction to the predictions for the inertial streamlines around both a rigid sphere\citep{Sub06a,Sub06b} and a spherical drop\citep{Kris18a,Kris18b}, and we comment on these computations later in the manuscript. While there have been other numerical investigations involving deformed drops in ambient linear flows\citep{Favelukis_2013,Favelukis_2014,Yang_2019}, all of these efforts are restricted to an ambient axisymmetric extensional flow where, as pointed out by \citet{Pavan_2023}, the constraint of axisymmetry precludes a deformation-induced alteration of the streamline topology\,(and the associated singular enhancement of scalar transport rates).

\subsection{Organisation of the paper}

In this paper, we analyze in detail the exterior streamline topology using the analytical velocity field, to $O(Ca)$, available from earlier literature, and  further validate the findings using BEM computations. In contrast to the studies above, but in accordance with the expected deviation from Stokesian reversibility, we find the original closed-streamline topology to indeed be destroyed by shear-induced drop deformation. We begin in Section \ref{Sec.2} by summarizing the Stokesian streamline topology around a spherical drop in an ambient planar linear flow\citep{Powell83,Kris18a}. This allows one to introduce two families of invariant surfaces, parameterized by the constants $C$ and $E$, whose intersections give the streamlines for $Re = Ca = 0$. The $C\!-\!E$ plane allows for a compact representation of the streamline topology by eliminating the coordinate that runs along the streamlines - each streamline for $Re = Ca = 0$ corresponds to an ordered pair $(C,E)$, and therefore, to a point on the $C\!\!-\!\!E$ plane. For small but finite $Re$ and/or $Ca$, $C$ and $E$ transform to quantities that vary slowly on a time scale of $O[\dot{\gamma}^{-1}\min(Re^{-1},Ca^{-1})]$. The equations governing them, at leading order, can then be derived from the velocity field, that includes the $O(R\!e)$ and $(C\!a)$ contributions, using the method of averaging\citep{Bajer90}. The solution to the averaged equations leads to the representation of streamlines as curves on the $C\!\!-\!\!E$ plane, for small but finite $Re$ and $Ca$. We first do this in Section \ref{inertia:CE} in the context of the simpler inertial streamline topology, originally characterized by \citet{Kris18b}; section \ref{inertia:physical} describes the inertial streamlines in physical space. In section \ref{Sec.4}, we move to the more complicated finite-$Ca$ exterior streamline topology. Section \ref{deformation:physical} describes the geometry of the streamlines in physical space for small but finite-$C\!a$, which is followed by their $C\!\!-\!\!E$-plane representations in section \ref{deformation:CE}. Section \ref{BEM} presents the results of boundary integral simulations that reinforce the findings of sections \ref{deformation:physical} and \ref{deformation:CE} which are based on the $O(Ca)$ velocity field; details with regard to validation of these computations appear in Appendix \ref{Appc}. In section \ref{ReCa:CE}, we briefly present the $C\!-\!E$ plane representations of streamlines, with both effects of inertia and drop deformation included, as a function of $Re/Ca$. Finally, in section \ref{sec:conclude}, we end with a summary of the main findings, and a mention of possible future extensions. Appendix \ref{AppA} defines the $\lambda$-dependent coefficients that appear in the expression for the $O(Re)$ and $O(Ca)$ corrections to the Stokesian velocity field; appendices \ref{AppB} and \ref{AppD} briefly discuss numerical issues encountered in integrating the averaged equations.

\section{Drop in canonical planar linear flows - Streamline topology}\label{Sec.2}

An ambient planar linear flow is given by $\bm{u}^\infty = \bm{\Gamma} \cdot \bm{x}$ 
where
\begin{align}
\bm{\Gamma} = \begin{bmatrix}
0 & 2 & 0\\
2\hat{\alpha} & 0 & 0\\
0 & 0 & 0
\end{bmatrix}, \label{1.2}
\end{align}
is the transpose of the velocity gradient tensor, the parameter $\hat{\alpha}$ being a measure of the strength of vorticity relative to extension
; as mentioned in the introduction, it ranges from $-1$ (solid-body rotation) to $1$ (planar extension), with $\hat{\alpha} = 0$ corresponding to simple shear flow. 
In what follows, we briefly present the results of \citet{Powell83} and \citet{Kris18a}, pertaining to the streamline topology outside a spherical drop in an ambient hyperbolic planar linear flow\,($0 < \hat{\alpha} \leq 1$), to lay the groundwork for our analysis of the deformed-drop streamline topology in section \ref{Sec.4}. The restriction to hyperbolic linear flows is because all exterior streamlines are closed for spherical drops in elliptic linear flows\,($-1 \leq \hat{\alpha} < 0$) regardless of $\lambda$, and therefore, rather trivially, all streamlines are expected to transform to spiraling ones on including irreversible effects arising from inertia or drop deformation. In contrast, the presence of a separatrix demarcating regions of closed and open exterior streamlines, for ambient hyperbolic linear flows, allows for non-trivial topological consequences. 

The components of the Stokesian velocity field for a spherical drop in an ambient planar linear flow, in a spherical coordinate system with polar axis aligned with the ambient vorticity, are given by\citep{Powell83}:
\begin{align}
 u_r &= (1 + \hat{\alpha}) r (A r^2 + B) \sin^2 \theta \sin 2\phi, \label{2} \\
 u_\theta &= \frac{(1 + \hat{\alpha})}{2} r B \sin 2 \theta \sin 2 \phi, \label{3}\\
 u_{\phi} &= (1 + \hat{\alpha})r \sin \theta (B \cos 2\phi - \beta), \label{4}
\end{align}
where $\beta = (1 - \hat{\alpha})/(1 + \hat{\alpha})$, and
\begin{eqnarray}
      B &=& 1 - \frac{\lambda}{(1 + \lambda)r^5}, \\
      A &=& \frac{5 \lambda}{2(1+\lambda)r^7}  - \frac{(5 \lambda) + 2}{2(1+\lambda)r^5}, \label{5}
\end{eqnarray}
for the exterior streamlines. The streamlines corresponding to (\ref{2}-\ref{4}) are the curves of intersection of the following two families of invariant surfaces\citep{Cox68,Powell83,Kris18a}:
\begin{align}
 x_2 &= \pm r \left[ \frac{\hat{\alpha}}{(1 + \hat{\alpha})} + E f^2(r) + \frac{\beta \lambda}{1 + \lambda} f^2(r) g(r)\right]^{1/2}, \label{7} \\
 x_3 &= r C f(r), \label{8}
\end{align}
characterized by the parameters $C$ and $E$, with
$f(r) = \left[ r^3 + \frac{3 \lambda}{2 (1 + \lambda) r^2} - \frac{5\lambda + 2}{2(1 + \lambda)} \right]^{-1/3}$ and 
$g(r) = \int_r^{\infty} \frac{f(y)}{y^3} dy$. 
From (\ref{7}) and (\ref{8}), the constant-$C$ and $E$ surfaces are seen to be surfaces of revolution about the $x_3$ and $x_2$ axes, respectively. Further, since $f(r) \rightarrow \infty$ for $r \rightarrow 1$ and $f(r) \rightarrow 0$ for $r \rightarrow \infty$, the exterior fluid domain corresponds to $C \in [0,\infty)$ with the drop surface, together with the flow-gradient\,($x_1-x_2$) plane, corresponding to the limiting surface $C = 0$. The aforementioned limiting values of $f$ also imply $E\in[E_0,\infty)$ for the exterior fluid domain, with the drop alone corresponding to the inner limiting surface $E = E_0 = -\frac{\beta \lambda g(1)}{1+\lambda}$. Furthermore, $E = E_{sep}$ is the axisymmetric separatrix surface that separates regions of closed\,($E_0 < E < E_{sep}$) and open\,($E > E_{sep}$) streamlines. This surface intersects the $x_1 x_3$-plane along a circle of degenerate fixed\,(saddle) points defined by $(r,\theta,\phi) \equiv [r_{sep}, \theta, (0,\pi)]$ with $r_{sep} = 
[\frac{\lambda(1 + \hat{\alpha})}{2\hat{\alpha}(1+\lambda)}]^{\frac{1}{5}}$; using this value of $r_{sep}$ in (\ref{8}) leads to $E_{sep} = \frac{-\hat{\alpha}}{(1+\hat{\alpha})f^2(r_{sep})} - \frac{\beta\lambda}{(1+\lambda)}g(r_{sep})$. Note that $r_{sep}$ diverges\,(with $E_{sep} \rightarrow 0$) for simple shear flow, so that the separatrix surface extends out to infinity along the $x_1x_3$-plane for this case alone, independent of $\lambda$.  

\begin{figure}
\centering
    \includegraphics[scale = 0.35]{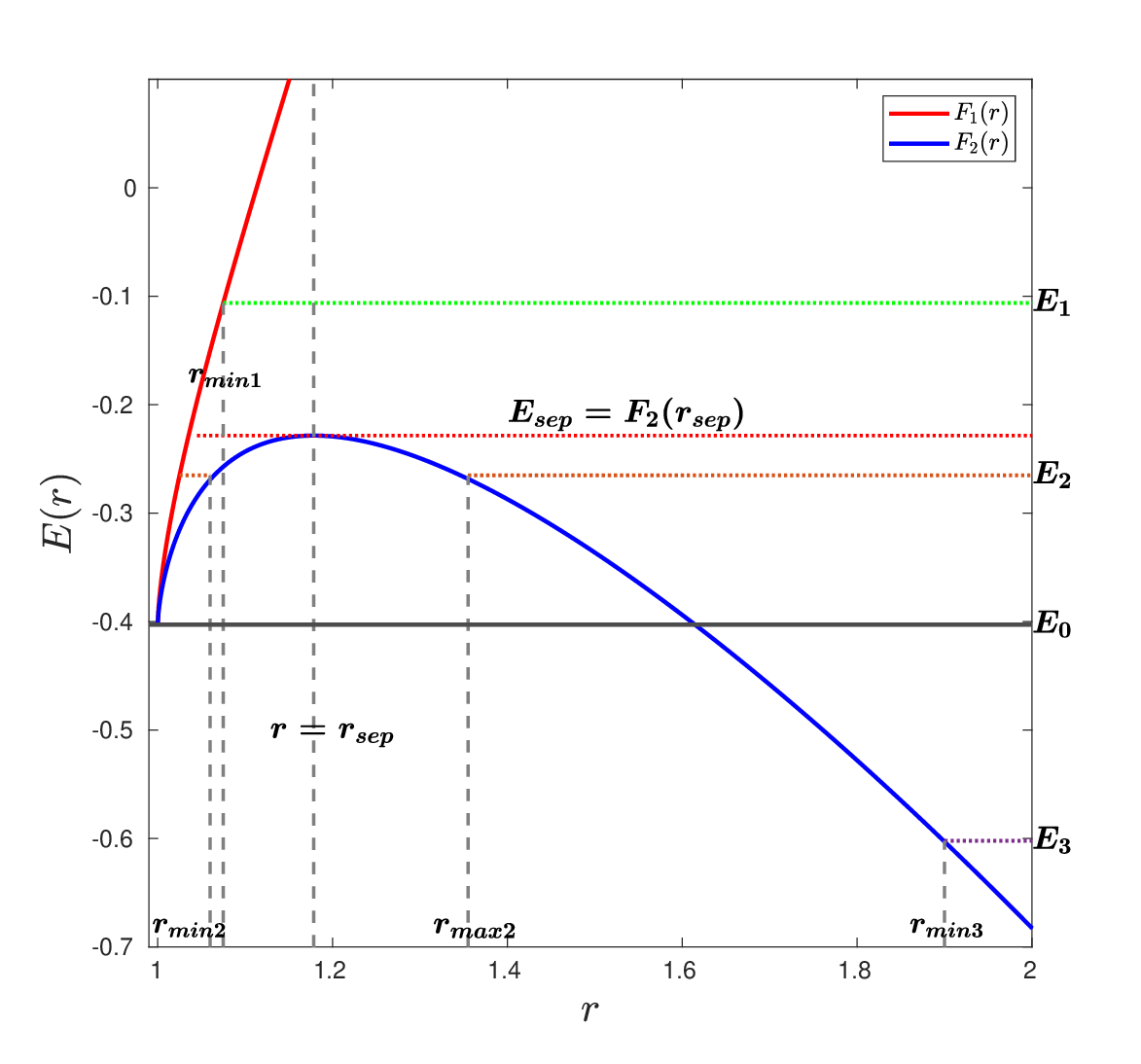}
    \includegraphics[trim= 0 -0.75cm 0 1cm,scale = 0.67]{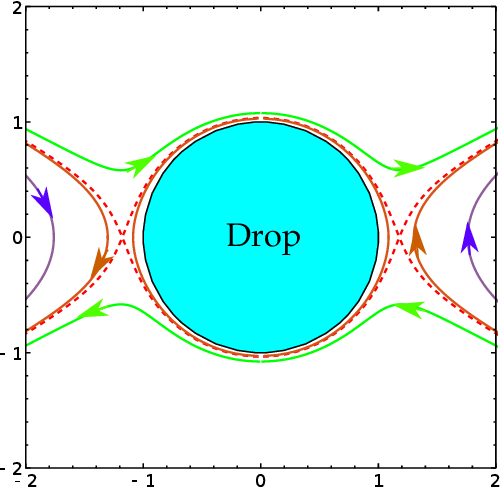}
\caption{(a) The functions $F_1(r)$\,(red) and $F_2(r)$\,(blue) versus $r$ for $\hat{\alpha}=0.25,\,\lambda=10$ with $C = 0$. The dashed red line with $E = E_{sep}$ corresponds to the in-plane separatrix, and demarcates open and closed streamline regions. The solid black line with $E = E_0$ separates the $E$-interval corresponding to both closed and open reversing streamlines\,($E_0 < E < E_{sep})$, from the one corresponding to only the latter\,($E< E_0$). Streamlines for three different $E$'s appear as horizontal segments; (b) The different classes of streamlines in the $x_1x_2$-plane, corresponding to the horizontal segments/lines in (a). A given streamline is depicted in both the $E-r$ and $x_1x_2$ planes using the same color: green for $E>E_{sep}$; orange for $E_0<E<E_{sep}$; purple for $E<E_0$.}
\label{fig:1}
\end{figure}

A given streamline, either open or closed, corresponds to an ordered pair $(C,E)$. The streamlines on the drop surface remain inaccessible, however, owing to the two families of surfaces becoming tangential for $r \rightarrow 1$, as evident from the drop being a limiting member of both families\,($C= 0$ or $E = E_0$). For all $\hat{\alpha}$ and $\lambda$, points along the $x_3$-axis, correspond trivially to zero-velocity streamlines, arising as the points of tangency of the surfaces $C = 1/f(x_3)$ and $E = \frac{-\hat{\alpha}}{(1+\hat{\alpha})f^2(x_3)} - \frac{\beta\lambda}{(1+\lambda)}g(x_3)$; the other zero-velocity streamlines correspond to the points of intersection of the separatrix surface and the $x_1x_3$-plane, as pointed out above. To characterize the nature of the streamline for general $C$ and $E$ in more detail, we first consider the case $C = 0$, corresponding to streamlines in the $x_1x_2$-plane; features of the streamlines outside this plane may be inferred by virtue of the axisymmetry of the constant-$E$ surfaces. The condition $0 \leq |x_2| \leq r$ leads then to the following constraints on $E$:
\begin{align}
\underbrace{\frac{-\hat{\alpha}}{f^2(r)(\hat{\alpha} + 1)} - \frac{\beta \lambda g(r)}{1 + \lambda}}  \leq \,&E\, \leq \underbrace{\frac{1}{f^2(r)(\hat{\alpha} + 1)} - \frac{\beta \lambda g(r)}{1 + \lambda}} \label{13a} \\
F_2 (r;\hat{\alpha}, \lambda)\hspace*{0.5in} &\hspace*{0.7in}F_1 (r, \hat{\alpha}, \lambda) \nonumber
\end{align}
The nature of the in-plane streamlines can now be understood based on the behavior of the functions $F_1$ and $F_2$, which are plotted against $r$ in Fig.\ref{fig:1}a, for $\hat{\alpha} = 0.25,\;\lambda = 10$. 
 In corresponding to a constant $E$, and in light of (\ref{13a}), the in-plane streamlines must appear in Fig.\ref{fig:1}a, either as horizontal segments bounded between the $F_1$ and $F_2$ curves, or as horizontal lines that extend to infinity starting from either of these curves. For the chosen $\hat{\alpha}$ and $\lambda$, $F_1$ is a strictly increasing function that starts from its minimum value of $E_0$ at $r = 1$. While $F_2$ starts from the same point, it is non-monotonic, attaining a maximum value of $E_{sep}$ at $r = r_{sep}$. The different classes of streamlines may now be organized as follows. Those with $E  > E_{sep}$ have a finite $r_{min}\,(= F_1^{-1}(E)$), but have $r_{max} = \infty$, and are open streamlines running from upstream to downstream infinity in the same sense as the streamlines of the ambient flow; for a given $E$\,($= E_1$ in Fig.\ref{fig:1}a), there are two such streamlines\,(the green curves in Fig.\ref{fig:1}b), with one obtained from the other via a reflection about the $x_1$-axis. There are two horizontal segments corresponding to each $E$ in the interval $E_0 \leq E < E_{sep}$ - the orange segments with $E = E_2$ in Fig.\ref{fig:1}a: (i) the first is bounded between the $F_1$ and $F_2$ curves, with both $r_{min}\,(= F_1^{-1}(E))$ and $r_{max}\,(= F_2^{-1}(E))$ being finite, and therefore, corresponds to a closed streamline\,(the orange closed curve in Fig.\ref{fig:1}b); (ii) the second starts from the $F_2$-curve, extending out to infinity, with $r_{min}\,(=F_2^{-1}(E))$ finite and $r_{max} = \infty$, and corresponds to a pair of open reversing streamlines\,(the open orange curves on either side of the drop in Fig.\ref{fig:1}b). The marginal value, $E = E_{sep}$, corresponds to the in-plane projection of the axisymmetric separatrix surface mentioned above, and consists of both closed and open\,(reversing) portions intersecting in a pair of saddle points given by $(\pm r_{sep},0,0)$; the separatrix projection appears as a dashed red curve in Fig.\ref{fig:1}b. As $E$ decreases below $E_{sep}$, the closed streamline moves towards the drop while the pair of open reversing streamlines move away, until at $E = E_0$, the former coincides with the drop surface. For $E < E_0$, one only has open reversing streamlines\,($E= E_3$ in Fig.\ref{fig:1}a; the purple curves in FIg.\ref{fig:1}b) that move off to infinity, along the positive and negative $x_1$-axes, as $E \rightarrow -\infty$. For the chosen $\hat{\alpha}$ and $\lambda$, the region of closed streamlines in Fig \ref{fig:1} is evidently finite in extent. For $\hat{\alpha} = 0$, $r_{sep} = \infty\,\forall\,\lambda$, and $F_2(r)$ becomes a monotonically increasing function, asymptoting to a maximum\,($E_{sep} =0$) for $r \rightarrow \infty$. Thus, for simple shear flow, the interval $E_0 < E < E_{sep}\,(= 0)$ corresponds only to closed streamlines, with the minimum, $E_0 =-\frac{\lambda g(1)}{1+\lambda}$, corresponding to the drop surface. 

\begin{figure}
\centering
 \includegraphics[trim= 0.1cm 0.25cm 0.5cm 0.75cm,clip,scale = 0.47]{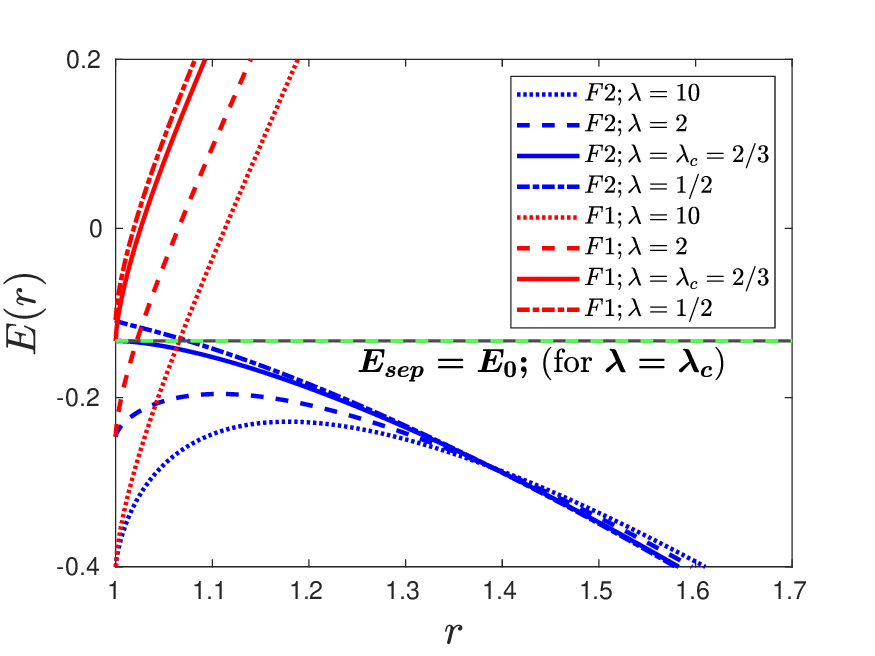}
 \includegraphics[trim= 0.1cm 0.25cm 0.5cm 0.75cm,,clip,scale = 0.47]{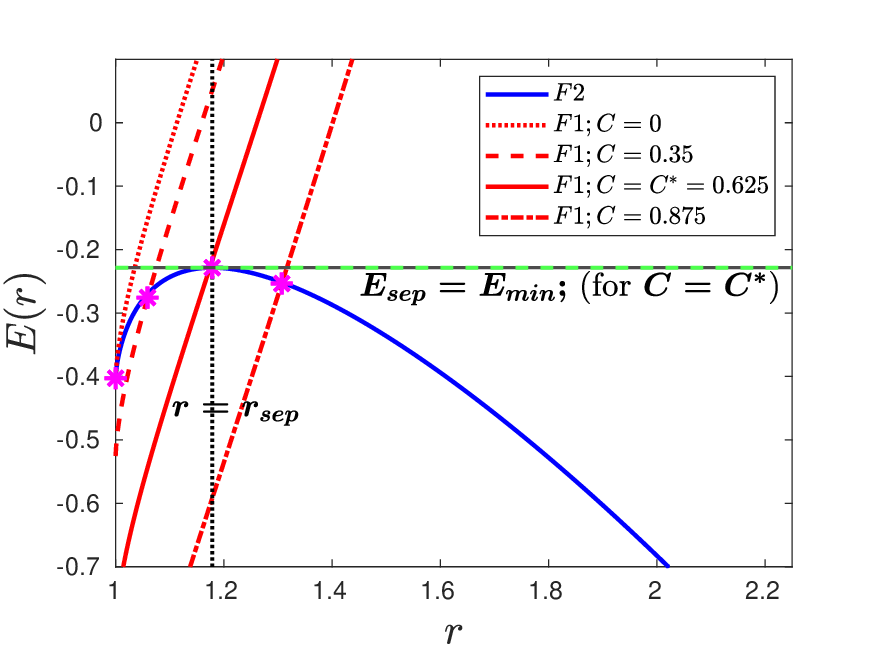} 
\caption{(a) The functions $F_1(r)$\,(red) and $F_2(r)$\,(blue) for $\hat{\alpha} = 0.25$, $C = 0$, for different $\lambda$; (b) the functions $F_1(r)$\,(red) and $F_2(r)$\,(blue) for $\hat{\alpha} = 0.25$, $\lambda = 10$ for varying $C$. The magenta stars denote the intersection points of $F_1$ and $F_2$, which change with $C$.}
\label{fig:2}
\end{figure}

Fig.\ref{fig:2}a plots $F_1$ and $F_2$ for the same $\hat{\alpha}\,(= 0.25)$ as in Fig.\ref{fig:1}, but for $\lambda$ varying from $10$ down to $0.5$. With decreasing $\lambda$, $F_2$ transitions from a non-monotonic to a strictly decreasing function of $r$ with a maximum at $r = 1$. The transition occurs across a threshold $\lambda = \lambda_c$, which may be obtained by setting $r_{sep} = 1$, giving $\lambda_c = \frac{2 \hat{\alpha}}{1- \hat{\alpha}}$; $\lambda_c = 2/3$ for $\hat{\alpha} = 0.25$. At $\lambda = \lambda_c$, $F_2$ is seen to have a zero slope at the origin. Although not shown, an analogous transition occurs for $\lambda$ fixed, and for $\hat{\alpha}$ increasing beyond the threshold $\hat{\alpha}_c = \frac{\lambda}{2 + \lambda}$. With $F_1$ monotonically increasing and $F_2$ monotonically decreasing, all streamlines in Fig.\ref{fig:2}a are horizontal lines starting from either of these curves and extending to infinity, and thence, are open. Thus, the aforementioned threshold viscosity ratio corresponds to a critical curve in the $\hat{\alpha} - \lambda$ plane, separating in-plane streamline patterns that include a closed-streamline region in the vicinity of the spherical drop, from those that do not. For $\hat{\alpha} = 1$, $\lambda_c = \infty$, and therefore, as mentioned in section \ref{sec:intro}, all in-plane streamlines remain open in planar extension regardless of $\lambda$. At the other extreme, $\lambda_c = 0$ for $\hat{\alpha} = 0$, so that a closed streamline region exists for all $\lambda$ in simple shear flow, being bounded by a separatrix curve that extends to infinity along the flow direction. This separatrix curve is in turn the projection of an axisymmetric separatrix surface that extends to infinity along the flow-vorticity plane, and includes an infinite volume of fluid within.

For $C \neq 0$, the modified constraint $0 \leq |x_2| \leq \sqrt{r^2 - x_3^2}$ leads to the same expression for $F_2$ as in (\ref{13a}), but with $F_1 (r, \hat{\alpha}, \lambda) = \frac{1}{f(r)^2(\hat{\alpha} + 1)} - \frac{\beta \lambda g(r)}{1 + \lambda} - C^2$. Thus, the $F_1$-curve is displaced downward by a constant amount. As shown in Fig.\ref{fig:2}b, for $\hat{\alpha} = 0.25$ and $\lambda = 10$, with increasing $C$, this displacement causes it to move across the maximum of the $F_2$-curve\,(that remains unchanged) to larger $r$, leading to the disappearance of the closed streamline region. The value of $C$, and thence that of $x_3$, at which the cross-over above happens, equals $r_{sep}$, consistent with the axisymmetry of the separatrix surface. Note again that $r_{sep} = \infty$ for simple shear flow, so the above cross-over does not happen at a finite $C$, consistent with the infinite extent of the separatrix surface already mentioned.

\begin{figure}
  \centering
  \includegraphics[width = 0.85\textwidth]{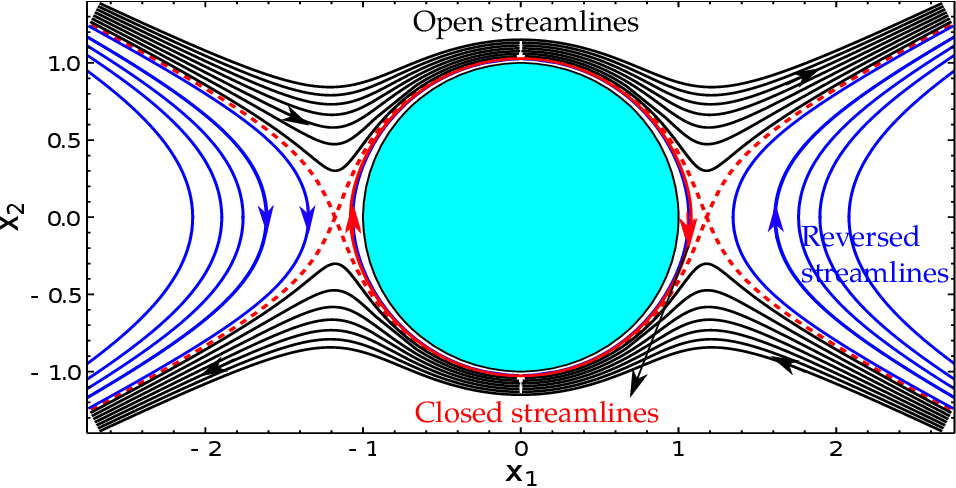}\\
  \includegraphics[scale = 0.65]{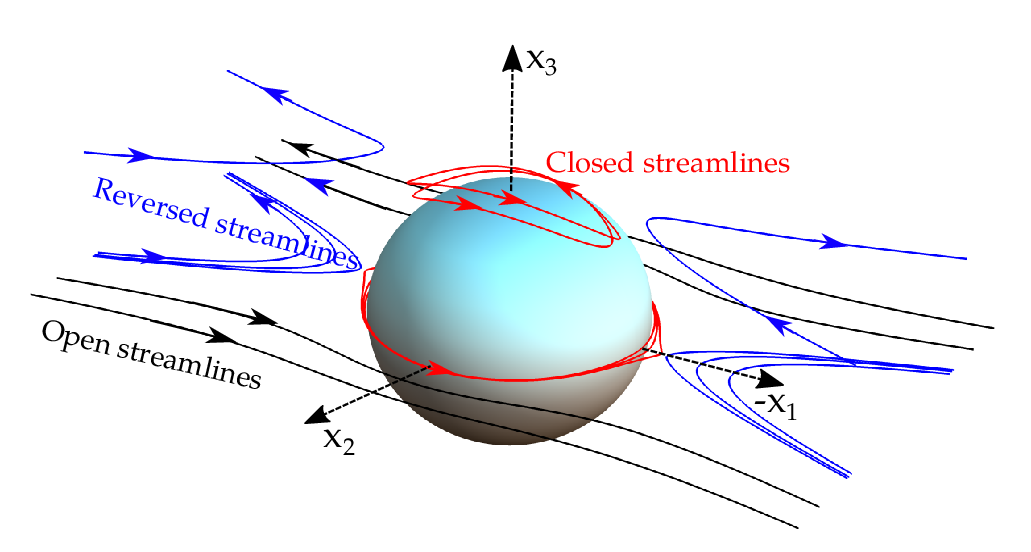}
  \caption{Stokesian streamlines around a drop for $\hat{\alpha} = 0.25,\,\lambda = 10$ (a) on the equatorial plane and (b) the complete three-dimensional topology. Note that for this $\hat{\alpha}$, the extent of closed streamlines is finite.}
\label{fig:5}
\end{figure}

\begin{figure}
  \centerline{\includegraphics[width = 0.85\textwidth]{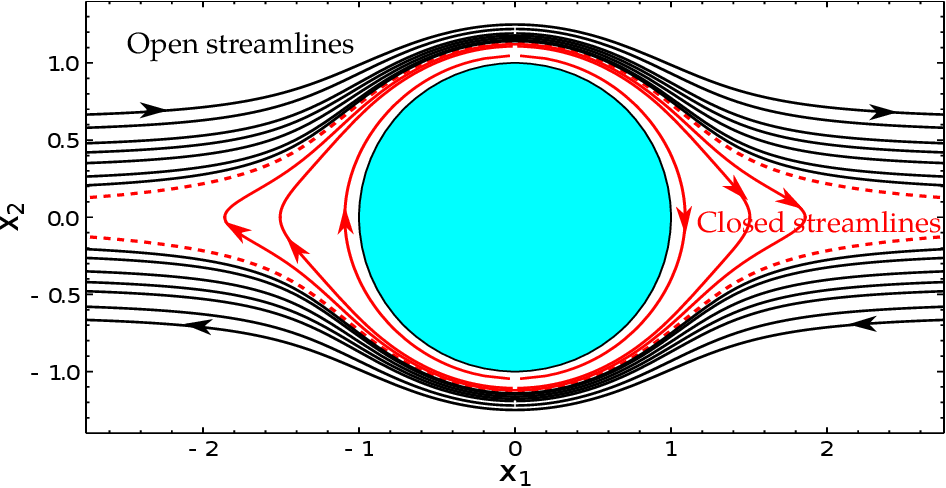}}
  \caption{Streamlines on the equatorial plane for $\hat{\alpha} = 0,\, \lambda = 10$. For this case, there are no reversed streamlines and the closed streamlines occupy a volume of infinite extent.}
\label{fig:6}
\end{figure}

\begin{figure}
    \centering
    \includegraphics[trim= 0.1cm 0.25cm 0.75cm 0.75cm,clip,scale = 0.375]{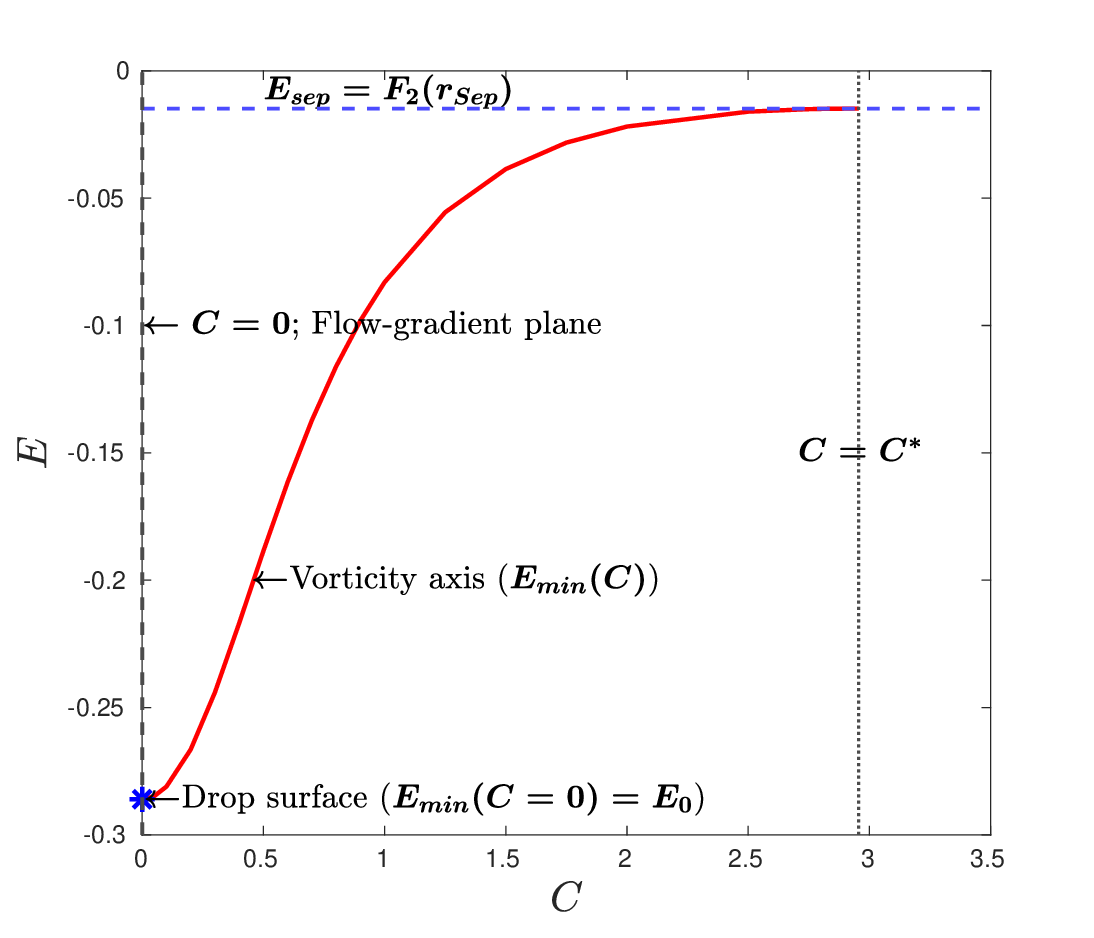}
    \includegraphics[trim= 0.1cm 0.25cm 0.75cm 0.75cm,clip,scale = 0.375]{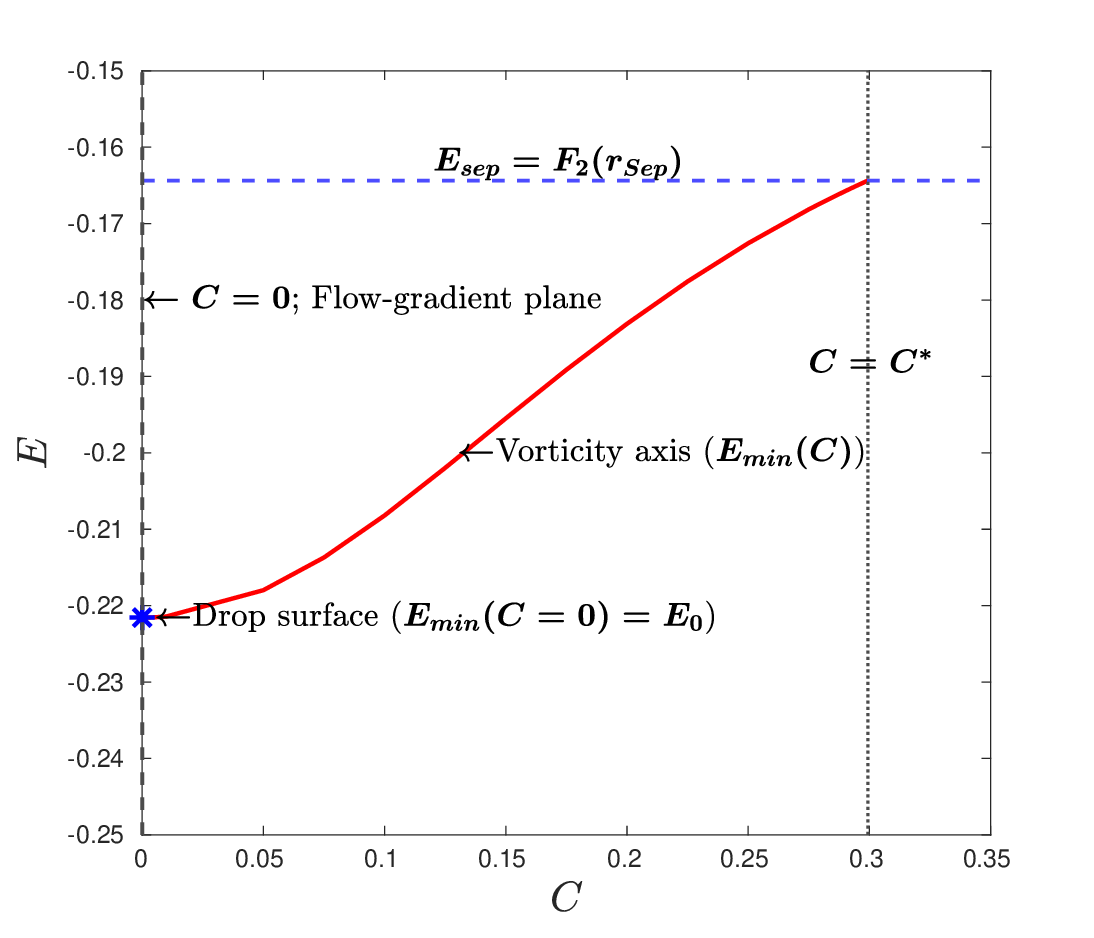}
    \caption{Region of closed streamlines in the $C - E$ plane for (a) $\hat{\alpha} = 10^{-3}, \lambda = 1 > \lambda_c$ and (b) $\hat{\alpha} = 0.25, \lambda = 10 > \lambda_c$, showing the various streamline topologies and their region of occurrence. }
\label{fig:7}
\end{figure}

Based on the discussion above, we show the in-plane streamline pattern and the 3D streamlines, for $\hat{\alpha}=0.25, \lambda = 10$, in Figs.\ref{fig:5}a and b. The in-plane streamline pattern for $\hat{\alpha} = 0$, for the same $\lambda$, is shown in Fig.\ref{fig:6}; the region of closed streamlines is now infinite in extent, with there being no open reversing streamlines. 
The aforementioned analysis suggests a compact representation of the Stokesian streamlines around a spherical drop as points in the $C-E$ plane, a representation that may then be used to interpret the altered streamline topology in the presence of inertia, drop deformation, suspending fluid viscoelasticity, etc. As is known from earlier rheological and transport rate calculations, the consequences of the altered streamline topology are most important in the region of closed streamlines, and therefore, we identify this region on the $C-E$ plane. This requires determining the $E$-interval, corresponding to closed streamlines, as a function of $C$. For $C = 0$, as evident from Fig.\ref{fig:1}a, $E \in [E_0,E_{sep}]$, with $E_0$ being the point of intersection of $F_1$ and $F_2$ that also happens to correspond to the smallest $r\,(=1)$. For non-zero $C$, the point of intersection moves to larger $r(>1)$, as shown in Fig\,\ref{fig:2}b, and may be determined by solving $F_1(r) = F_2(r)$  - closed streamlines exist only for $r$'s greater than that corresponding to this point of intersection, and until $r = r_{sep}$. One obtains the corresponding $E$\,($E_{min}(C)$, say; $E_0 = E_{min}(0)$) using (\ref{13a}), and the $E$-interval therefore is $[E_{min}(C),E_{sep}]$.  This interval decreases with increasing $C$, vanishing at $C = C^*$\,(say) when $E_{min}(C^*) = E_{sep}$. $C^*$ remains finite for non-zero $\hat{\alpha}$, diverging for $\hat{\alpha} \rightarrow 0$. The region of closed streamlines in the $C - E$ plane is shown in Figs.\ref{fig:7}a and b for a pair of $(\hat{\alpha},\lambda)$ combinations. The region is bounded by the interval $[0,C^*]$ on the $C$-axis, the interval $[E_0,E_{sep}]$ on the $E$-axis, and the curve $E_{min}(C)$. Almost all points on $E_{min}(C)$ correspond to the minimal closed streamlines that are the points of tangency of the two surfaces involved. Only the point at which the $E$-axis and $E_{min}(C)$ meet corresponds to the entire drop surface\,($r=1$); this abrupt transition from a point to a surface occurs on account of the surfaces $E= E_{min}(0)$ and $C=0$ surfaces coinciding in this limit, as mentioned earlier.


\section{Spherical drop in canonical planar hyperbolic flows - Effect of inertia on the streamline topology} \label{Sec.3}

In this section, we briefly review the work of \citet{Kris18b} on the inertial streamline topology around a spherical drop in an ambient hyperbolic planar linear flow, and then use the method of averaging to obtain the $C\!-\!E$ plane representations of the inertial streamlines.

\subsection{Exterior Streamlines in physical space} \label{inertia:physical}

The $O(Re)$ correction to the Stokesian velocity field induced by a rigid sphere or a spherical drop, in an ambient linear flow, has been available since the work of \citet{Peery66}. However, the resulting qualitative alteration of the Stokesian streamline topology for ambient planar linear flows, and the implications for scalar transport, were recognized much later; by \cite{Sub06a, Sub06b} for a rigid sphere, and by \citet{Kris18b} for a spherical drop. The velocity field in the latter case, for small but finite $Re$, is:
\begin{equation}
 \bm{u(x)} = \bm{u}^{(0)}(\bm{x}) + Re\, \bm{u}^{(1)}(\bm{x}) + O(Re^{3/2}), \label{3.1}
\end{equation}
where $\bm{u}^{(0)}$ has been defined in (\ref{2}-\ref{4}), and $\bm{u}^{(1)}$ is given by: 
\begin{align}
 \begin{split}
  \bm{u}^{(1)}(\bm{x}) &= u_1(r,\lambda) (\bm{\Gamma}:\bm{xx})^2\bm{x} + u_2(r,\lambda) (\bm{\Gamma}:\bm{xx}) \bm{\Gamma.x} + u_3(r,\lambda)(\bm{\Gamma}:\bm{xx})\bm{\Gamma}^\dag.\bm{x} \\
  &+ u_4(r,\lambda)(\bm{\Gamma}.\bm{x})(\bm{\Gamma.x})\bm{x} + u_5(r,\lambda)(\bm{\Gamma}.\bm{x})(\bm{\Gamma}^\dag.\bm{x})\bm{x} + u_6(r,\lambda)(\bm{\Gamma}^\dag.\bm{x})(\bm{\Gamma}^\dag.\bm{x})\bm{x} \\
  &+ u_7(r,\lambda)\bm{\Gamma.}(\bm{\Gamma.x}) + u_8(r,\lambda)\bm{\Gamma}^\dag.(\bm{\Gamma}^\dag.\bm{x})
  + u_9(r,\lambda) \bm{\Gamma}^\dag.(\bm{\Gamma.x}) \\
  &+ u_{10}(r,\lambda)\bm{\Gamma.}(\bm{\Gamma}^\dag.x) +u_{11}(r,\lambda)\left(\bm{\Gamma} : \bm{\Gamma}^\dag + \bm{\Gamma} : \bm{\Gamma} \right)\bm{x}.\label{3.2}
 \end{split}
\end{align}
As mentioned above, the above expression was first derived by \citet{Peery66}, and later corrected in \citet{Raja11}; for convenience, the $u_i(r,\lambda)$'s have been tabulated in Appendix \ref{AppA}1. (\ref{3.2}) has a regular character, being a solution of the inhomogeneous Stokes equations. In contrast, the next term of $O(Re^{3/2})$ has a singular character\citep{Zhang11}, being driven by the limiting linear-flow form of the velocity field in the outer region\,(characterized by scales of $O(aRe^{-1/2})$). The expansion (\ref{3.1}) is only valid in the inner region, that is, for $r \ll a Re^{-\frac{1}{2}}$, since it does not satisfy the decay criterion at infinity\,($\bm{u}^{(1)}$ in (\ref{3.2}) is independent of $r$ for large $r$). This is, however, not a limitation in the present context, since the most important inertia-induced alterations of the streamline topology are within the Stokesian closed-streamline region, and the separatrix surface enclosing this region only extends to radial distances of order unity for $\hat{\alpha}$ not too close to zero. For $\hat{\alpha} \rightarrow 0$, corresponding to ambient linear flows in the neighborhood of simple shear, one needs to include the $O(Re^{\frac{3}{2}})$ contribution, so that the inertial separatrix envelope, that separates spiraling from the open reversing streamlines, is finite in extent, being bounded by a pair of saddle points at a distance of $O(Re^{-\frac{3}{10}})$ along the positive and negative $x_1$-axes\,(note that these points still lie in the inner region\citep{Sub06b}, although only marginally so). We only include the $O(Re)$ term here, so the region of spiraling streamlines is infinite in extent along the flow direction for $\hat{\alpha} = 0$.

\begin{figure}
\centering
 \includegraphics[trim= 2cm 0cm 0 0cm,scale = 0.45]{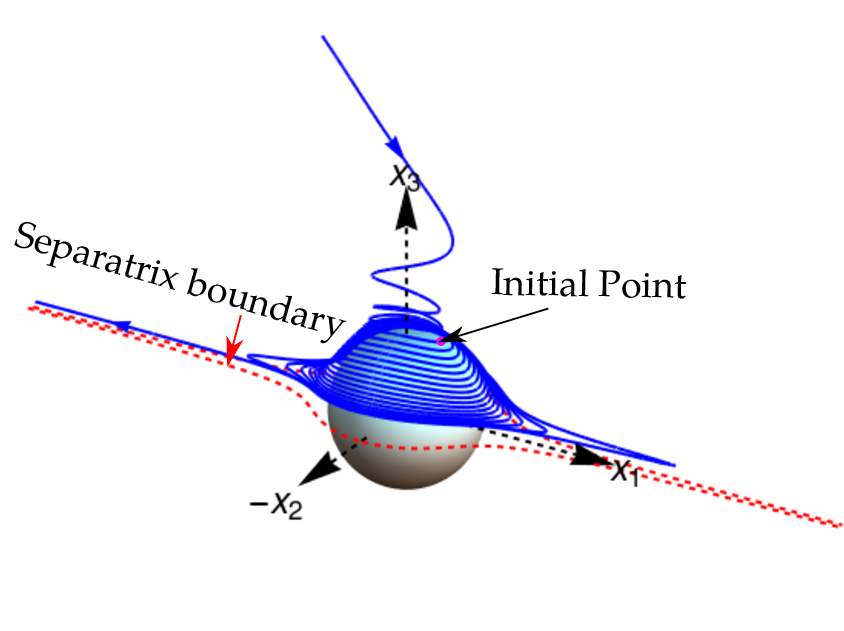}
 \hspace{0.2in}
 \includegraphics[trim= 0.25cm -1cm 0 1cm,scale = 0.45]{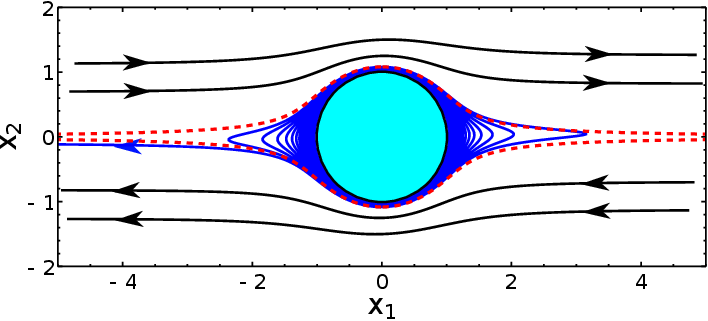}
 \includegraphics[scale = 0.45]{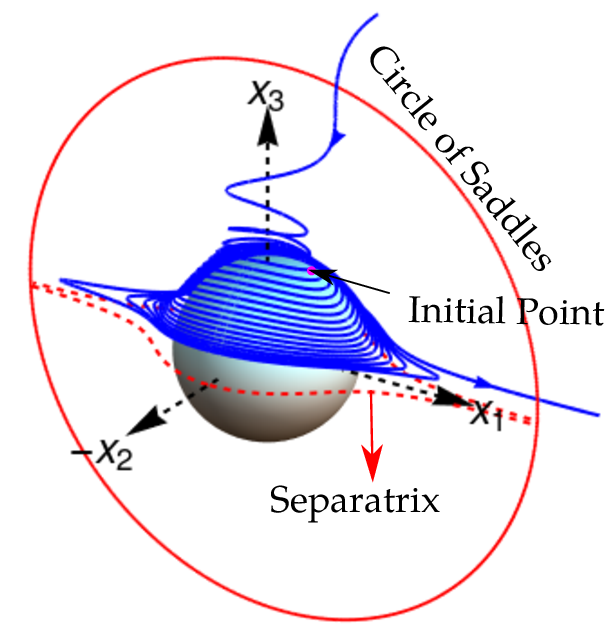}
 \hspace{0.2in}
 \includegraphics[trim= -1.75cm -2cm 0 1cm,scale = 0.45]{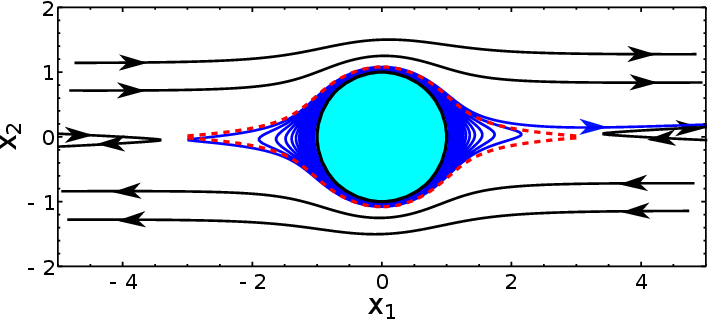}
 \includegraphics[trim= 2cm 0cm 0 0cm,scale = 0.3]{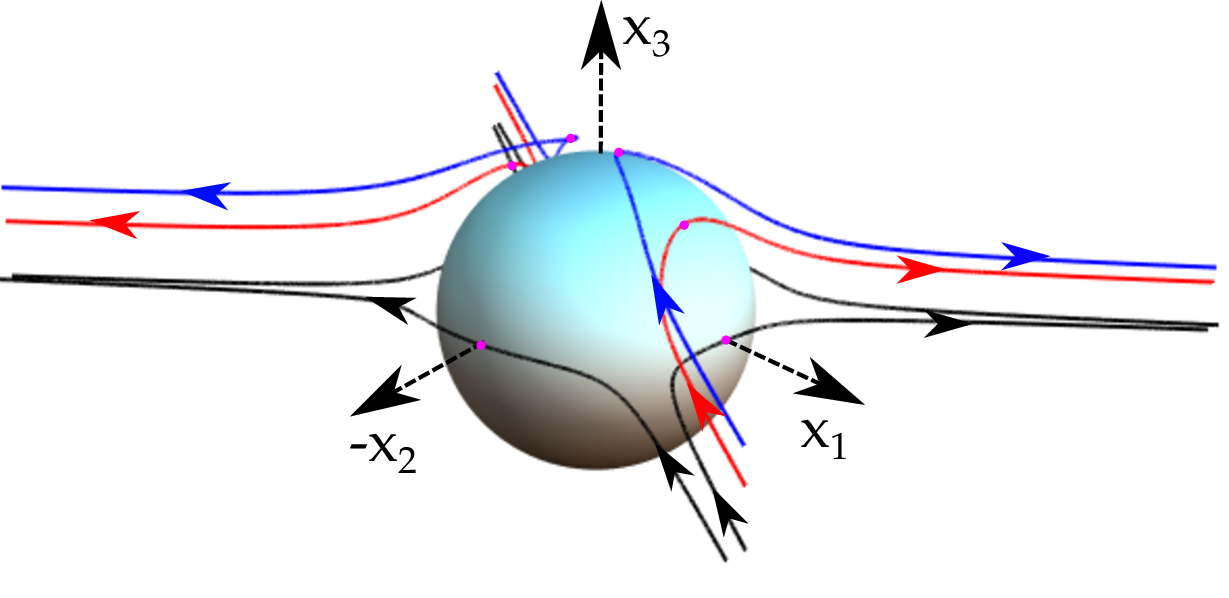}
 \includegraphics[trim= -1.5cm -0.5cm 0 1cm,scale = 0.345]{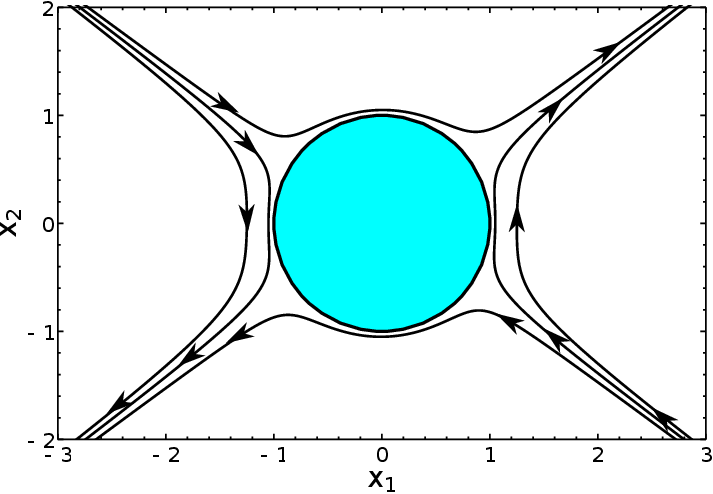}
\caption{Exterior streamlines around a spherical drop for $Re = 0.5, \lambda = 2$\,($\hat{\alpha}_c = 0.5$); (a) and (b) correspond to $\hat{\alpha} = 0$, (c) and (d) to $\hat{\alpha} = 10^{-3}$, and (e) and (f) to $\hat{\alpha} = 0.55$. Figures (a),(c) and (e) show the 3D streamlines, and figures (b), (d) and (f) show those in the $x_1-x_2$ plane. The magenta points mark the `initial' points from which one carries out integrations forward and backward in time.}
\label{fig:8}
\end{figure}

We now plot the exterior streamlines by numerically integrating (\ref{3.1}). In Figs.\ref{fig:8}a-f, we have chosen $\lambda = 2$ and $Re = 0.5$, and have plotted the exterior inertial streamlines for $\hat{\alpha} = 0$\,(Figs.\ref{fig:8}a and b), $10^{-3}$\,(Figs.\ref{fig:8}c and d) and $0.55$\,(Figs.\ref{fig:8}e and f); the final $\hat{\alpha}$ being greater than $\hat{\alpha}_c\,(=0.5)$, there is no Stokesian closed streamline region in this case. The subfigures on the left\,(Figs.\ref{fig:8}a, c and e) show the 3D streamlines, while those on the right\,(Figs.\ref{fig:8}b, d and f) show the ones in the $x_1x_2$-plane. Since the first two rows correspond to $\hat{\alpha} < \hat{\alpha}_c$, the in-plane separatrix, that bounds the Stokesian closed streamlines, is also shown as a dashed red curve in these figures. In Figs.\ref{fig:8}a and b, corresponding to simple shear, this separatix extends to infinity along the $x_1$-axis, while it has a finite extent in Figs.\ref{fig:8}c and d, terminating on the fixed-point circle in the $x_1x_3$-plane with $r_{sep} \approx 3.1964$\,(the solid red curve in Fig.\ref{fig:8}c). From the in-plane streamline patterns in Figs.\ref{fig:8}b and d, it is apparent that the existence of the Stokesian separatrix organizes the finite-$Re$ streamlines into three groups: (i) open streamlines that go from upstream to downstream infinity in the same sense as the ambient flow, but unlike their Stokesian counterparts, are fore-aft asymmetric about the $x_2$-axis; (ii) streamlines that spiral around the drop before heading off downstream, with the spiraling being increasingly tight as $Re$ decreases; (iii) for cases where the separatrix has a finite extent, as in Figs.\ref{fig:8}c and d, the spiraling streamlines are contained between regions of open reversing streamlines that, unlike their Stokesian counterparts, are no longer symmetric about the $x_1$-axis. Figs.\ref{fig:8}a and c show that, for the chosen $\hat{\alpha}$ and $\lambda$, the 3D spiraling streamlines come in along the $x_3$-axis, before going off to infinity sufficiently close to the $x_1x_2$-plane, consistent with the nature of the in-plane spiraling streamlines in Figs.\ref{fig:8}b and d. In Figs.\ref{fig:8}e and f, there are no spiraling streamlines. 
The inertial streamline configuration above is similar to that originally identified around a rigid sphere in an ambient planar linear flow for small $Re$\citep{Sub06b}; for simple shear flow, the open and reversing streamlines have also been determined numerically for finite $Re$, and the computations show that the region of spiralling streamlines rapidly shrinks in extent with increasing $R\!e$\citep{Morris2004}. One of the differences between a rigid sphere and a drop manifests in the geometry of the spiraling streamlines. While each turn of a near-surface spiraling streamline is almost circular for a rigid sphere, that for a spherical drop conforms to a Jeffery orbit\,(a spherical ellipse); see \cite{Kris18b}.

\begin{figure}
\centering
\includegraphics[scale = 0.6]{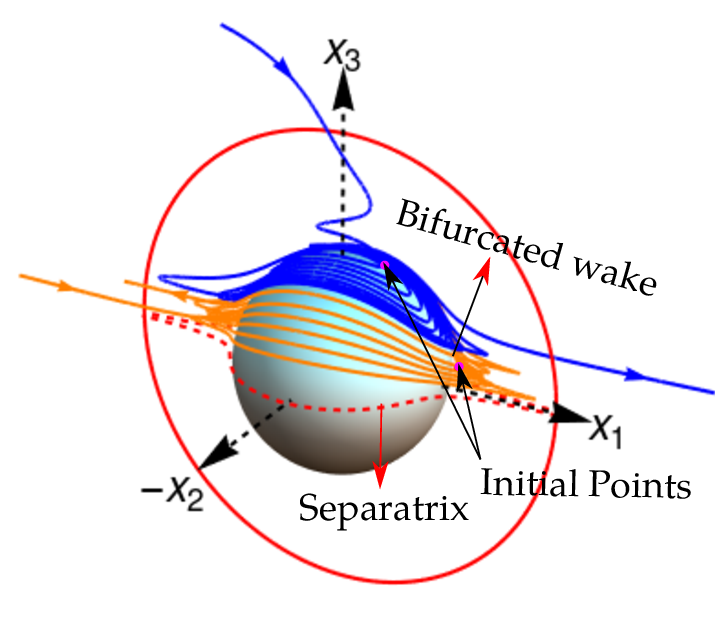}
\includegraphics[trim= 0cm -1.5cm 0 1cm,scale = 0.45]{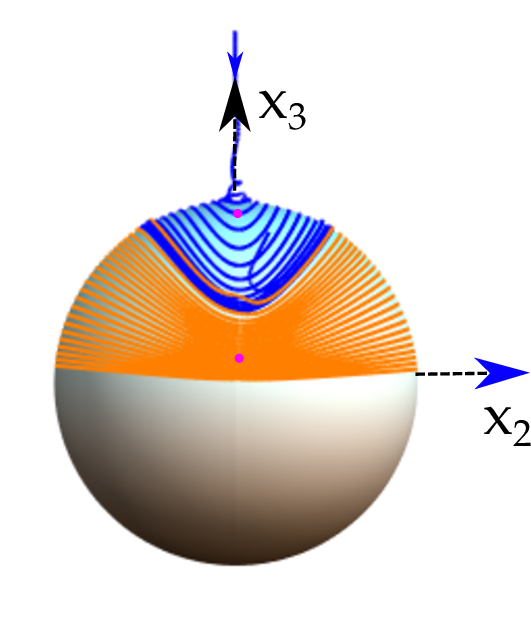}
\caption{Exterior Streamlines around a spherical drop for $Re = 0.5,\lambda = 0.1\,(< \lambda_{bif}), \hat{\alpha} = 10^{-3}$, corresponding to the bifurcated-wake regime (a) in 3D and (b) projected onto the $x_2-x_3$ plane. The streamlines spiral in along both the vorticity axis, and close to the flow-gradient plane, escaping to infinity along an intermediate direction.}
\label{fig:9}
\end{figure}

The spiraling-streamline topology described above corresponds to most of the region for $\lambda > \lambda_c$, but \citet{Kris18b} showed that there exists a narrow interval, $\lambda_{c} \leq \lambda < \lambda_{bif}$, where the inertial streamlines exhibit a different topology\,(Fig.6a in their paper). Within this interval, which is restricted to the range $0 \leq \hat{\alpha} \lesssim \hat{\alpha}_{bif}\,(=0.35)$, 
  the streamlines spiral in both along the $x_3$-axis, and in the vicinity of the $x_1x_2$-plane, before going off to infinity at an intermediate angular location in the vicinity of the drop surface. This is illustrated in Figs.\ref{fig:9}a and b for $Re = 0.5$, $\hat{\alpha} = 10^{-3}$ and $\lambda = 0.1\,(< \lambda_{bif} \approx 0.35$). Note that the Jeffery-orbit-like character of the individual turns, of a spiralling streamline, is more readily evident in Fig.\ref{fig:9}, than in Fig.\ref{fig:8}, owing to the lower $\lambda$.

 While the nature of individual inertial streamlines is evident from Figs.\ref{fig:8} and \ref{fig:9}, one can only afford to plot a few streamlines in a single figure for purposes of clarity. With the intent of representing the whole streamline pattern, we move towards a compact representation of the finite-$Re$
 streamlines on the $C-E$ plane; this representation will also be used to examine the streamline topology for small but finite $Ca$. 

\subsection{The $C-E$ representation of the exterior streamlines} \label{inertia:CE}

There are two time scales associated with a tightly spiraling inertial streamline for small $Re$: (i) the fast time scale of $O(\dot{\gamma}^{-1})$ characterizing the circulation around the closed Stokesian streamlines, and that is also the time in which the spiraling inertial streamline executes a single turn; (ii) the asymptotically longer time scale of $O(\dot{\gamma}^{-1}Re^{-1})$ in which the inertia-induced drift\,(across closed Stokesian streamlines) displaces a fluid element through a distance of $O(a)$. While $C$ and $E$ were constants of motion in the Stokes limit, for small but finite $Re$, they vary on the aforementioned slow time scale. The equations governing this slow variation may be derived using the method of averaging (\citet{Bajer90}). To do so, we start with the definitions in (\ref{7}) and (\ref{8}), rewritten as:
\begin{align}
 C &= \frac{\cos \theta}{f(r)} \label{3.4} \\
 E &= \frac{1}{f^2(r)} \left(\sin^2 \theta \sin^2 \phi- \frac{\hat{\alpha}}{(1+\hat{\alpha})} - \frac{\beta \lambda}{1 + \lambda} f^2(r)g(r) \right) \label{3.5}
\end{align}
where $x_2$ and $x_3$ have been replaced by their expressions in spherical coordinates 
Differentiating (\ref{3.4}-\ref{3.5}) with respect to time, and using $\frac{d \bm{x}}{d t} = \bm{u}^{(0)}\bm{(x)} + Re \; \bm{u}^{(1)}\bm{(x)}$, one obtains:
\begin{align}
 \frac{dC}{dt} = \mathcal{F}_0(r,\theta,\phi) + Re \; \mathcal{F}_1(r,\theta,\phi), \label{3.7}\\
 \frac{dE}{dt} = \mathcal{G}_0(r,\theta,\phi) + Re \; \mathcal{G}_1(r,\theta,\phi), \label{3.8}
\end{align}
where
\begin{align}
 \mathcal{F}_0(r,\theta,\phi) &= \frac{1}{f^2(r)} \left( \sin \theta f(r) u_{\theta}^{(0)} + \cos \theta f'(r) u_{r}^{(0)} \right), \label{3.9} \\
 \mathcal{F}_1(r,\theta,\phi) &= \frac{1}{f^2(r)} \left(\sin \theta f(r) u_{\theta}^{(1)} + \cos \theta f'(r) u_{r}^{(1)} \right), \label{3.10} \\
 \mathcal{G}_0(r,\theta,\phi) &= u_\theta^{(0)} I_1 + u_r^{(0)} I_2 + u_\phi^{(0)} I_3, \label{3.11} \\
 \mathcal{G}_1(r,\theta,\phi) &= u_\theta^{(1)} I_1 + u_r^{(1)} I_2 + u_\phi^{(1)} I_3, \label{3.12}
\end{align}
with $I_{1} = f^2(r) \sin^{2}\phi \sin 2 \theta$, $I_{2} = 2 f(r) f'(r) (\frac{\hat{\alpha}}{1 + \hat{\alpha}} - \sin^{2}\phi \sin^{2}\theta) + \frac{\beta \lambda f^5(r)}{(1 + \lambda) r^{3}}$ and $I_{3} = f^{2}(r) \sin^{2}\theta \sin 2 \phi$. On using (\ref{2}-\ref{4}) for $u_l^{(0)}$ ($l \equiv r, \theta, \phi$), $\mathcal{F}_0$ and $\mathcal{G}_0$, turn out to be identically zero, consistent with $C$ and $E$ being invariants of the Stokesian field. (\ref{3.7}-\ref{3.8}) therefore reduce to:
\begin{align}
 \frac{d C}{dt} = Re \; \mathcal{F}_1(r,\theta,\phi), \label{3.16}\\
 \frac{dE}{dt} = Re \; \mathcal{G}_1(r,\theta,\phi), \label{3.17}
\end{align}
where the expressions for $u_l^{(1)}$, involved in the definitions of $\mathcal{F}_1$ and $\mathcal{G}_1$, are given by (\ref{3.2}). Combining $Re$ with $t$ on the LHS to give $\hat{t} = Re \; t$, the slow time variable in the method of averaging framework, and regarding $C$ and $E$ as being functions of $\hat{t}$ alone, at leading order in $Re$, (\ref{3.16}) and (\ref{3.17}) may be written in the form:
\begin{align}
 \frac{d C}{d\hat{t}} = \mathcal{F}_1[r(t;C,E),\theta(t;C,E),\phi(t;C,E)], \label{3.16}\\
 \frac{dE}{d\hat{t}} = \mathcal{G}_1[r(t;C,E),\theta(t;C,E),\phi(t;C,E)]. \label{3.17}
\end{align}
The RHS now explicitly recognizes that $r$, $\theta$ and $\phi$ also depend on the fast time variable, $t$, with this dependence arising, at leading order, from motion (approximately)\,along a Stokesian streamline corresponding to a given $C$ and $E$. One obtains the averaged equations by integrating both sides of (\ref{3.16}-\ref{3.17}) over a single fast time period:
\begin{align}
 \frac{d C}{d\hat{t}} = \bar{\mathcal{F}_1}(C,E) \; = \; \frac{1}{t_{period}} \int_0^{t_{period}} \mathcal{F}_1 \; dt, \label{3.18} \\
 \frac{dE}{d\hat{t}} = \bar{\mathcal{G}_1}(C,E) \;= \; \frac{1}{t_{period}} \int_0^{t_{period}} \mathcal{G}_1 \; dt, \label{3.19}
\end{align}
where the LHS of (\ref{3.18}) and (\ref{3.19}) are unaffected by the averaging since $C$ and $E$ are independent of $t$. The fast-time integration above corresponds to a fixed $\hat{t}$, and thence, to fixed $C$ and $E$ at leading order, with $t_{period}$ being the period of circulation around a given Stokesian streamline $(C,E)$. One may therefore write:
\begin{align}
 \begin{split}
 t_{period}(C,E) &= 4 \int_{r_{min}(C,E)}^{r_{max}(C,E)} (u_r^{(0)})^{-1} dr, \\
 &= 4 \int_{r_{min}(C,E)}^{r_{max}(C,E)} \frac{r^2 f(r) \; dr}{\left[ \frac{\hat{\alpha}}{(1 + \hat{\alpha})f(r)^2} + \beta \left( E + \frac{\beta \lambda}{1 + \lambda} g(r) \right) - f^2(r) \left(E + \frac{\beta \lambda}{1 + \lambda} g(r) \right)^2 \right]^{1/2}},
 \end{split} \label{3.25}
\end{align}
where the four-fold symmetry of the Stokesian closed streamlines is used in converting the time integration to one over the radial coordinate; this also implies that the $\theta$ and $\phi$ in the integrands of (\ref{3.18}) and (\ref{3.19}) are now expressed in terms of $r$ using (\ref{3.4}) and (\ref{3.5}).
The integration limits in (\ref{3.25}) correspond to the minimum and maximum radial distances of the Stokesian streamline which always occur at $\phi = \frac{\pi}{2}$ and $\phi = 0$, respectively, leading to the following relations for $r_{min}$ and $r_{max}$: 
\begin{align}
 \left[ \frac{\hat{\alpha}}{(1 + \hat{\alpha})} + E f^2(r_{min}) + \frac{\beta \lambda}{1 + \lambda} f^2(r_{min}) g(r_{min})\right] &= 1 - C^2 f^2(r_{min}), \label{3.22} \\
 \left[ \frac{\hat{\alpha}}{(1 + \hat{\alpha})} + E f^2(r_{max}) + \frac{\beta \lambda}{1 + \lambda} f^2(r_{max}) g(r_{max})\right] &= 0, \label{3.23}
\end{align}
which are again derived from the definitions of $C$ and $E$. Using all relevant expressions in (\ref{3.18}-\ref{3.19}), one finally obtains the following 2D autonomous system in explicit form:
\begin{align}
 \frac{dC}{d \hat{t}} &= -\frac{\int_{r_{min}}^{r_{max}} \left( \sin \theta f(r) u_{\theta}^{(1)} + \cos \theta f'(r) u_{r}^{(1)} \right)f(r)^{-2} (u_r^{(0)})^{-1} dr}{\int_{r_{min}}^{r_{max}} (u_r^{(0)})^{-1} dr}, \label{3.20} \\
 \frac{dE}{d \hat{t}} &= \frac{\int_{r_{min}}^{r_{max}} \left( u_\theta^{(1)} I_1 + u_r^{(1)} I_2 + u_\phi^{(1)} I_3 \right) (u_r^{(0)})^{-1} dr}{\int_{r_{min}}^{r_{max}} (u_r^{(0)})^{-1} dr}, \label{3.21}
\end{align}
whose solutions are $Re$-independent trajectories in the $C-E$ plane. 

The above derivation only applies within the region of closed streamlines that exists for $Re = 0$. Inertia perturbs the open streamlines in a regular sense, leading to an $O(Re)$ difference between the upstream and downstream branches of any open streamline; there is no notion of multiple time scales in this case. We therefore numerically integrate (\ref{3.20}) and (\ref{3.21}) only in the region of the $C-E$ plane corresponding to the Stokesian closed streamline region - these regions have already been depicted earlier in Figs.\ref{fig:7}a and b. The integration involves starting from a given $(C,E)$ within the said region, and then proceeding forward and backwards in time until one approaches any of the boundaries shown in Fig.\ref{fig:7}\,($E = E_{min}(C)$, $E=E_{sep}$ or $C =0$); note that one has to solve the transcendental relations (\ref{3.22}) and (\ref{3.23}), for $r_{min}$ and $r_{max}$, at each time step in this integration.

\begin{figure}
\centering
\includegraphics[trim = {0.1cm 0.2cm 0.65cm 0.65cm}, clip,scale = 0.44]{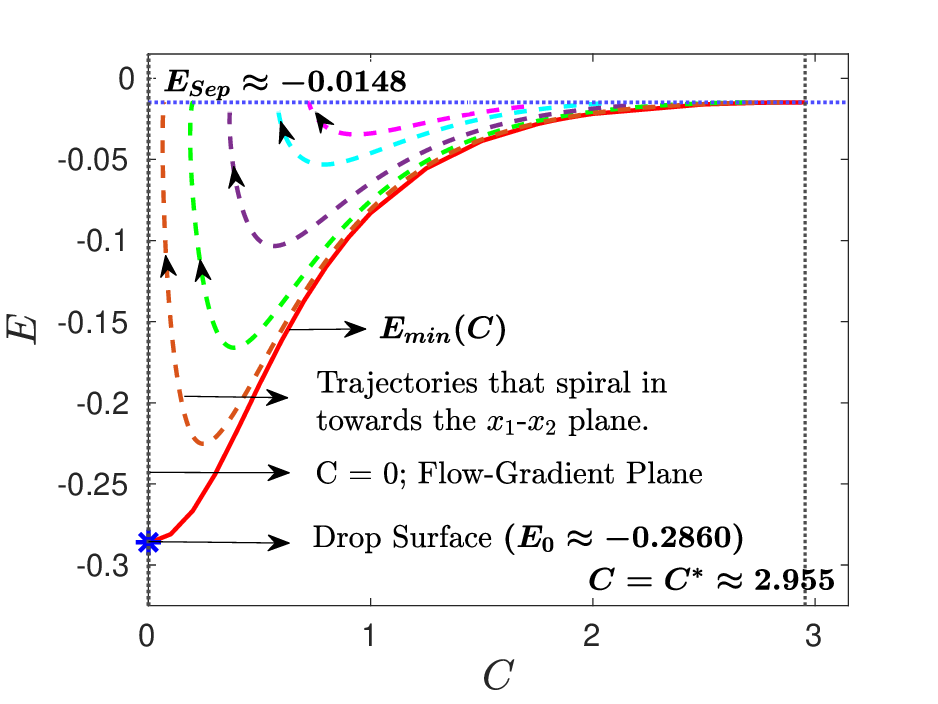}
\includegraphics[trim = {0.1cm 0.2cm 0.65cm 0.65cm}, clip,scale = 0.44]{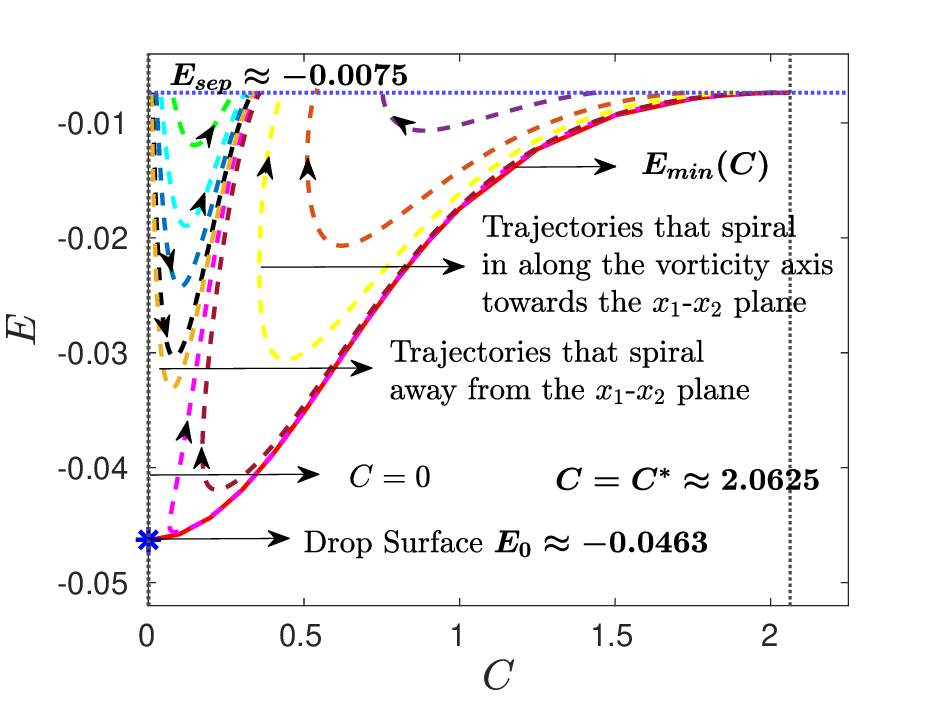}
\caption{Representations of the inertial streamlines on the $C-E$ plane, for small but finite $R\!e$, using the method of averaging, for $\hat{\alpha} = 10^{-3}$: (a) $\lambda = 1\,(> \lambda_{bif})$ - the single-wake regime; (b) $\lambda = 0.1\,(< \lambda_{bif})$ - the bifurcated-wake regime.}
\label{fig:10}
\end{figure}

The $C\!-\!E$-plane representations of inertial streamlines, obtained from solving  (\ref{3.20}) and (\ref{3.21}), 
are shown in Figs.\ref{fig:10}a and b for $\hat{\alpha} = 10^{-3}$, and for $\lambda = 1$ and $\lambda = 0.1$, respectively. For $\hat{\alpha} = 10^{-3}$, $\lambda_{bif} \approx 0.35 $, and the aforesaid figures therefore correspond to the single and bifurcated wake regimes\citep{Kris18b}, respectively. In both figures, the trajectories are seen to originate from and terminate at the separatrix\,($E = E_{sep}$), which is consistent with an inertial streamline transitioning from an open to a tightly spiraling curve only on crossing the separatrix surface into the Stokesian closed streamline region, and eventually, transitioning back to an open streamline on exiting this region. The solution trajectories in Fig.\ref{fig:10}a are always directed towards smaller $C$, consistent with inertial streamlines spiraling in along the vorticity axis towards the $x_1x_2$-plane. In contrast, the C-E plane in Fig.\ref{fig:10}b is divided into two regions, consisting of trajectories that move towards larger and smaller $C$, again consistent with the reversal in direction of spiraling in the bifurcated-wake regime\,(see Fig.\ref{fig:9}). The averaged trajectories in Figs.\ref{fig:10}a and b do not cross the separatrix. They only asymptote to $E = E_{sep}$ in the limit $\hat{t} \rightarrow \infty$ - the velocity of an averaged trajectory, ($\frac{dC}{d\hat{t}},\frac{dE}{d\hat{t}}$), goes to zero for $E \rightarrow E_{sep}$, owing to the divergence of the period of rotation with approach towards the separatrix; the nature of this divergence is analyzed in more detail in Appendix \ref{AppB}. This confinement of the averaged trajectories to the closed-streamline region is consistent with the breakdown of the averaged-equations-based description outside it.

\begin{figure}
\centering
 \includegraphics[trim = {0.1cm 0.3cm 0.5cm 0.5cm}, clip,scale = 0.41]{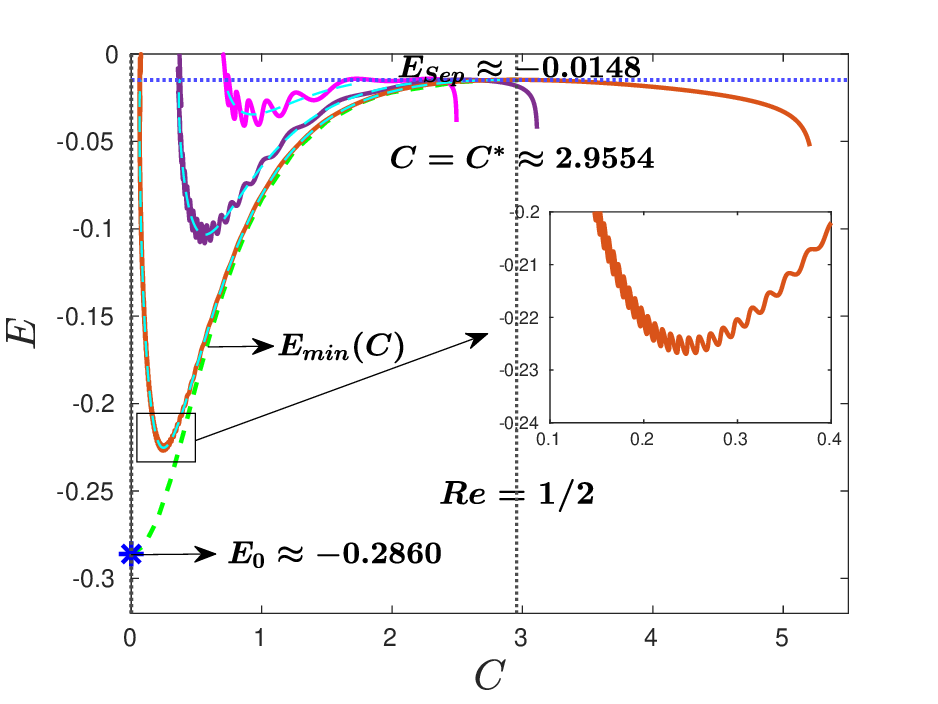}
 \hspace{0.01in}
 \includegraphics[trim = {0.1cm 0.3cm 0.5cm 0.5cm}, clip,scale = 0.41]{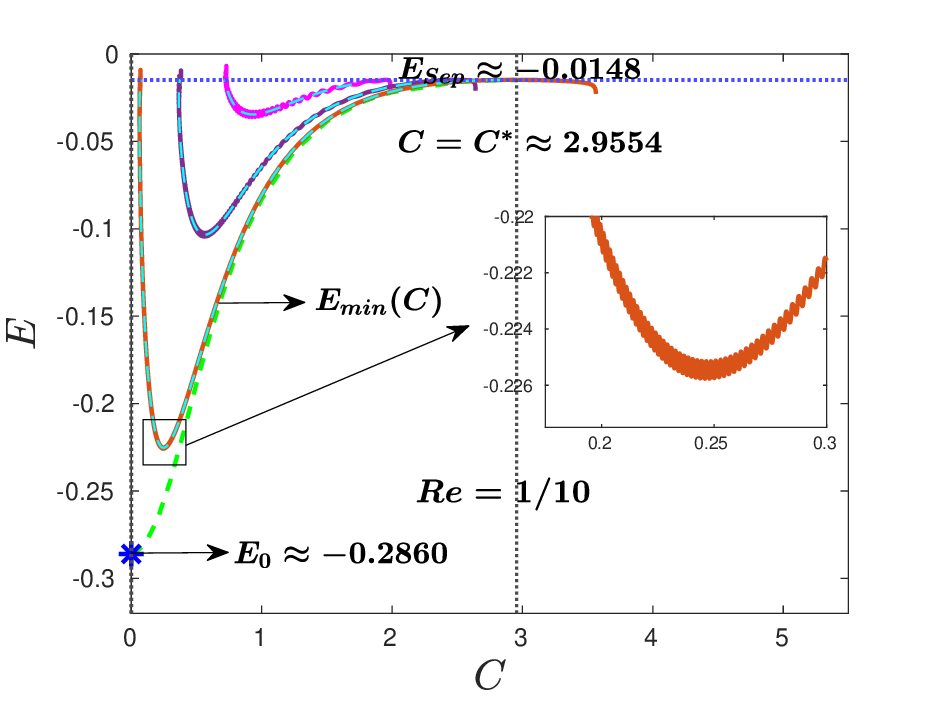}
 \includegraphics[trim = {0.1cm 0.3cm 0.5cm 0.5cm}, clip,scale = 0.41]{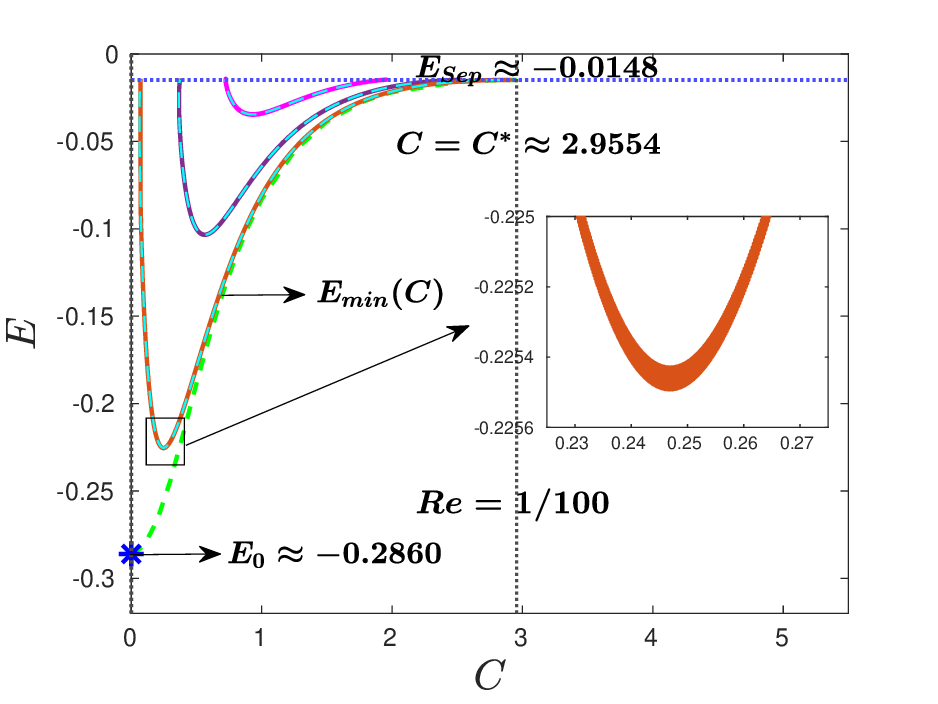}
\caption{Comparison of the averaged\,(dashed lines) and full-solution\,(solid lines) trajectories, on the $C-E$ plane, for $\hat{\alpha} = 10^{-3},\lambda = 1\,(> \lambda_{bif} \approx 0.35)$: (a) $Re = 0.5$, (b) $Re = 0.1$ and (c) $Re = 0.001$. The insets present a magnified view of the fast-oscillations present in the full-solution trajectories.}
\label{fig:12}
\end{figure}

Next, in Figs.\ref{fig:12}a-c, we compare the averaged trajectories in the $C-E$ plane, to the full-solution trajectories obtained from directly integrating\,(numerically) the inertial velocity field in (\ref{3.1}), for $\hat{\alpha} =10^{-3}$, $\lambda = 1$; the Cartesian coordinate representations of the latter trajectories, obtained via this integration, are converted to ones in the $C\!-\!E$ plane using (\ref{7}-\ref{8}). 
The full-solution trajectories are a function of $R\!e$, and exhibit a two-time-scale structure with fast oscillations superposed on a slow drift; only the latter drift is accounted for in the $R\!e$-independent leading order averaged approximation obtained above. While the amplitude of the fast oscillations scales as $O(Re)$, and therefore increases with increasing $Re$, there is nevertheless very good agreement between the averaged and full trajectories for all three $Re$'s examined. The amplitude of the aforementioned oscillations increases with an increase in $Re$, while the frequency decreases with approach towards the separatrix; the separation between the fast and slow scales decreases in both cases: due to a faster inertial drift in the former instance, and due to a longer period of circulation in the latter. Unlike the averaged trajectories, the full-solution trajectories in Figs.\ref{fig:12}a and b enter the closed-streamline region by crossing $E_{min}(C)$, and eventually exit it by crossing the separatrix $E = E_{sep}$. Further, these trajectories continue to move along vertical lines\,(that is, constant-$C$ lines) outside of the closed streamline region. Now, once an inertial streamline exits the closed-streamline domain, it heads off to infinity without further oscillations, similar to a Stokesian open streamline, and would therefore asymptote to a particular $C$ and $E$. This would imply that the corresponding trajectory in the $C-E$ plane must terminate at a finite point for $\hat{t} \rightarrow \infty$. The discrepancy between the expected and observed nature of the full-solution trajectories, arising from the continued vertical movement outside the closed-streamline region, is due to (\ref{3.2}) becoming invalid for distances greater than $O(Re^{-1/2})$. As already mentioned in the text following (\ref{3.2}), $\bm{u}^{(1)}$ is independent of $r$ for $r \rightarrow \infty$, and it is this lack of decay that manifests as a continuous change in $E$ in Figs.\ref{fig:12}a and b. 

\section{A weakly deformed drop in canonical hyperbolic linear flows - Effect of drop deformation} \label{Sec.4}
In this section, we analyse the streamline topology outside a weakly deformed drop in ambient planar hyperbolic flows, including the important limiting case of simple shear flow. As already mentioned in the introduction, drop deformation leads to interfacial forces that drive the drop towards sphericity even if the ambient shear were to be reversed. Therefore, drop deformation evidently breaks Stokesian reversibility, and similar to inertia, one expects an altered streamline topology for non-zero $Ca$. We shall restrict ourselves to illustrating this alteration for $Ca \ll 1$, $\lambda \sim O(1)$, and will only briefly discuss the case of large viscosity\,($\lambda \gg 1$, $\lambda Ca \sim O(1)$) in the conclusions. While the discussion of the streamline topology below is largely based on the $O(Ca)$ velocity field, in light of our findings contradicting the earlier efforts of \citet{Torza71}, \citet{Kennedy94} and \citet{Komrakova14}, we also present the results of boundary element simulations which agree well with analytical predictions. The exterior velocity field, to $O(Ca)$, is given by:
\begin{align}
    \bm{u} = \bm{u}^{(0)}(\bm{x}) + Ca\, \bm{u}^{(1)}(\bm{x}) + O(Ca^2), \label{smallCa:expansion}
\end{align}
where $\bm{u}^{(0)}$, as before, is given by (\ref{2}-\ref{4}), and $\bm{u}^{(1)}$ is given by\citep{Greco02}:
\begin{align}
\begin{split}
 \bm{u}^{(1)} = &c_1(r,\lambda) (\bm{E:}\bm{x}\bm{x})^2 \bm{x} + c_2(r) (\bm{E:}\bm{x}\bm{x}) \bm{E.}\bm{x} + c_3(r,\lambda) (\bm{E:E})\bm{x} + \\
 &c_4(r,\lambda) (\bm{E.E:}\bm{x}\bm{x})\bm{x} + c_5(r,\lambda) \bm{E.E.}\bm{x} + c_6(r,\lambda) (\bm{A}_2\bm{:x}\bm{x})\bm{x} +\\
 &c_7(r,\lambda) \bm{A}_2\bm{.x}, \label{4.25main}
 \end{split}
\end{align}
with $\bm{A}_2 = 2(\bm{E} \cdot \bm{\Omega} - \bm{\Omega} \cdot \bm{E}) + 4\bm{E} \cdot \bm{E}$ being the second Rivlin-Ericksen tensor; the $c_i(r,\lambda)$'s are defined in Appendix \ref{AppA}2. Note that (\ref{4.25main}) has essentially the same tensorial form as the inertial correction in (\ref{3.2}), as must be the case since both expressions have a regular character, being quadratic functionals of the ambient velocity gradient. Unlike the inertial correction, (\ref{4.25main}) does satisfy the far-field decay requirement, and therefore (\ref{smallCa:expansion}) remains valid everywhere in the exterior domain for sufficiently small $Ca$. 

\subsection{Exterior streamlines in physical space} \label{deformation:physical}

\begin{figure}
\centering
 \includegraphics[scale = 0.45]{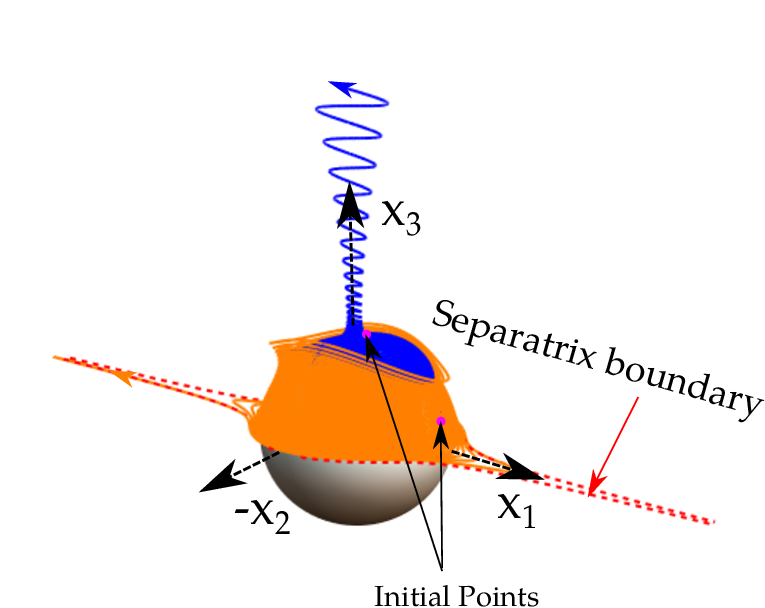}
 \hspace{0.2in}
 \includegraphics[trim= 1cm -1cm 0 1cm,scale = 0.45]{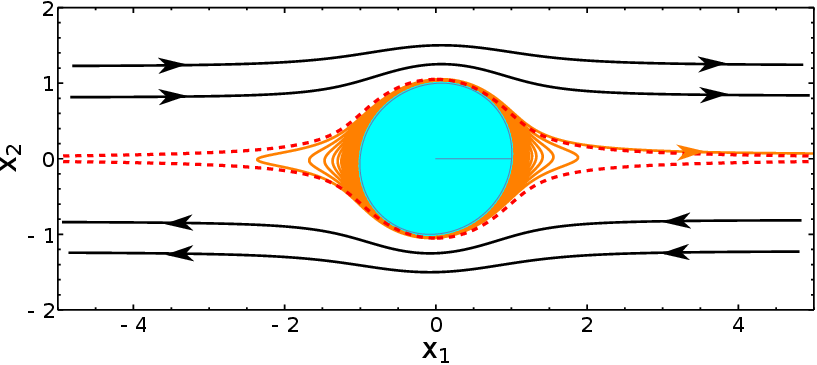}
 \includegraphics[trim= 0cm 0cm -1cm -1cm,scale = 0.4]{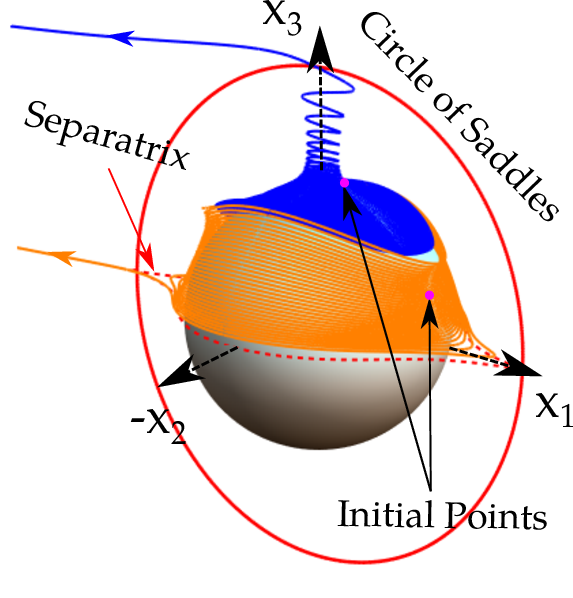}
 \hspace{0.2in}
 \includegraphics[trim= -1.75cm -2cm 0 1cm,scale = 0.45]{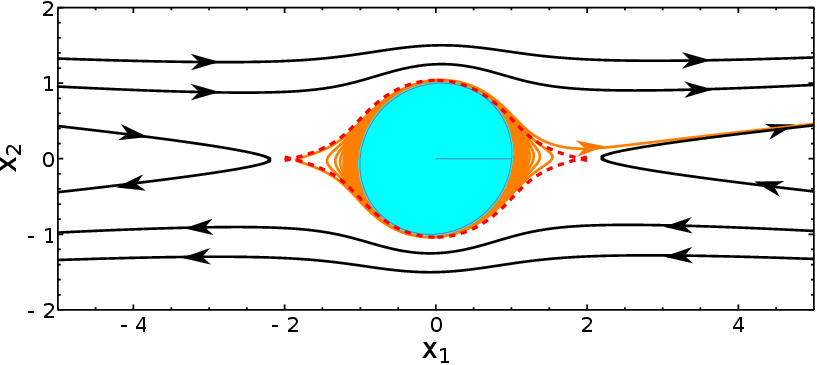}
 \includegraphics[trim= 2cm 0cm 0 0cm,scale = 0.4]{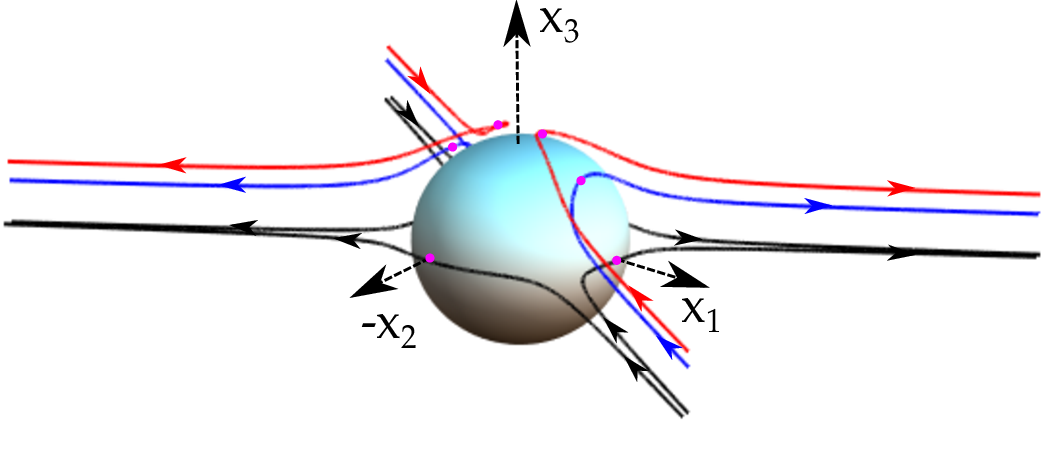}
 \includegraphics[trim= -1.5cm -0.5cm 0 1cm,scale = 0.345]{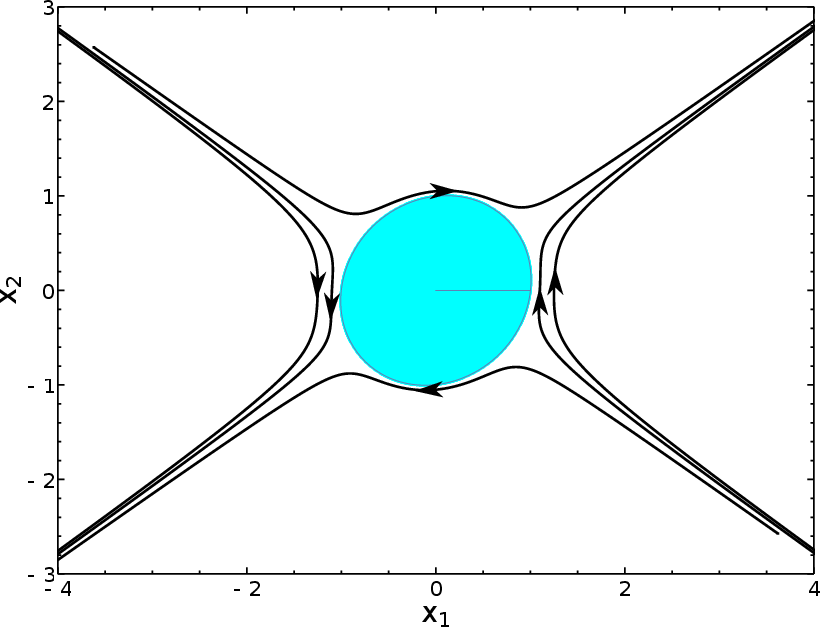}
\caption{Exterior streamlines around a deformed drop for $Ca = 0.0333\,(1/30), \lambda = 1$\,($\hat{\alpha}_c = 1/3$); (a) and (b) correspond to $\hat{\alpha} = 0$, (c) and (d) to $\hat{\alpha} = 10^{-2}$, and (e) and (f) to $\hat{\alpha} = 0.5\,(> \hat{\alpha}_c)$. Figures (a),(c) and (e) show the 3D streamlines, and figures (b), (d) and (f) show those in the $x_1-x_2$ plane. The magenta points mark the `initial' points from which one carries out forward and backward integrations in time..} \label{fig:Ca13O}
\end{figure}

We now characterize the finite-$Ca$ streamline topology arising from (\ref{smallCa:expansion}), proceeding in the same manner as for the inertial case in section \ref{Sec.3}, by fixing a $\lambda$ and plotting the streamlines for various $\hat{\alpha}$. Thus, in Figs.\ref{fig:Ca13O}a-f, we have chosen $\lambda = 1$, $Ca = 0.0333$, and have plotted the exterior streamlines for $\hat{\alpha} = 0$\,(Figs.\ref{fig:Ca13O}a and b), $10^{-2}$\,(Figs.\ref{fig:Ca13O}c and d) and $0.5$\,(Figs.\ref{fig:Ca13O}e and f); the final $\hat{\alpha}$ is greater than $\hat{\alpha}_c\,(=1/3)$, so there is no zero-$Ca$ closed streamline region in this case. From Figs.\ref{fig:Ca13O}a and c, the streamlines are seen to come in from infinity towards the drop surface, along an intermediate direction, and then spiral off to infinity either along the $x_3$-axis\,(the blue streamlines), or as they approach the $x_1-x_2$ plane\,(the orange streamlines); the latter being consistent with the (outward)\,direction of spiraling within this plane, shown in Figs.\ref{fig:Ca13O}b and d. Thus, the spiraling streamlines appear to be time-reversed versions of those seen in Fig.\ref{fig:9} for the inertial case in the bifurcated-wake regime. The zero-$Ca$ separatrix surface again organizes the finite-$Ca$ streamlines into three groups: (i) fore-aft asymmetric\,(about the $x_2$-axis) open streamlines that go from upstream to downstream infinity in the same sense as the ambient ones; (ii) streamlines that spiral around the drop, heading off to infinity either along the $x_3$ axis, or sufficiently close to the $x_1-x_2$ plane, and that resemble the time-reversed versions of the inertial spiraling streamlines in the bifurcated-wake regime; and (iii) for $\hat{\alpha} \neq 0$, with a zero-$Ca$ Stokesian separatrix that is finite in extent in the $x_1-x_3$-plane, the spiraling streamlines above are contained between regions of fore-aft asymmetric\,(about the $x_1$-axis) open reversing streamlines. Note that, for $\hat{\alpha} > \hat{\alpha}_c$, the absence of a separatrix surface for $Ca = 0$ implies that all finite-$Ca$ streamlines fall into categories (i) and (iii); see Figs.\ref{fig:Ca13O}e and f. 
\begin{figure}
\centering
 \includegraphics[scale = 0.4]{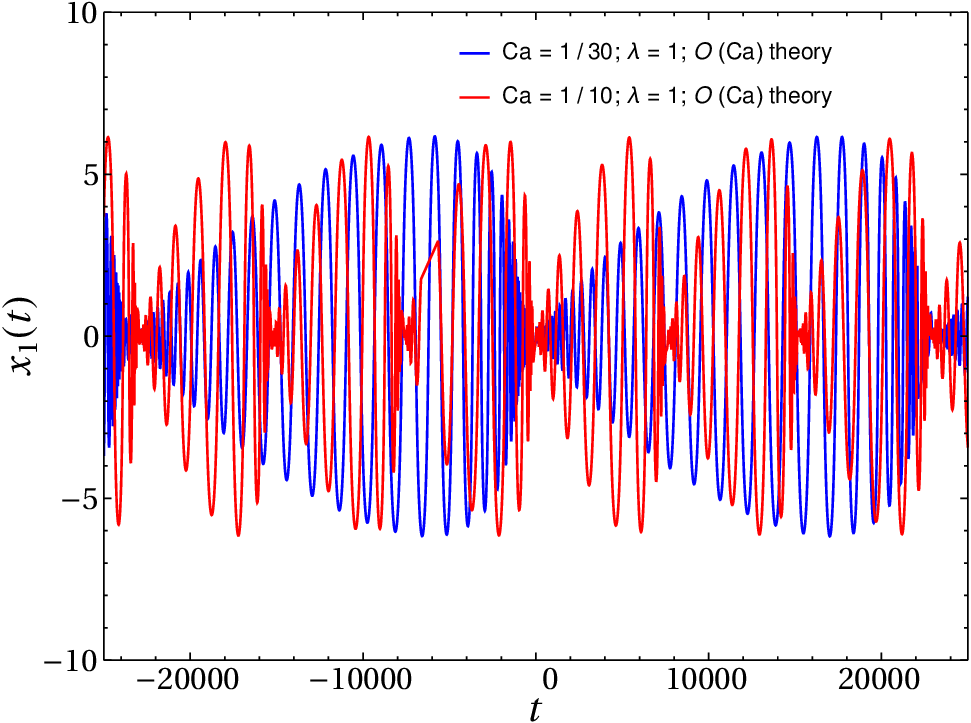}
 \includegraphics[scale = 0.4]{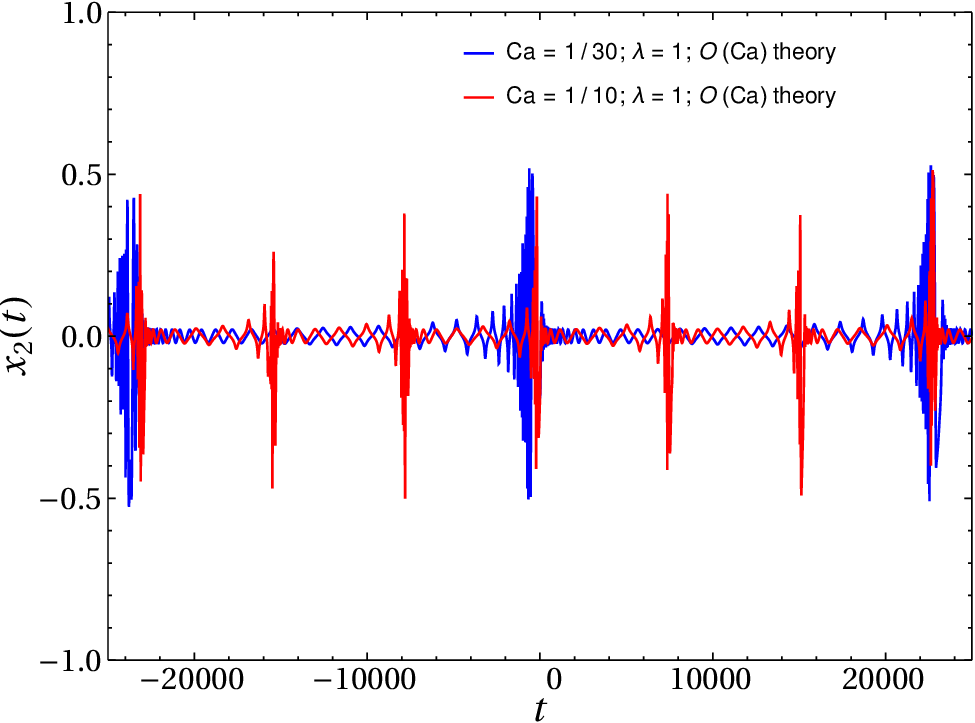}
 \includegraphics[scale = 0.4]{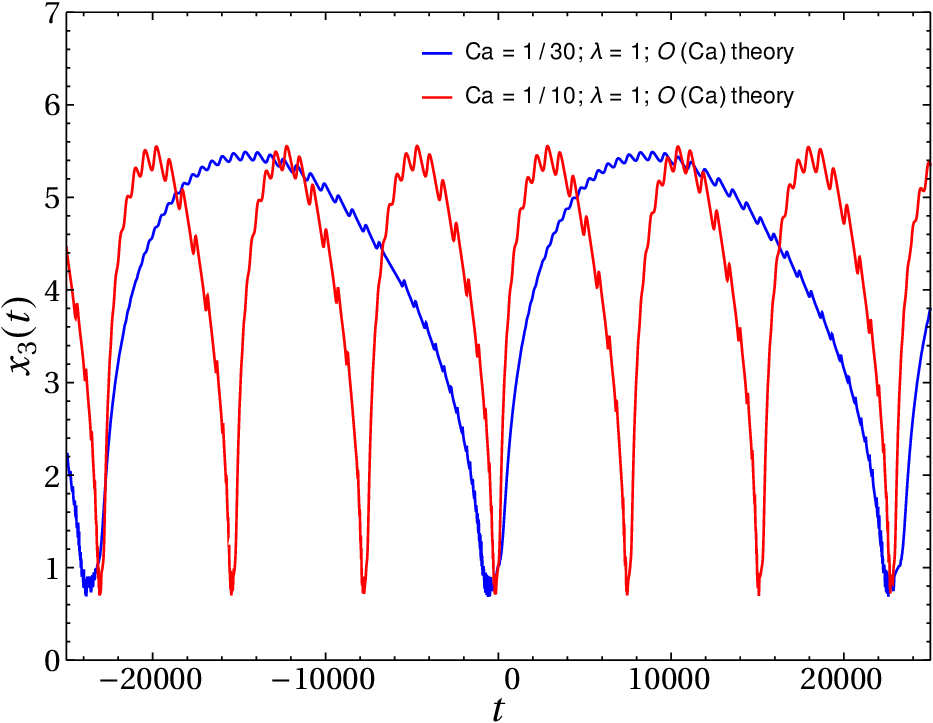}
\caption{The Cartesian coordinates of finite-$Ca$ spiralling streamlines, that spiral out along the vorticity axis for $t \approx 0$,  plotted over  times longer than $O(Ca^{-1}\dot{\gamma}^{-1})$; $\hat{\alpha} = 0$, $\lambda = 1$, with $Ca = 0.0333\,(1/30)$\,(blue)  and $Ca = 0.1$\,(red). A short-time depiction of the streamline, for $Ca = 0.0333$, appears\,(in blue) in Fig.\ref{fig:Ca13O}a. For both $Ca$, the streamlines exhibit a quasi-periodic behavior arising from dense winding around an invariant torus.}
\label{fig:15}
\end{figure}

\begin{figure}
\centering
 \includegraphics[scale = 0.35]{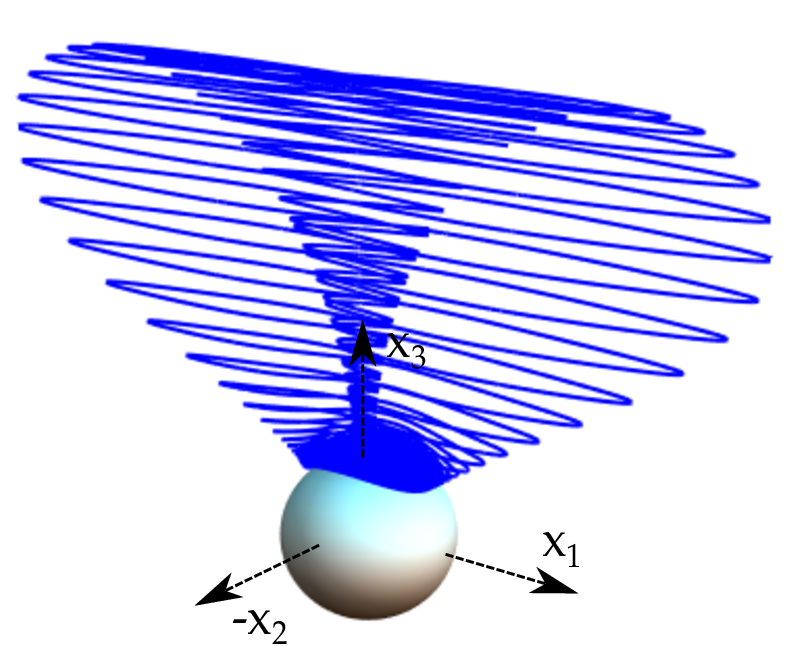}
 \includegraphics[scale = 0.35]{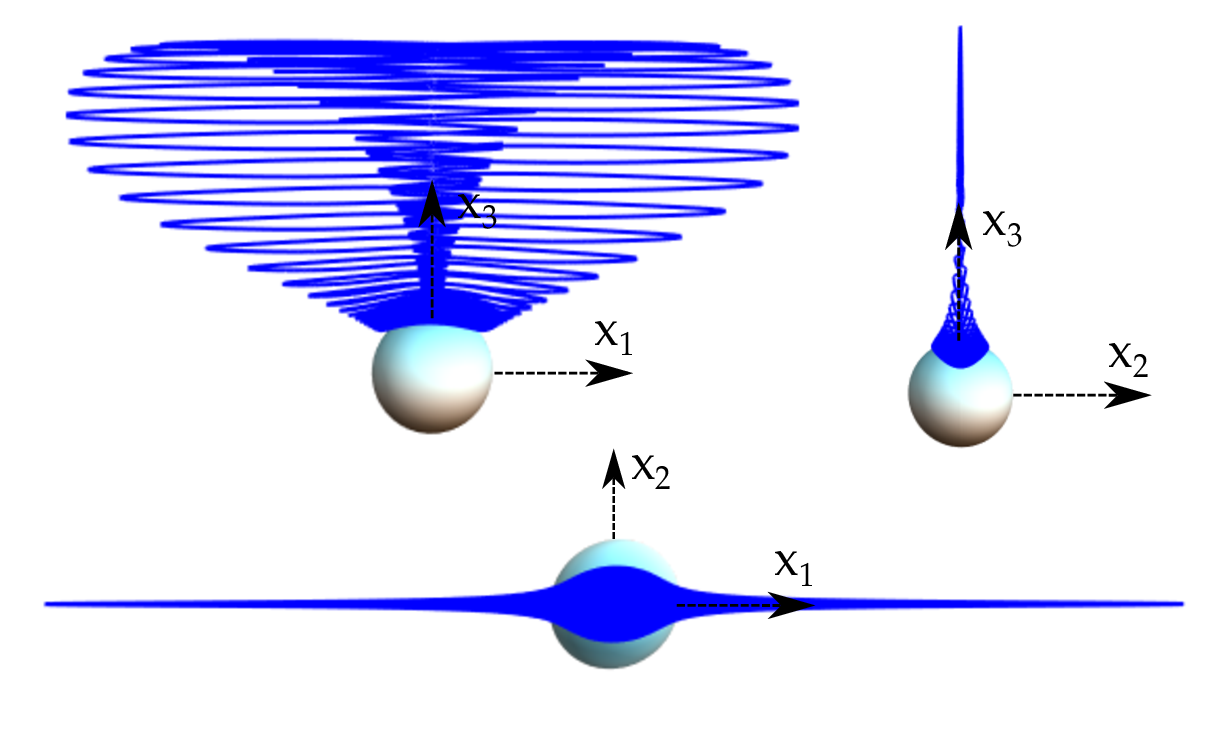} 
\caption{(a) Depiction of the dense winding of a finite-$Ca$ spiralling streamline\,($Ca = 0.0333$), around an invariant torus; this streamline was shown\,(in blue) earlier in Fig.\ref{fig:Ca13O}a, for shorter times, where it had started to spiral out along the $x_3$-axis. Plots (b-d) show projections of the streamline onto the three Cartesian planes.}
\label{fig:13N}
\end{figure}

The description of the streamline topology given above is incomplete, however. Not all finite-$Ca$ streamlines, that spiral out along the $x_3$-axis, head off to infinity. For $\hat{\alpha} = 0$, on integrating for longer times, rather surprisingly, a streamline that initially spirals out along the $x_3$-axis is found to turn around and come back towards the $x_1-x_2$ plane, tracing out a torus in the process, and eventually ends up winding densely on this invariant torus. 
This unexpected doubly-periodic behavior of a finite-$Ca$ streamline, with widely separated periods of $O(\dot{\gamma}^{-1})$ and $O(\dot{\gamma}^{-1}Ca^{-1})$, is illustrated in Fig.\ref{fig:15} via plots of all three Cartesian coordinates as functions of time for $Ca = 0.0333$ and $Ca = 0.1$. Next, Fig.\ref{fig:13N}a presents a 3D view of a particular invariant torus for $\hat{\alpha} = 0$; projections of this invariant torus onto the principal Cartesian planes are shown in Fig.\ref{fig:13N}b-d. The torus is seen to be highly distorted with its largest span in the $x_1-x_3$ plane, and being far narrower in the other two planes. The nature of this distortion is the same for all invariant tori, and is consistent with the shape of the zero-$Ca$ separatrix surface for simple shear flow.

\begin{figure}
\centering
\includegraphics[scale = 0.45]{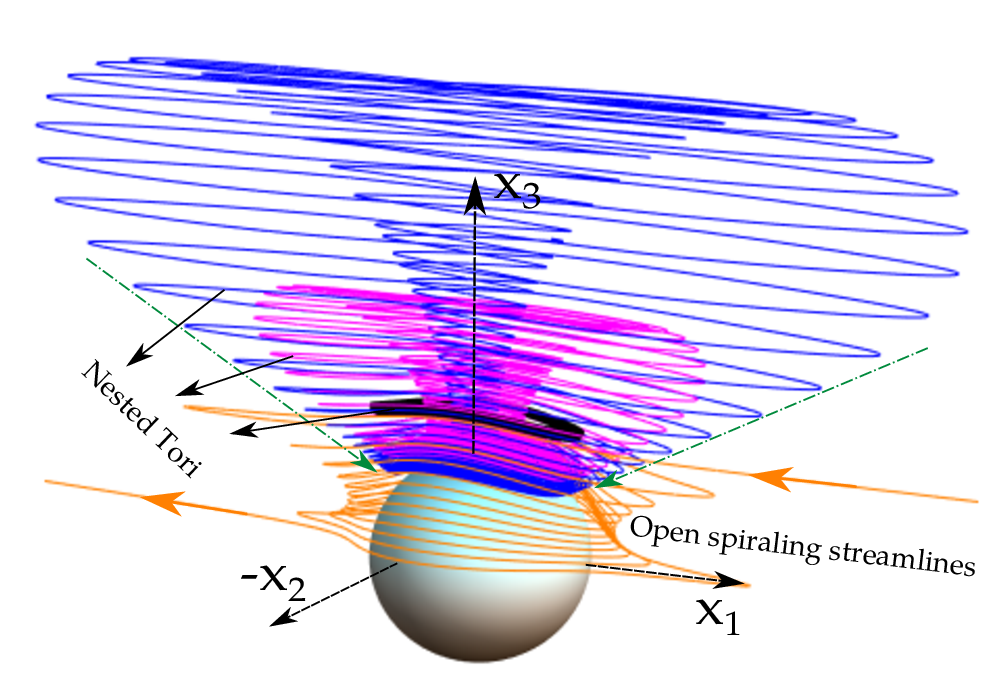}
\includegraphics[scale = 0.35]{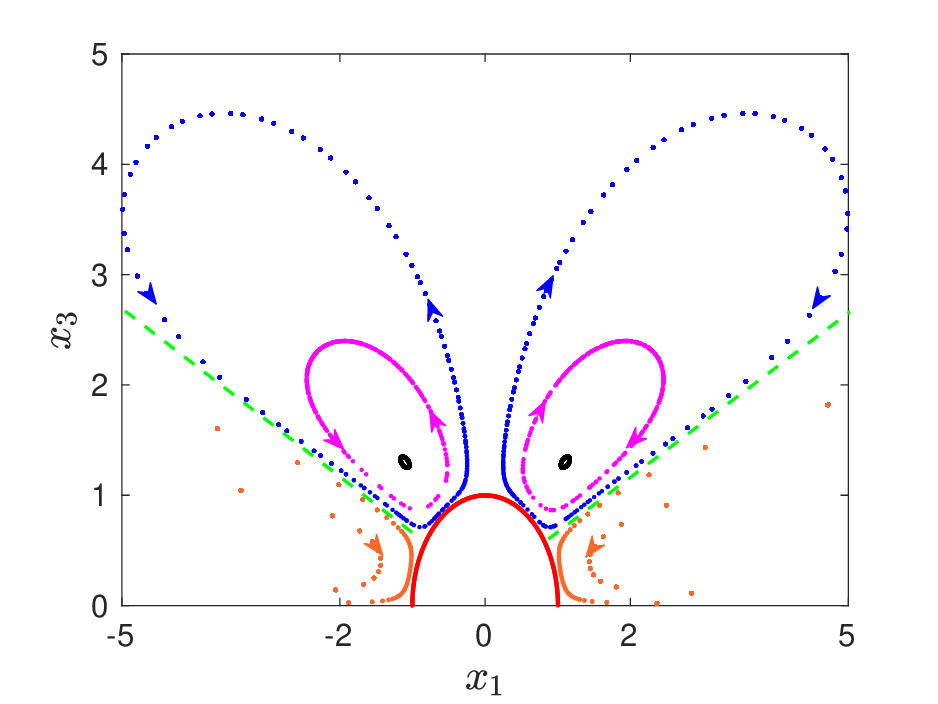}
\caption{(a) The exterior streamline topology for $\hat{\alpha} = 0$, $\lambda = 1$ and $Ca = 0.0333$: Three of the tori, including the innermost singular curve\,(in black), that form part of the nested tori structure for non-zero $Ca$, are highlighted; the orange streamline spirals in towards the drop from infinity, while remaining outside of the nested tori configuration, eventually escaping to infinity in the vicinity of the flow-gradient\,($x_1\!-\!x_2$) plane. (b) Projections of the three invariant tori, and of open streamlines spiralling towards the flow-gradient plane, onto the flow-vorticity\,($x_1\!-\!x_3$) plane. The projections have been generated by plotting the points of intersection of the finite-$Ca$ streamlines, with this plane; the projection of the drop appears as a red semi-circle.}
\label{fig:14}
\end{figure}

Fig.\ref{fig:14}a clarifies the global streamline topology for $\hat{\alpha} = 0, \lambda = 1$ and $Ca = 0.0333$ via the simultaneous depiction of multiple finite-$Ca$ streamlines, which helps highlight the configuration of nested invariant tori that arises for small but finite $Ca$. The outermost torus extends to infinity in the $x_1\!-\!x_3$ plane, its spatial extent in this plane being indicated by the pair of dash-dotted lines\,(in green). In contrast, the innermost torus is a closed curve looping around the $x_3$-axis, and is, in fact, one of the closed streamlines\,(perturbed by $O(Ca)$) around the original undeformed drop. The projections of spiralling streamlines on the nested tori, generated via points of intersection of each streamline with the $x_1\!-\!x_3$ plane, appear in Fig.\ref{fig:14}b. Since each of these streamlines winds indefinitely on an invariant torus, the points of intersection, taken over sufficiently many windings, would densely fill out a curve for sufficiently long times. Based on its projection, the outermost torus is seen to only occupy a fraction of the deformed drop surface\,(that appears as a red semi-circle). The remainder of the drop is covered by streamlines that spiral in towards the drop on the outside of the outermost torus, and then along the drop surface towards the $x_1\!-\!x_2$ plane, before moving off to infinity at a location closer to this plane. An example of this streamline is the orange curve in Fig.\ref{fig:14}a, whose points of intersection with the $x_1\!-\!x_3$ plane have been indicated\,(again in orange) in Fig.\ref{fig:14}b. In contrast to the densely winding streamlines above, the points are much more sparsely distributed on account of each open streamline only spending a finite time in the neighborhood of the drop.

\begin{figure}
\centering
\includegraphics[trim = {1cm 2cm 0.5cm 1cm}, clip,scale = 0.5]{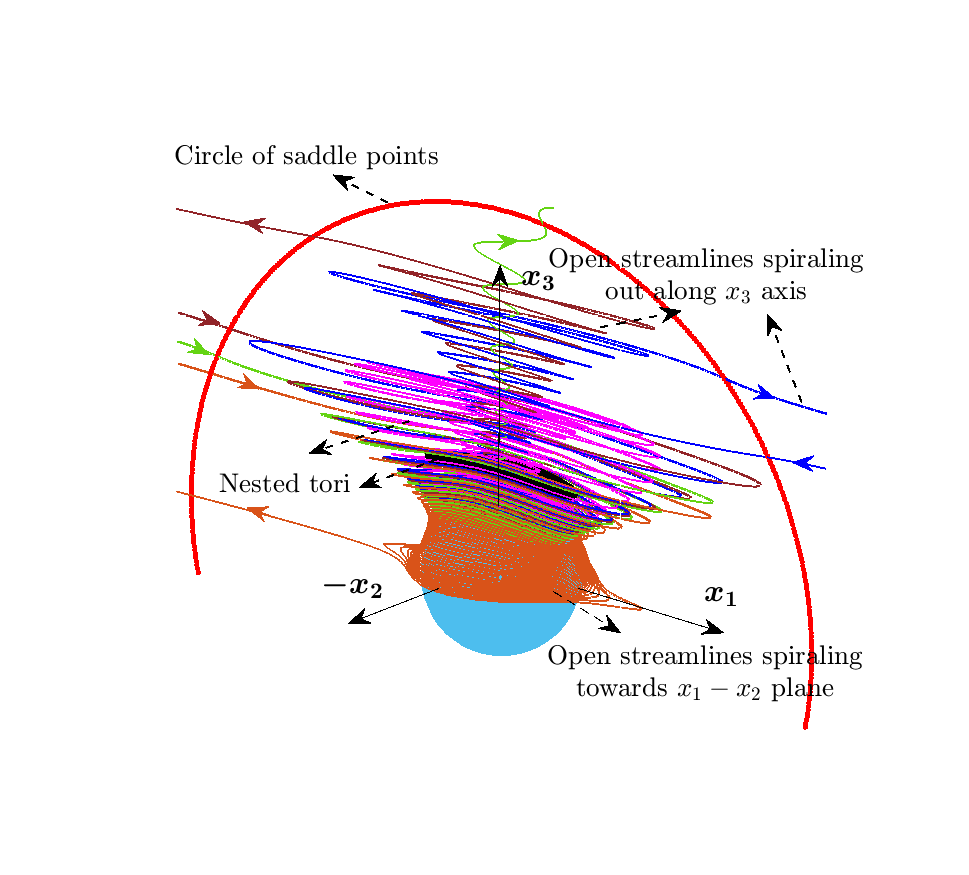}
\includegraphics[scale = 0.37]{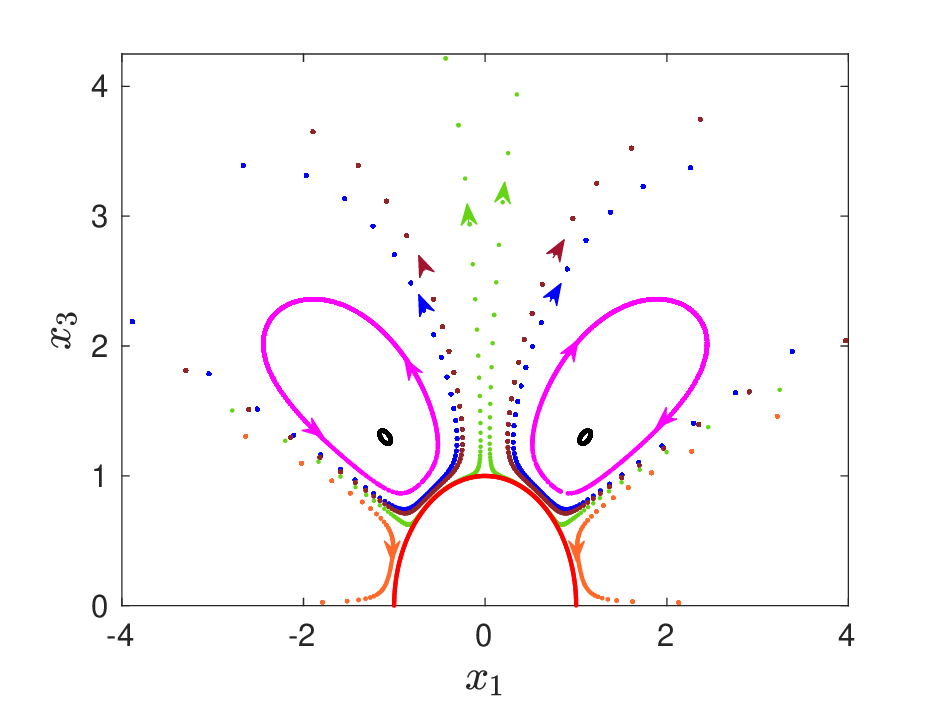}
\caption{(a) The exterior streamline topology for $\hat{\alpha} = 10^{-4}$, $\lambda = 1$ and $Ca = 0.0333$; the finite separatrix surface intersects the $x_1 - x_3$ plane in the red circle shown. The pink streamline winds around on an invariant torus, with the black curve being a good approximation of the innermost singular torus. The remaining streamlines spiral in towards the drop surface, and then either towards the $x_1-x_2$-plane\,(orange curve), or out along the vorticity axis\,(green, brown and black curves). (b) depicts the points of intersection of the curves in (a) with the $x_1-x_3$ plane. Similar to Fig.13b, the tori only occupy a fraction of the drop surface.}
\label{fig:16}
\end{figure}

The finite-$Ca$ streamline topology for a non-zero $\hat{\alpha}\,(=10^{-4})$, again for $\lambda = 1$, $Ca = 0.0333$, is depicted in Fig.\ref{fig:16}a. The structure remains qualitatively similar to that for $\hat{\alpha} = 0$, in that there still exists a configuration of nested invariant tori\,(the pink and black streamlines). However, the finite extent of the original zero-$Ca$ separatrix surface, and the saddle points at its rim\,(along a circle in the $x_1-x_3$ plane), imply that the outermost torus is no longer infinite in extent. As a result, not all finite-$Ca$ streamlines, that begin spiraling outward along the $x_3$-axis, turn back and continue to wind around invariant tori for all time. A subset of these streamlines, that start sufficiently close to the $x_3$ axis\,(this neighborhood gets progressively smaller for decreasing $\hat{\alpha}$, becoming vanishingly small for $\hat{\alpha} \rightarrow 0$), approach the aforementioned fixed-point circle sufficiently closely when starting to spiral back towards the $x_1-x_2$ plane. This close approach causes the streamlines to exit the spiraling zone at some point, and head off to downstream infinity similar to an open streamline; the time reversed version of this argument implies that such streamlines should also have come into the spiralling zone, from infinity, at a finite time in the past. The green, brown and blue streamlines in Figs.\ref{fig:16}a and b exhibit this behavior. Thus, there emerges a new class of streamlines\,(relative to those that exist for $\hat{\alpha} = 0$) that appear to spiral around on an invariant torus, but eventually transform to open streamlines upon close approach to the saddle-point circle. Related to this is the emergence of a `gap' between the outermost torus and the deformed drop surface, for non-zero $\hat{\alpha}$, as highlighted via the nested-tori projections in Fig.\ref{fig:16}b. The existence of this gap implies that, unlike  $\hat{\alpha} = 0$, streamlines that come in from infinity, and reach the immediate vicinity of the drop surface, can now spiral out to infinity in one of two ways: either by spiraling towards the  $x_1-x_2$ plane, and then to downstream infinity, similar to $\hat{\alpha} = 0$\,(the orange streamline in Fig.\ref{fig:16}a); or, by spiraling away from this plane, through the aforementioned gap, and then continuing to spiral on the outside of the nested-tori configuration, before escaping to infinity due to a close approach towards the fixed-point circle\,(the green, brown and blue streamlines mentioned above).

In summary, drop deformation does destroy the region of closed streamlines for any non-zero $Ca$. The resulting finite-$Ca$ streamline topology appears more complex than its finite-$Re$ counterpart\,(in either the single or bifurcated wake regimes). This complexity is highlighted by the existence of finite-$Ca$ streamlines that wind densely around a configuration of nested invariant tori, this configuration being embedded within a region of spiraling streamlines. The presence of such a structure is counter-intuitive, given its structural instability\,(in the sense of being susceptible to small perturbations) from the dynamical systems perspective. 

\subsection{$C-E$ representation of the exterior streamlines} \label{deformation:CE}

\begin{figure}
\centerline{\includegraphics[trim = {0.45cm 0.1cm 1cm 1cm}, clip, scale = 0.385]{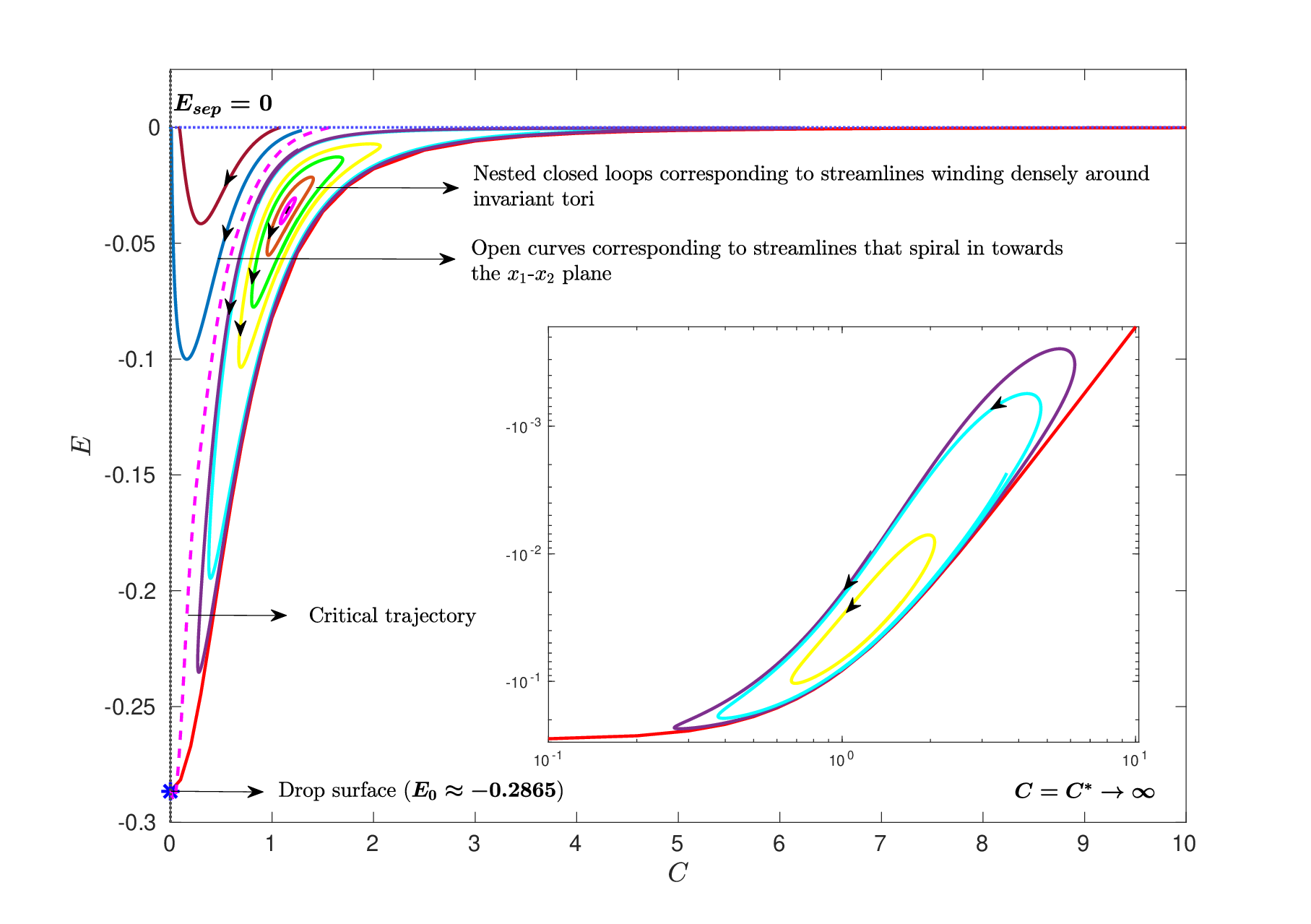}}
\caption{The $C-E$ plane representations of streamlines for small but finite $C\!a$; $\hat{\alpha} = 0$, $\lambda = 1\,(> \lambda_{c}= 1/3)$. The dashed pink curve is an approximate depiction of the critical trajectory that separates streamlines spiraling towards the flow-gradient plane\,(and then, to infinity) from those that wind densely around nested invariant tori.}
\label{fig:17}
\end{figure}

The representations of the finite-$Ca$ streamlines on the $C-E$ plane, at leading order, may again be obtained by solving a system of ODEs, derived using the method of averaging in a manner analogous to the inertial case. Thus, the governing equations are formally identical to (\ref{3.20}-\ref{3.21}), except that $u_r^{(1)}$, $u_{\theta}^{(1)}$ and $u_{\phi}^{(1)}$ are now obtained from (\ref{4.25main}); the slow time variable in these equations given by $\hat{t}=Ca\,t$. 
Fig.\ref{fig:17} depicts the averaged trajectories on the $C-E$ plane for $\hat{\alpha} = 0, \lambda = 1$. The zero-$C\!a$ closed-streamline region clearly consists of two distinct parts. The trajectories on the right form a system of nested closed curves which extends right up to the separatrix\,($E_{sep} = 0$) - see inset. These closed curves correspond to the configuration of nested invariant tori in Fig.\ref{fig:14}. Note that each closed curve corresponds to the entire set of streamlines winding around a given torus. In contrast, the trajectories on the left are open; they start from the separatrix, extend downwards to lower $E$ and smaller $C$, before turning around and proceeding in the direction of increasing $E$, again ending up on the separatrix in the limit of infinite time. They resemble time-reversed versions of inertial trajectories, in the analogous region of the $C-E$ plane, for the bifurcated-wake regime\,(see Fig.\ref{fig:10}b). In Fig.\ref{fig:14}, these trajectories correspond to open streamlines outside of the invariant tori, that spiral in towards the $x_1-x_2$ plane, before going off to infinity. The dashed pink curve in Fig.\ref{fig:17} is an approximate depiction of the critical trajectory that demarcates the two classes of trajectories above, and extends from the separatrix to the drop surface\,($E = E_0, C = 0$). In fact, the outermost closed curve may be interpreted as a composite one, consisting of the aforementioned critical curve and the red curve $E_{min}(C)$, connected at the terminal points ($E = E_0, C = 0$); corresponding to the drop, and ($E =E_{sep},C\rightarrow \infty$); corresponding to the point at $x_3 = \infty$ on the separatrix surface. In physical space, this corresponds to the outermost torus being interpreted as a composite of the part of the separatrix surface whose projection\,(onto the $x_1 -x_3$ plane) is the pair of green dash-dotted lines in Fig.\ref{fig:14}b, the enclosed portion of the drop surface\,(with projection being the portion of the red curve between the pair of green lines), and the vorticity axis.

\begin{figure}
\centerline{\includegraphics[trim = {0.45cm 0.8cm 1cm 1cm}, clip, scale = 0.385]{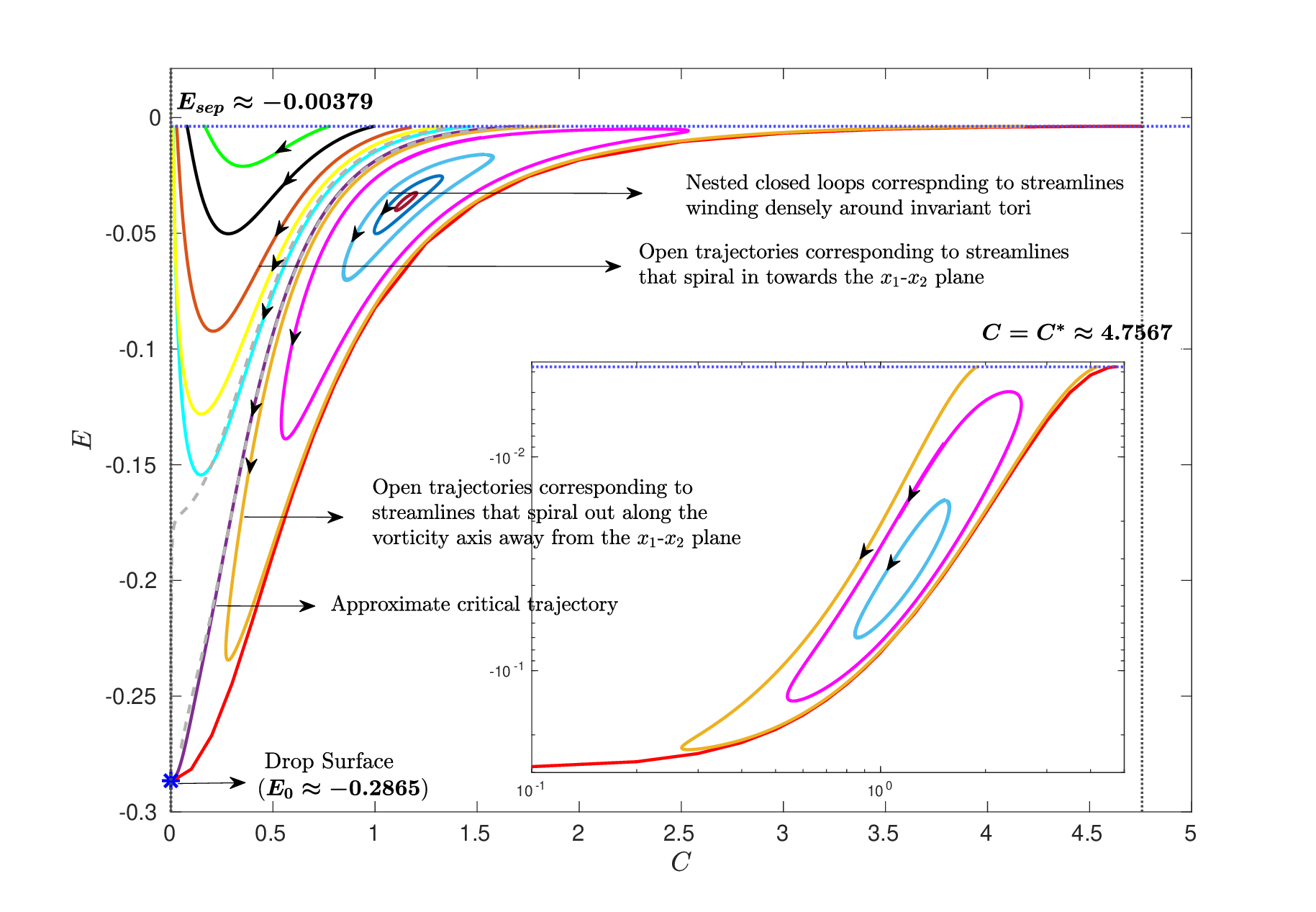}}
\caption{The $C-E$ plane representation, for small but finite $C\!a$, for $\hat{\alpha} = 10^{-4}$, $\lambda = 1\,(> \lambda_{c})$. The nearly coincident dashed gray and violet curves are approximations\,(with differing resolutions: $N_G = 300$ and $2000$) of the critical trajectory that demarcates two groups of streamlines: the first group consists of those that spiral towards the $x_1-x_2$ plane\, and then to infinity); the second group includes those that wind densely around nested invariant tori, and those that spiral in towards the drop and then out to infinity along the $x_3$-axis. The dashed gray\,($N_G = 300$) and cyan\,($N_G =2000$) curves correspond to differing-resolution depictions of an open trajectory in the neighborhood of the critical trajectory.}
\label{fig:18}
\end{figure}

Fig.\ref{fig:18} depicts the averaged trajectories for $\hat{\alpha} = 10^{-4}, \lambda = 1$. Not all trajectories on the right of the critical curve\,(shown in magenta and gray-dashed) are now closed. While the closed trajectories still correspond to invariant tori in physical space, there are open trajectories running around the configuration of nested closed curves, and that start and end on the separatrix\,($E_{sep} = -0.00379$) - an example being the orange curve shown both in the main plot and the inset. These correspond to the streamlines in Fig.\ref{fig:16} that spiral in towards the drop outside of the invariant tori, and then to infinity along the $x_3$-axis. Unlike in Fig.\ref{fig:17}, the outermost torus, corresponding to the largest closed curve, must now be finite in extent, and must separate the aforementioned groups of trajectories. Further, it must touch the separatrix at a finite $C$, with finite-time uniqueness constraints implying that this point be singular\,(for instance, a corner). Our attempts at identifying the nature of this singular curve have, however, been frustrated by convergence difficulties in the vicinity of the separatrix, which are discussed in Appendix \ref{AppD}. While trajectories on the left of the critical curve are still all open, and therefore must start and end on the separatrix similar to Fig.\ref{fig:17}, convergence difficulties beset the ones
immediately to the left of the critical curve. For a given finite numerical resolution, there appears to be a \,(LHS)\,neighborhood of the critical curve where trajectories, rather than turning around and heading up towards the separatrix, head down towards the drop instead; the neighborhood shrinks with increasing resolution,  as illustrated by the differing character of the cyan and dashed-gray curves in Fig.\ref{fig:18} which are trajectories passing through a given initial point corresponding to different numerical resolutions. This convergence difficulty is absent for the inertial case, and therefore, appears to be an artifact of drop deformation.

\begin{figure}
\includegraphics[scale = 0.35]{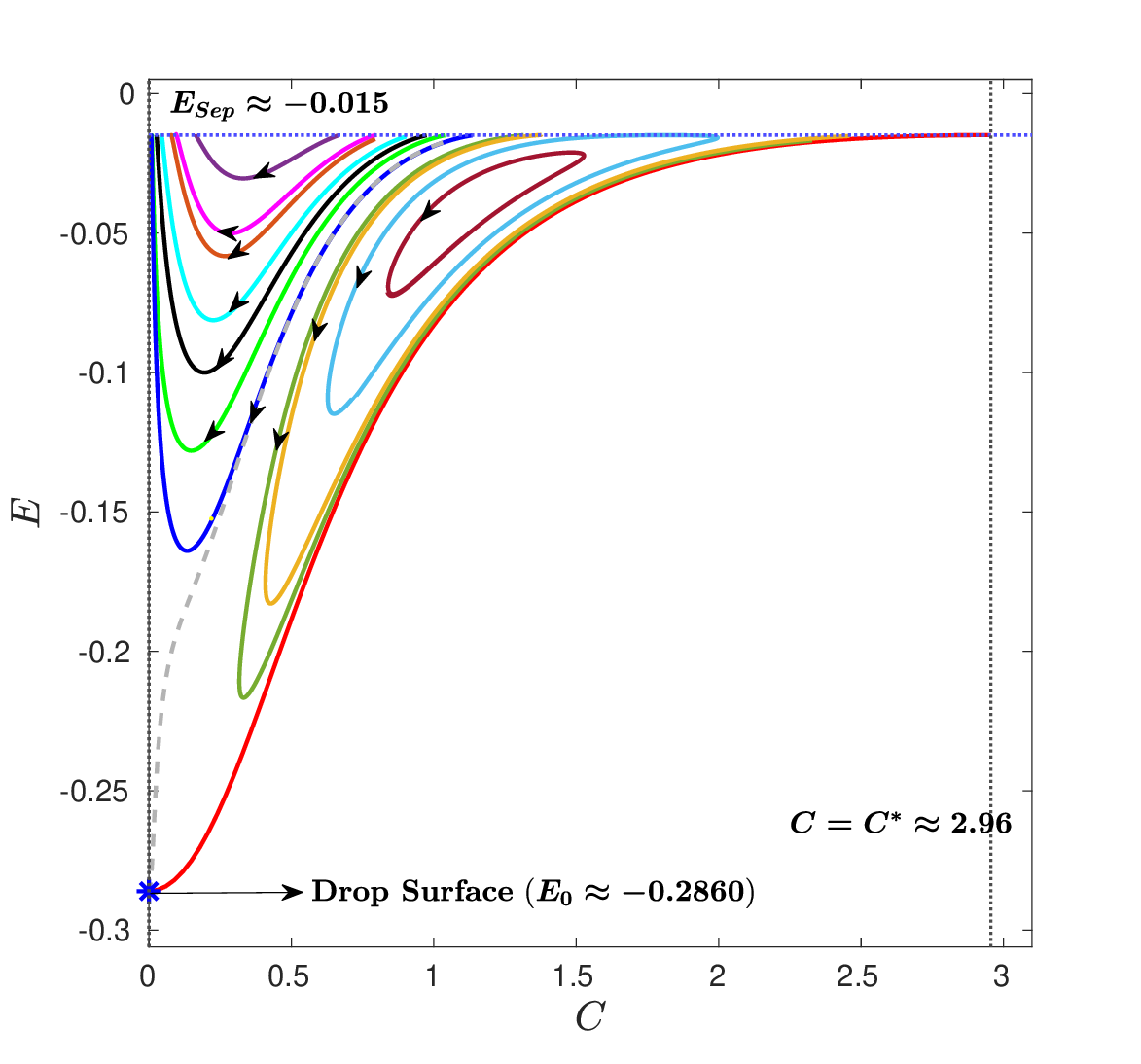}
\includegraphics[scale = 0.35]{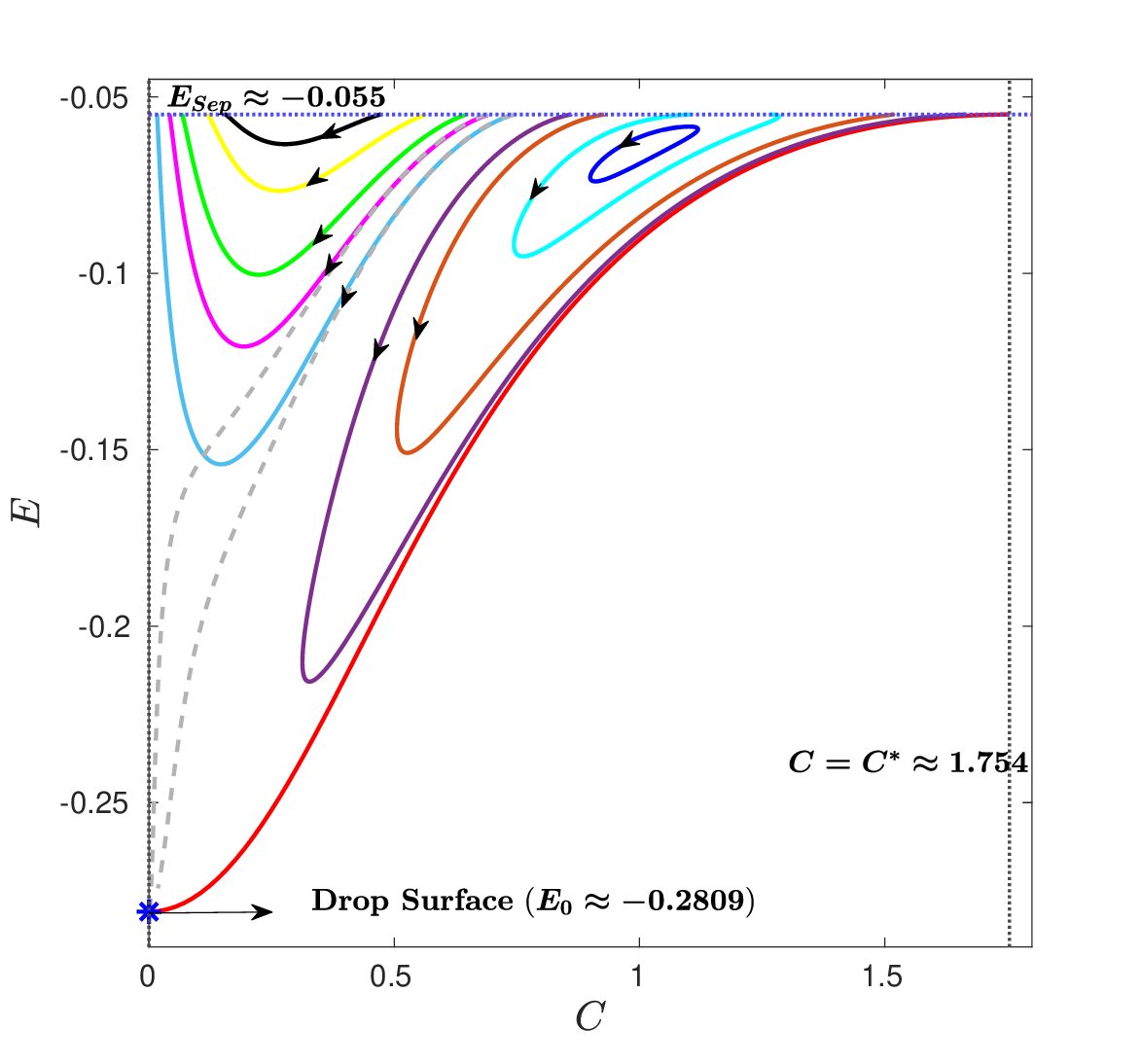}
\includegraphics[scale = 0.35]{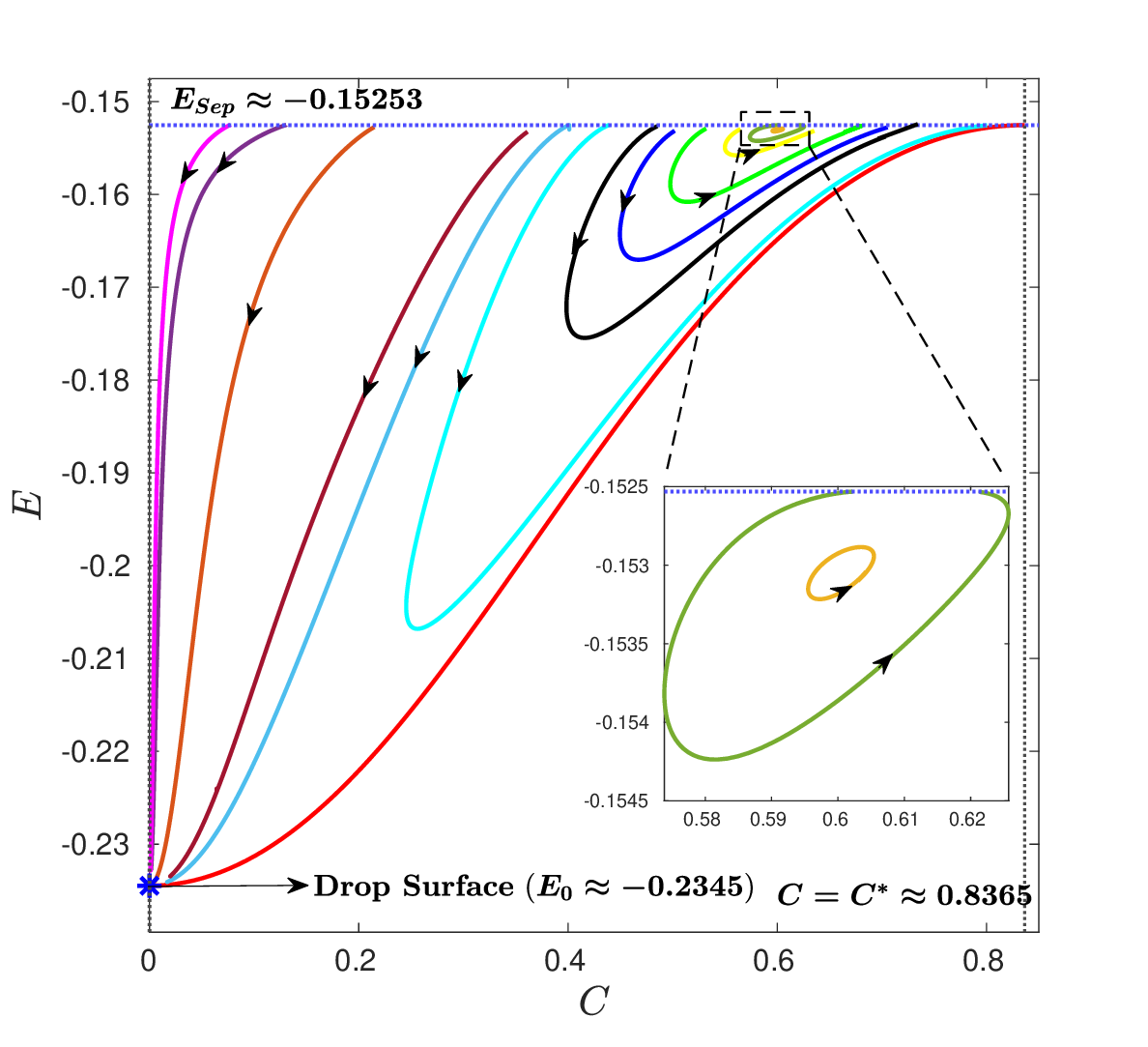}
\includegraphics[scale = 0.35]{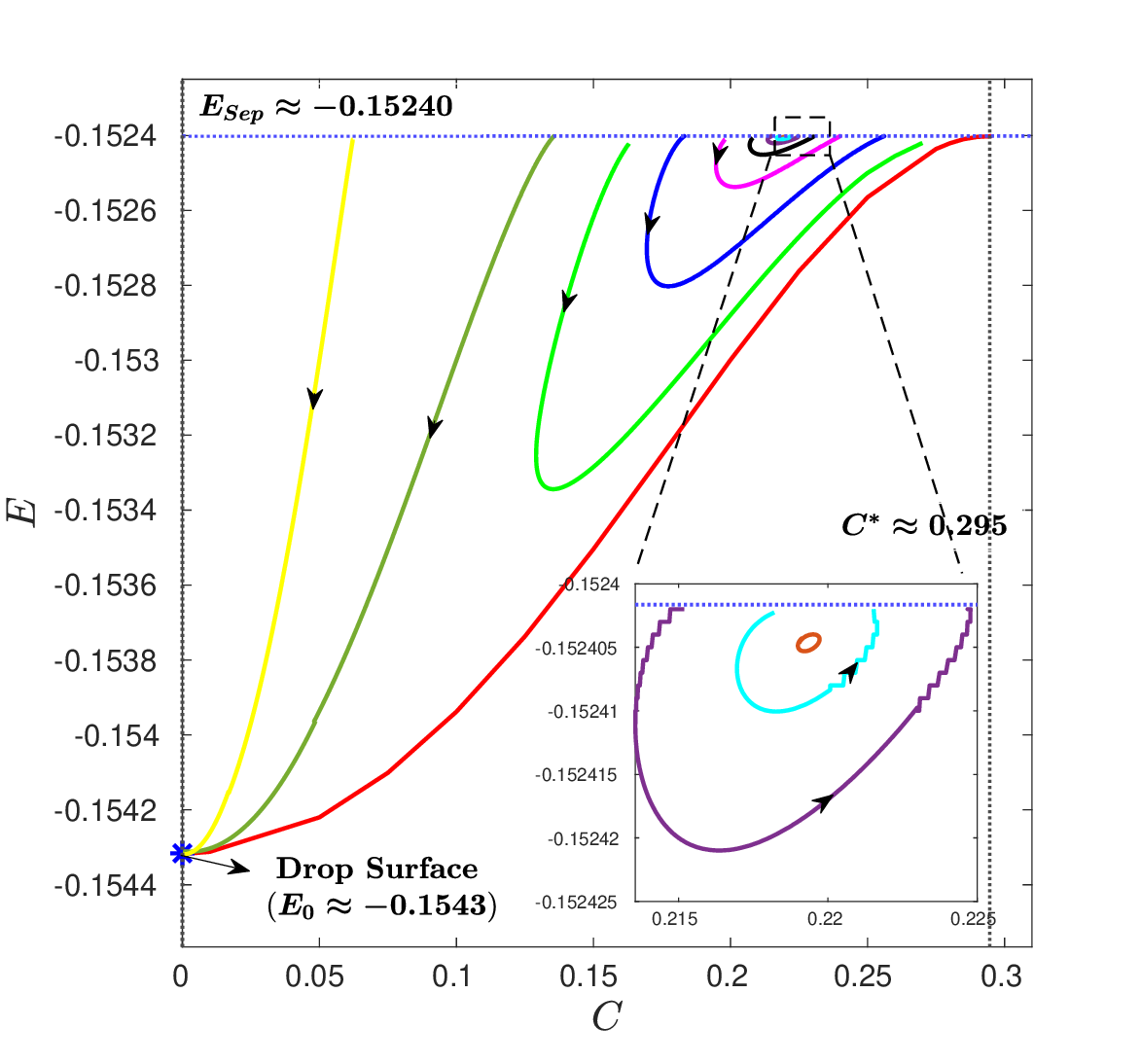}
\caption{The $C-E$ plane representations for (a) $\hat{\alpha} = 10^{-3}$, (b) $\hat{\alpha} = 10^{-2}$, (c) $\hat{\alpha} = 10^{-1}$ and (d) $\hat{\alpha} = 0.3$; $\lambda = 1$, $\hat{\alpha}_c = \frac{1}{3}$. 
 A system of nested closed curves exists up until $\hat{\alpha} = \hat{\alpha}_c$, as indicated by insets in (c) and (d); although, this system decreases in size for $\hat{\alpha} \rightarrow \hat{\alpha}_c$, while also approaching the separatrix. The dashed gray curves in all subfigures correspond to lower-resolution trajectories\,($N_G = 300$) that (erroneously)\,head towards the drop, as opposed to the separatrix; increasing the resolution\,(to $N_G = 2000$) restricts this error to an increasingly small neighborhood of the critical curve.}
\label{fig:18A}
\end{figure}

Figs.\ref{fig:18A}a-d show the $C-E$ plane representations for larger $\hat{\alpha}$, up until $\hat{\alpha} = 0.3$ which is just less than $\hat{\alpha}_c = \frac{1}{3}$. Despite decreasing in size with increasing $\hat{\alpha}$, the configuration of nested closed curves on the right appears to persist for $\hat{\alpha} \rightarrow \hat{\alpha}_c^-$; this persistence being illustrated by the insets in Figs.\ref{fig:18A}c and d. The reduction in size implies that a greater fraction of trajectories on the right are open ones that start\,(at a lower $C$) and end\,(at a higher $C$) on the separatrix. For a fixed resolution, the drop-deformation-induced convergence artifact leads to an increasing fraction of trajectories on the left heading towards the drop, instead of the separatrix\,(as opposed to only the critical curve). However, a comparison of the colored trajectories, to the corresponding (gray)\,dashed ones shows that increasing the resolution does lead to improved representations.


\begin{figure}
\includegraphics[scale = 0.375]{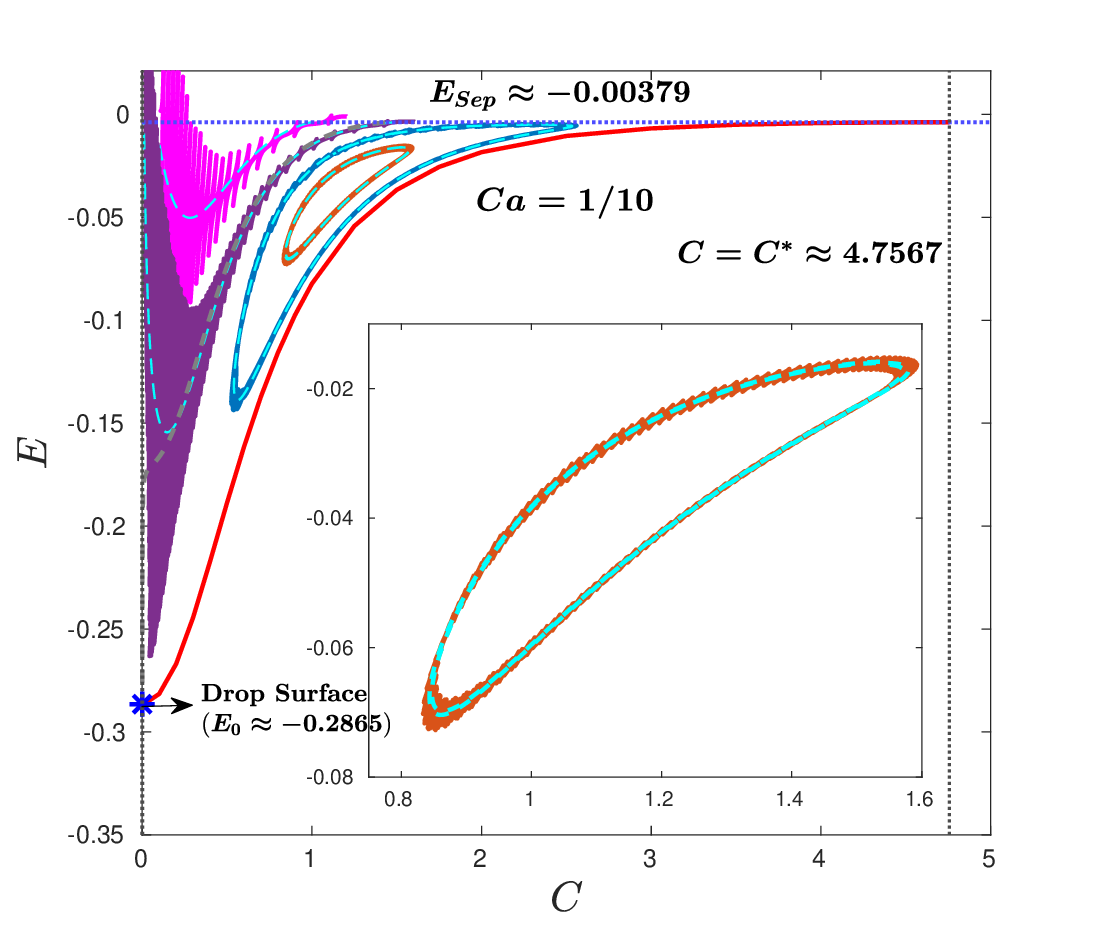}
\includegraphics[scale = 0.375]{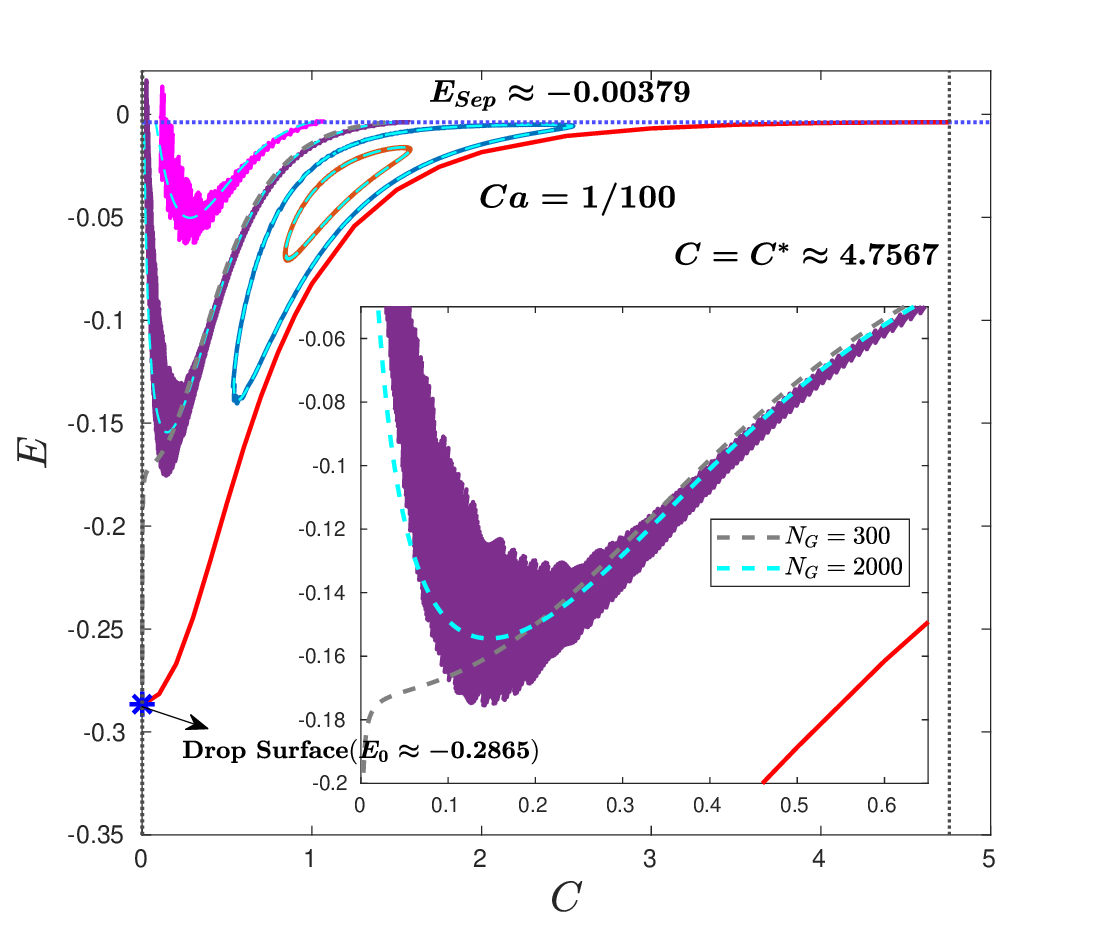}
\caption{The $C-E$ plane for $\hat{\alpha} = 10^{-4}$, $\lambda = 1$. Trajectories from the averaging analysis are compared with the full-solution trajectories for (a) $C\!a = 0.1$ and (b) $C\!a = 0.01$; note the larger amplitude of the fast oscillations compared to the inertial case\,(for same $R\!e$ values) in Fig.\ref{fig:12}. Gray and blue dashed curves correspond to lower\,($N_G = 300$) and higher\,($N_G= 2000$) resolution averaged trajectories}
\label{fig:20}
\end{figure}

In Fig.\ref{fig:20}, we compare the averaged trajectories to the full-solution trajectories, obtained from integration of the velocity field given in (\ref{smallCa:expansion}), for $\hat{\alpha} = 10^{-4}, \lambda = 1$; details regarding this comparison have already been given for the inertial case\,(see Fig.\ref{fig:12}). Notwithstanding the fast oscillations, the full-solution trajectories compare well with the averaged ones for both $C\!a = 0.1$\,(Fig.\ref{fig:20}a) and $0.01$\,(Fig.\ref{fig:20}b), over most of the $C-E$ plane. As for the inertial case, the full-solution trajectories do cross the separatrix, spiralling out to infinity close to the $x_1-x_2$ plane. Unlike the inertial case, this escape to infinity in physical space leads to the trajectories in the $C-E$ plane converging to a finite point; this because, as noted earlier, the $O(C\!a)$ velocity field decays sufficiently rapidly at large distances. Two further points are worth noting. First, for comparable $C\!a$ and $R\!e$, the full-solution trajectories in the former case exhibit much larger amplitude of fast wiggles - compare Figs.\ref{fig:12}a and \ref{fig:20}b. Second, the full-solution trajectories respect the deformation of the drop; thus as emphasized via the inset in Fig.\ref{fig:20}b, these trajectories invariably turn around and cross the separatrix, although the corresponding averaged trajectory\,(when close enough to the critical curve) heads towards the drop owing to insufficient numerical resolution\,(the dashed gray curves in Figs.\ref{fig:20}a and b).

\subsection{Boundary Element Simulations} \label{BEM}
The exterior streamline topology detailed in the earlier two subsections is nontrivial for two reasons. The first, as already mentioned in the introduction, is that our findings go against the belief prevailing in the literature\citep{Torza71,Kennedy94,Komrakova14}. The second is that, from a dynamical systems perspective, one expects the nested-tori configuration to be structurally unstable, and thereby susceptible to additional perturbations which include deformation-induced corrections of an order higher than the $O(Ca)$ correction examined in sections \ref{deformation:physical} and \ref{deformation:CE}.  
This latter reason motivated us to perform BEM simulations, following along the lines of \citet{Kennedy94} and \citet{Pozrikidis02}.
BEM simulations have, of course, been extensively used in the earlier literature for the study of drop deformation and breakup, and we therefore directly write down the equation governing the interfacial velocity field for a finite-$C\!a$ drop in an ambient planar linear flow:
\begin{align}
\begin{split}
\bm{u}(\bm{x}) = 2 C\!a\,\bm{u}^{\infty}(\bm{x}) &- \frac{1}{2\pi} \int_D \bm{G}(\bm{x - x_0}).(\bm{\nabla_{\bm{x_0}}.n})\bm{n}(\bm{x_0}) \; dA(\bm{x_0}) \\
&+ \frac{1 - \lambda}{(1 + \lambda)4\pi} \int_D \bm{T}(\bm{x} - \bm{x_0}).\bm{u}(\bm{x_0}).\bm{n}(\bm{x_0}) \; dA(\bm{x_0}),
\end{split} \label{4.29}
\end{align}
in terms of the single and double layer potentials involving the Oseen-Burgers tensor\,($\bm{G}(\bm{x})$) and the associated stress tensor\,($\bm{T}(\bm{x})$), respectively, as kernels. In (\ref{4.29}), $\bm{x}$ and $\bm{x}_0$ are points on the drop surface, $\bm{n}(\bm{x}_0)$ is the outer unit normal, with $\bm{\nabla_{\bm{x_0}} \cdot n}$ denoting (twice)\,the mean interfacial curvature; $\bm{u}^{\infty} =\bm{\Gamma} \cdot \bm{x}$, with $\bm{\Gamma}$ as defined in (\ref{1.2}), denotes the ambient flow. While finding $\bm{u}(\bm{x})$ for arbitrary $\lambda$ requires the solution of an integral equation, the double layer potential contribution vanishes for $\lambda = 1$, leading to a simple explicit algebraic relation for the velocity field.

Following \citet{Pozrikidis02}, a serial code was written to solve (\ref{4.29}). Simulations begin with an initially spherical drop with its surface discretised into $N$ elements. The discretization library from the open source package {\it BEMLIB}\citep{Pozrikidis02} is used, the discretization being achieved via an inscribed octahedron or icosahedron whose faces are then successively tessellated - each tessellation involves subdividing a given triangular face into four smaller ones. Thus, one has $N = 8 \mbox{x} 4^n$\,(octahedron) or $20 \mbox{x} 4^n$\,(icosahedron) surface elements for an order $n$ of the tessellation. In Appendix \ref{Appc}, we present a detailed validation of our simulations via a comparison with theoretical results available for weakly deformed drops. An important conclusion of this exercise is that, for small but finite $C\!a$, capturing the altered streamline topology requires a higher resolution than that needed to obtain converged `bulk' quantities such as the Taylor deformation parameter. Our simulations therefore use up to $N = 8192$, a value much higher than the earlier simulations of \citet{Kennedy94} with $N = 512$. 

In the discretized form, (\ref{4.29}) reduces to $N$ explicit relations for the nodal velocities for $\lambda = 1$, and to a system of $N$ coupled linear equations for $\lambda \neq 1$, The latter  is solved iteratively using the Gauss-Siedel algorithm until convergence is achieved to within a tolerance of $10^{-6}$, yielding the velocity field at the end of a single time step, which is then used to update the drop shape. A fourth order Runge-Kutta method is used to march forward in time, using the solution at the previous instant as the initial guess for the next time step. The time step used is dependent on the (shape)\,relaxation time which is a function of $C\!a$ and $\lambda$. For the parameters used, steady states were typically attained within two to three relaxation times.
\begin{figure}
\centering
 \includegraphics[scale = 0.3]{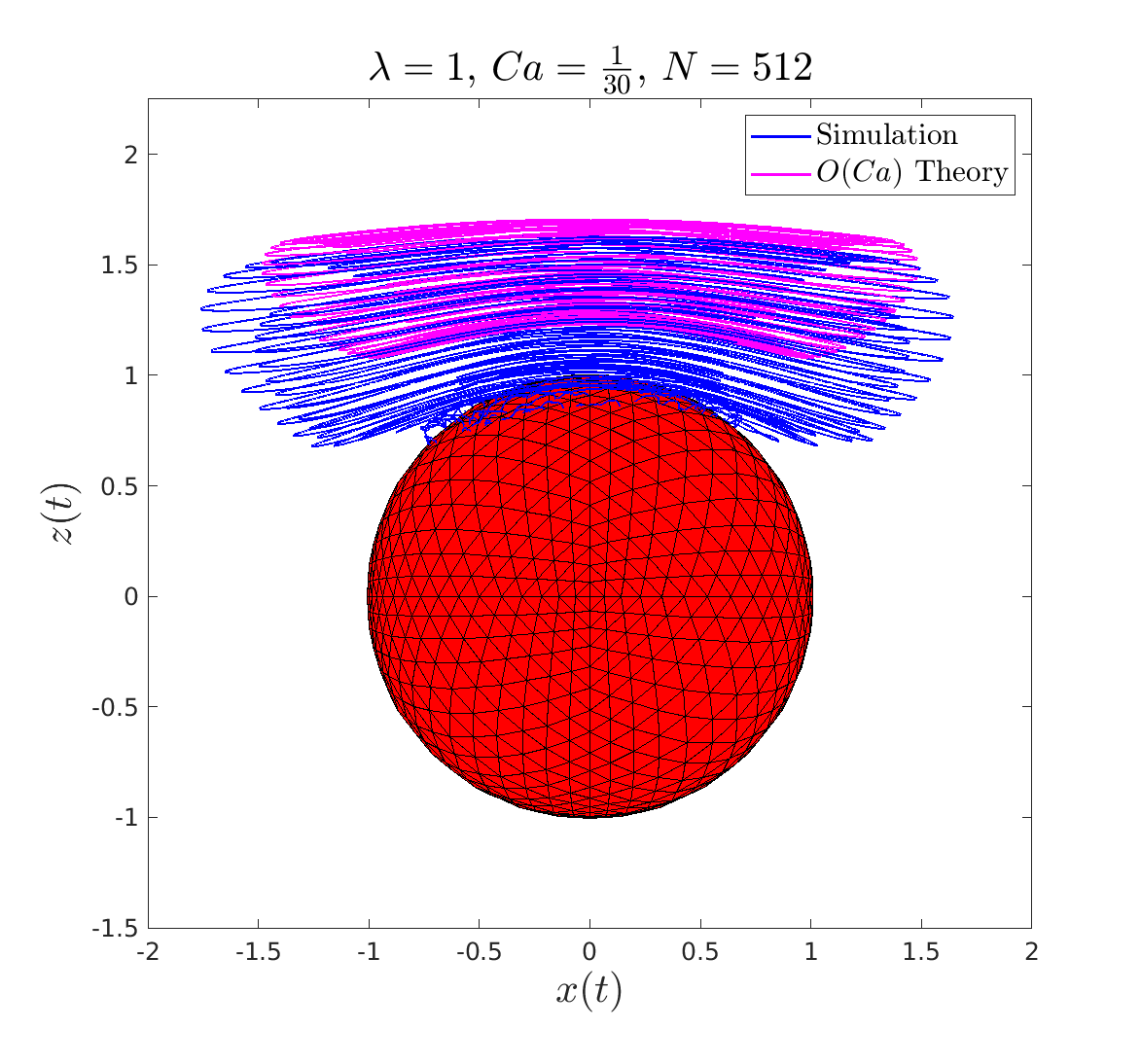}
 \includegraphics[scale = 0.3]{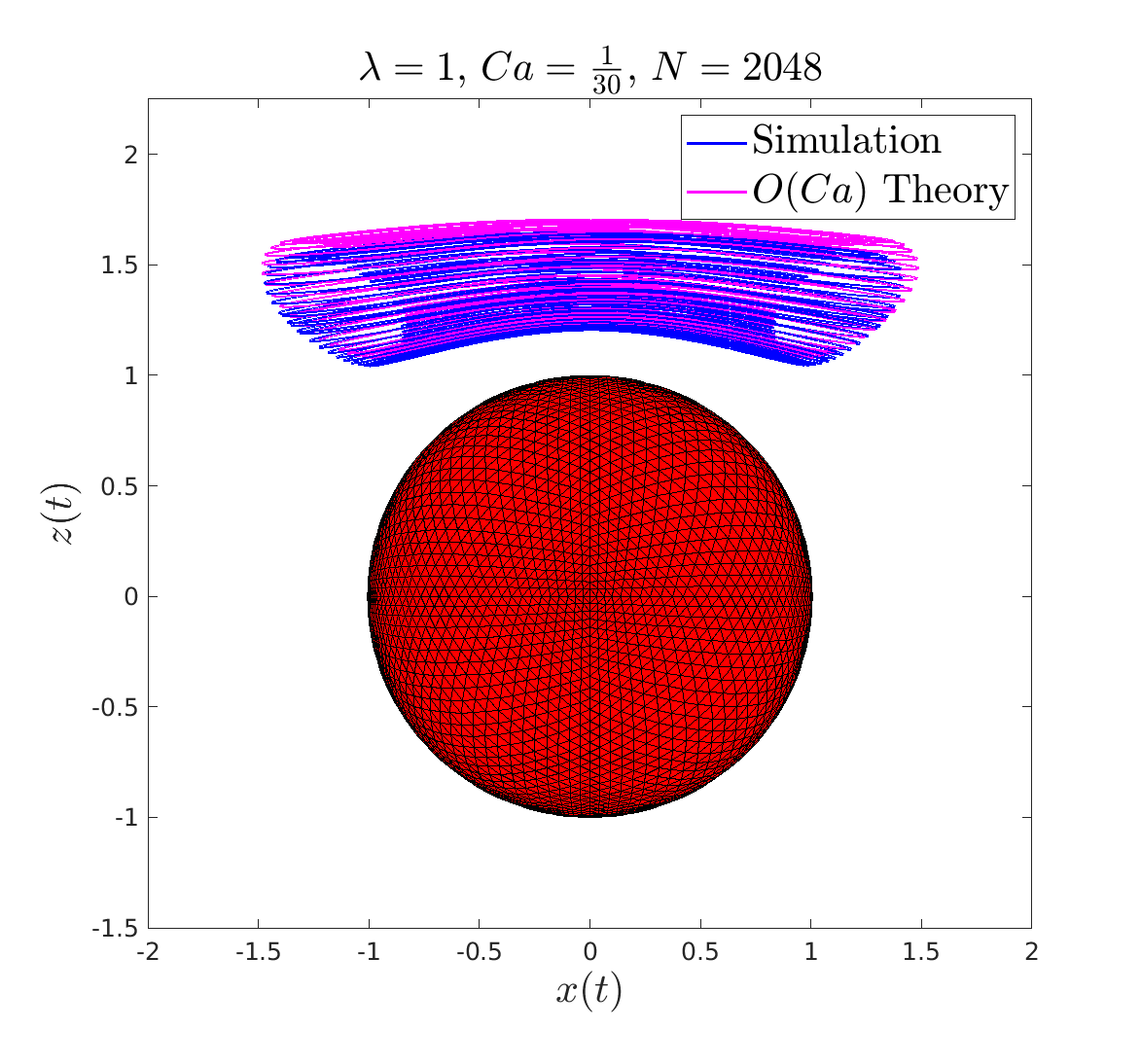}
 \includegraphics[scale = 0.3]{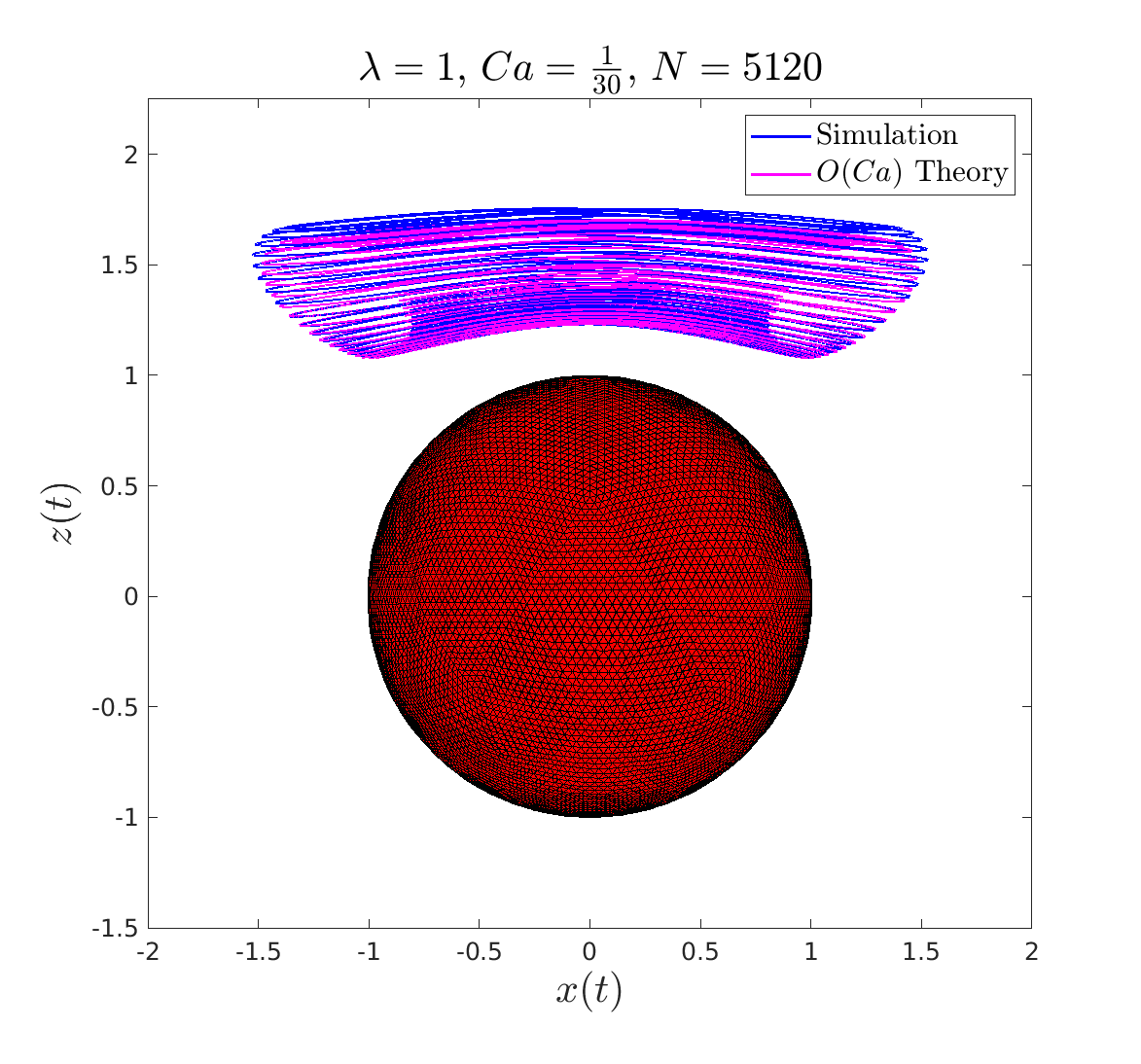}
 \includegraphics[scale = 0.3]{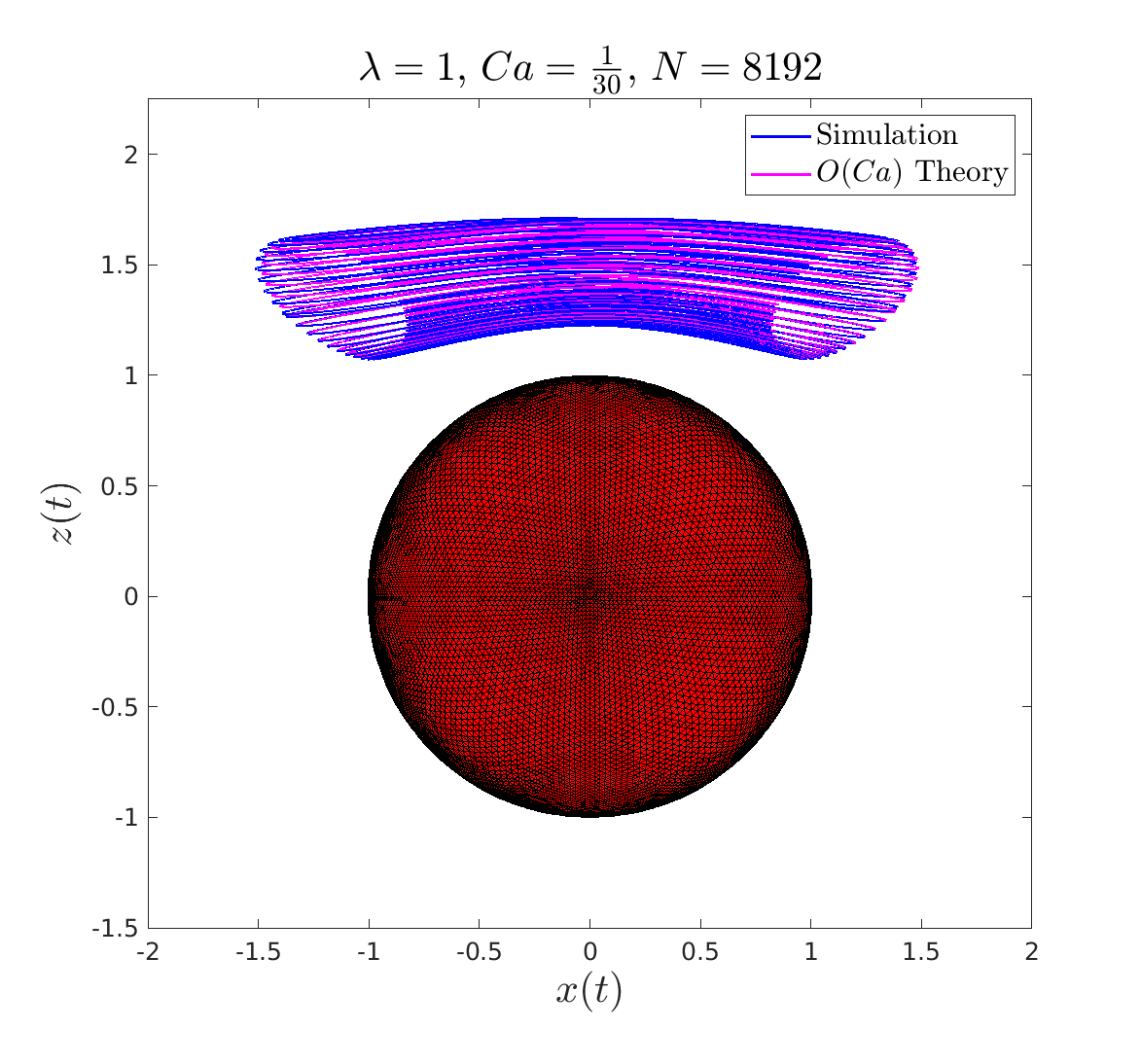}
\caption{Comparison between a BEM-streamline, and that predicted based on the $O(Ca)$ velocity field, for $\lambda = 1$, $Ca = 0.0333$. Both streamlines start from $(0.1910,0,1.0833)$, and eventually end up winding densely around an invariant torus.}
\label{fig:22}
\end{figure}

The large values of $N$, and the need to obtain streamlines after determination of the steady drop shape, imply that the BEM simulations are very time consuming. We therefore restrict the comparison between the BEM-streamlines, and those based on the analytical $O(Ca)$ field,  to $\lambda = 1$, and simple shear flow\,($\hat{\alpha} = 0$); in contrast, the validation exercise in Appendix \ref{Appc} extends over a range of $\lambda$ values. Further, the streamline comparison is focused on the existence of invariant tori, and this motivates the choice of simple shear flow above where the nested-tori configuration has the largest spatial extent\,(with the outermost torus extending to infinity in the flow-vorticity plane). It is also worth noting that one cannot choose an arbitrarily small $Ca$ for the comparison since, as discussed in \cite{Pavan_2023}, one expects the BEM-streamlines to be representative of those for a smooth drop only when the size of a single surface element is smaller than the pitch of a spiralling finite-$Ca$ streamline - this imposes a lower bound on $Ca$ for a fixed $N$. 
\begin{figure}
\centering
 \includegraphics[scale = 0.3]{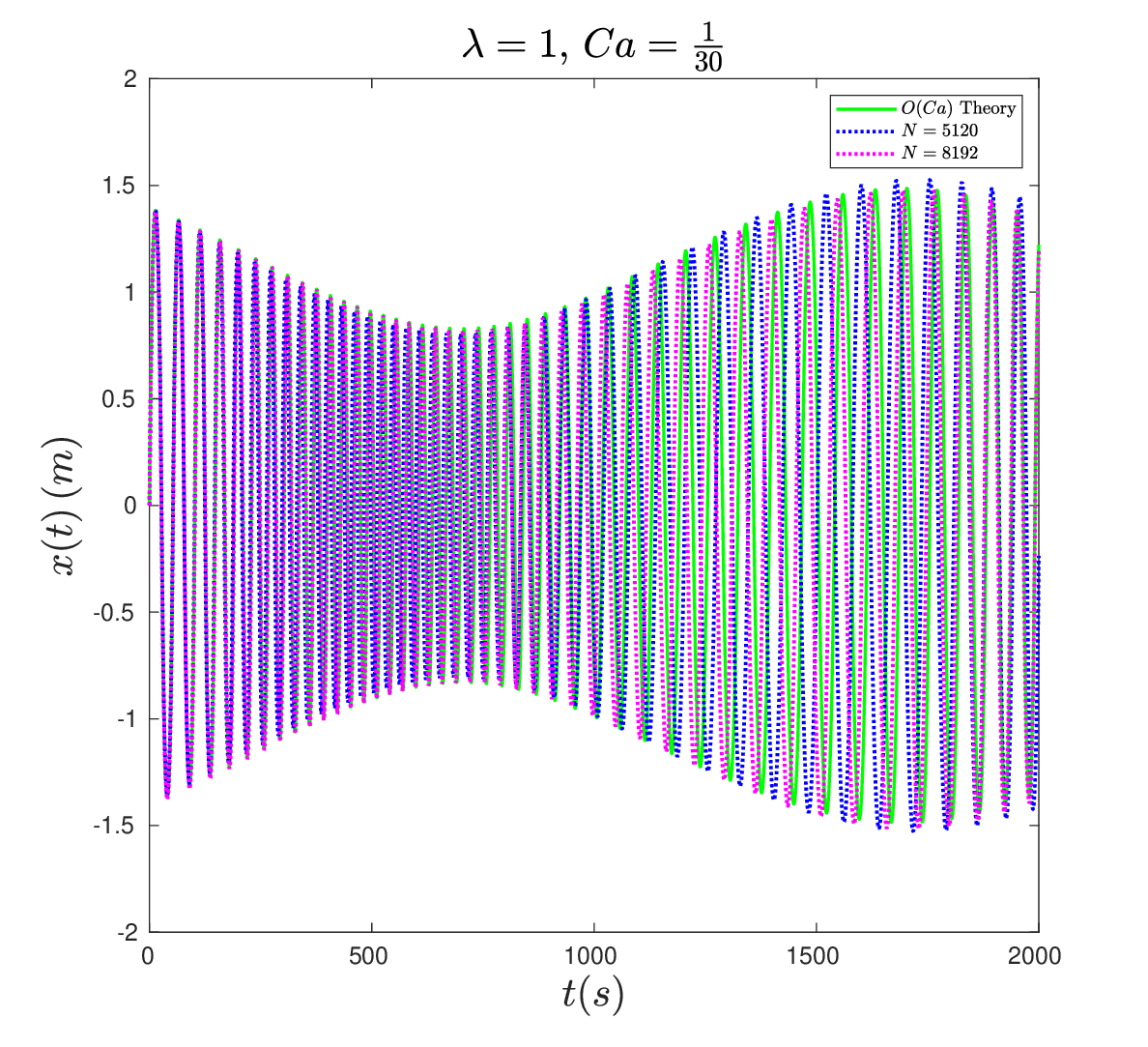}
 \includegraphics[scale = 0.3]{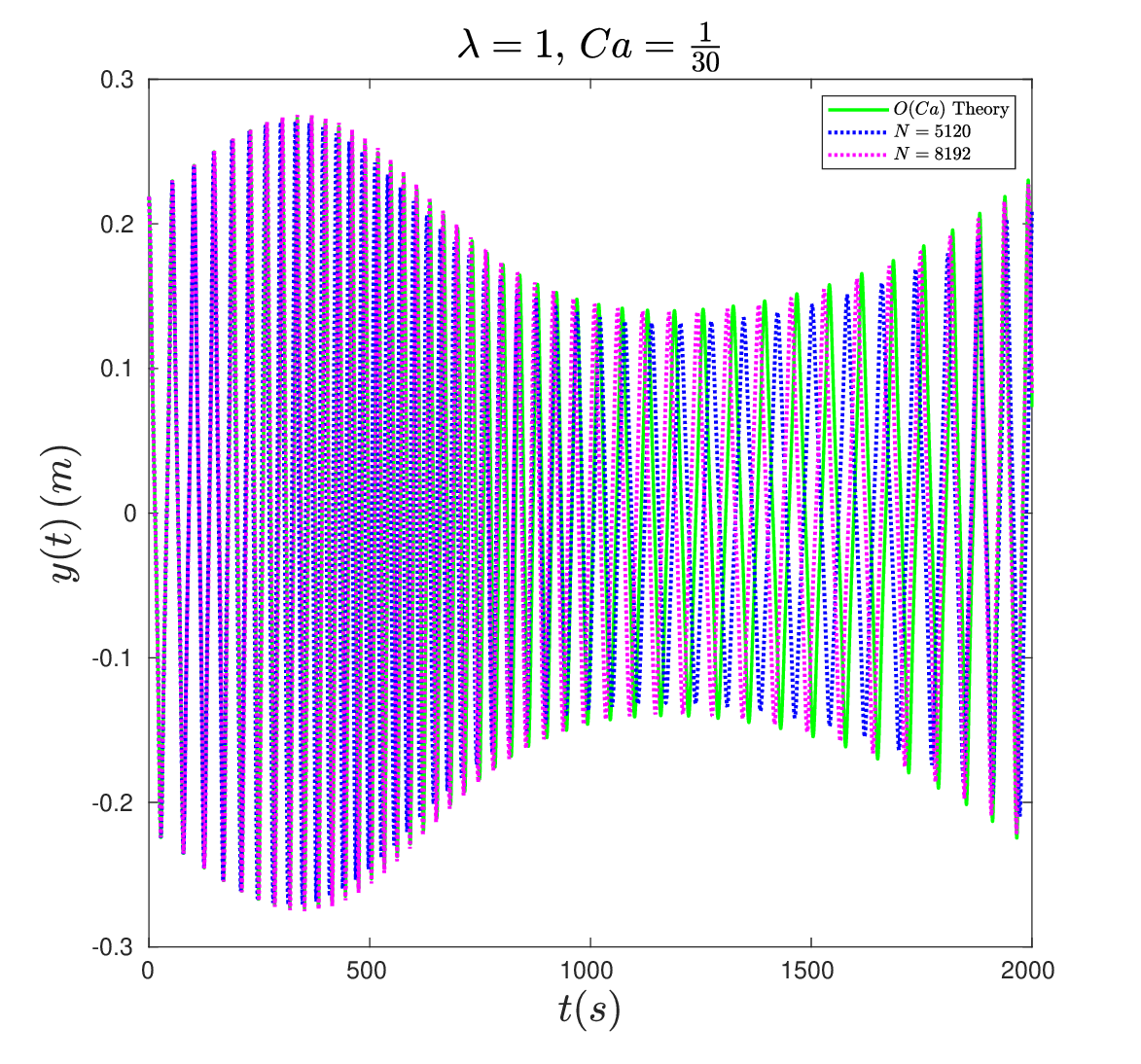}
 \includegraphics[scale = 0.3]{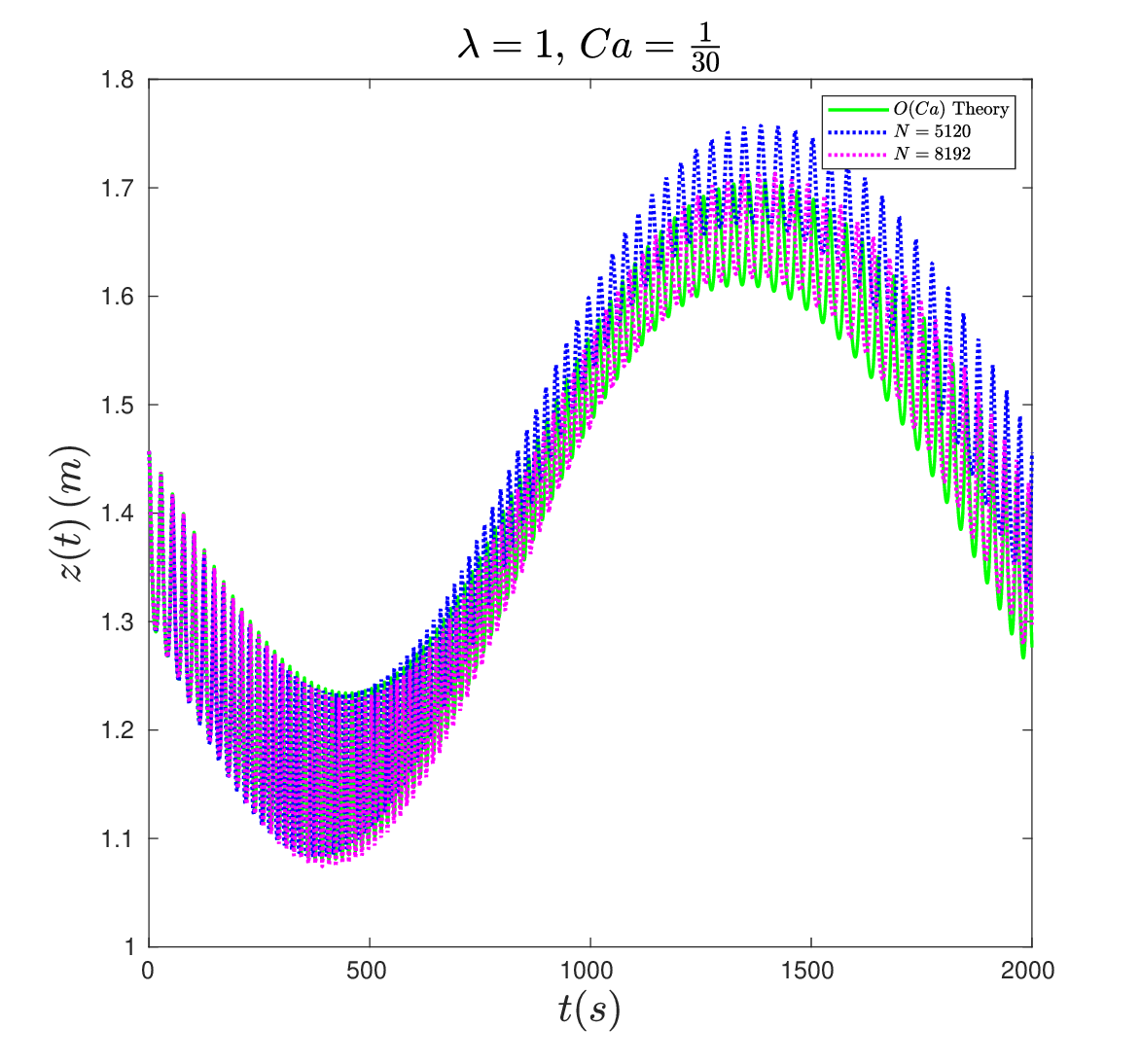}
\caption{The Cartesian coordinates of streamlines starting from $(0.1910,0,1.0833)$, and winding around an invariant torus, for $Ca = 0.0333$, $\lambda = 1$. The three curves correspond to BEM-streamlines for $N = 5120$ and $8192$, and the one obtained using the $O(Ca)$ velocity field. }
\label{fig:23}
\end{figure}

Fig.\ref{fig:22} compares the tori swept by the BEM and $O(Ca)$ streamlines in the exterior of a deformed drop in simple shear flow for $Ca = 0.0333$\,(not $Ca = 0.01$ for reasons mentioned above), and with $N$ ranging from $512$ to $8192$. Both streamlines start from the point $(x_0,y_0,z_0) = (0.1910,\,0,\,1.0833)$, which lies close to the innermost torus. The latter choice is deliberate since this enables the fluid element to complete an entire circuit of the torus in a reasonable time. Initial locations closer to the outermost torus lead to progressively longer circuit times, leading to a numerical drift, and in turn requiring a larger $N$ for agreement with the analytical prediction. It is clear from Fig.\ref{fig:22} that one needs $N\gtrsim 5120$ for quantitative agreement, a value much larger than that required to obtain a converged Taylor deformation parameter\,(see Appendix \ref{Appc}).
Fig.\ref{fig:23} shows a comparison between the same pair of streamlines via plots of the respective Cartesian coordinates as a function of time, again emphasizing the large $N$ needed for quantitative agreement. In fact, the plot of the $x_3$-coordinate shows that the BEM-streamline drifts away from the $O(Ca)$ one even for $N = 5120$, and that one requires $N = 8192$ for agreement over an entire circuit. Importantly, however, Figs.\ref{fig:22} and \ref{fig:23} reinforce the existence of a nested tori configuration for small but finite $C\!a$, and thereby, the conclusion that drop deformation indeed alters the Stokesian closed streamline topology.

\subsection{The $C\!\!-\!\!E$ representation of exterior streamlines for finite $Re$ and $Ca$} \label{ReCa:CE}
\begin{figure}
\includegraphics[scale = 0.35]{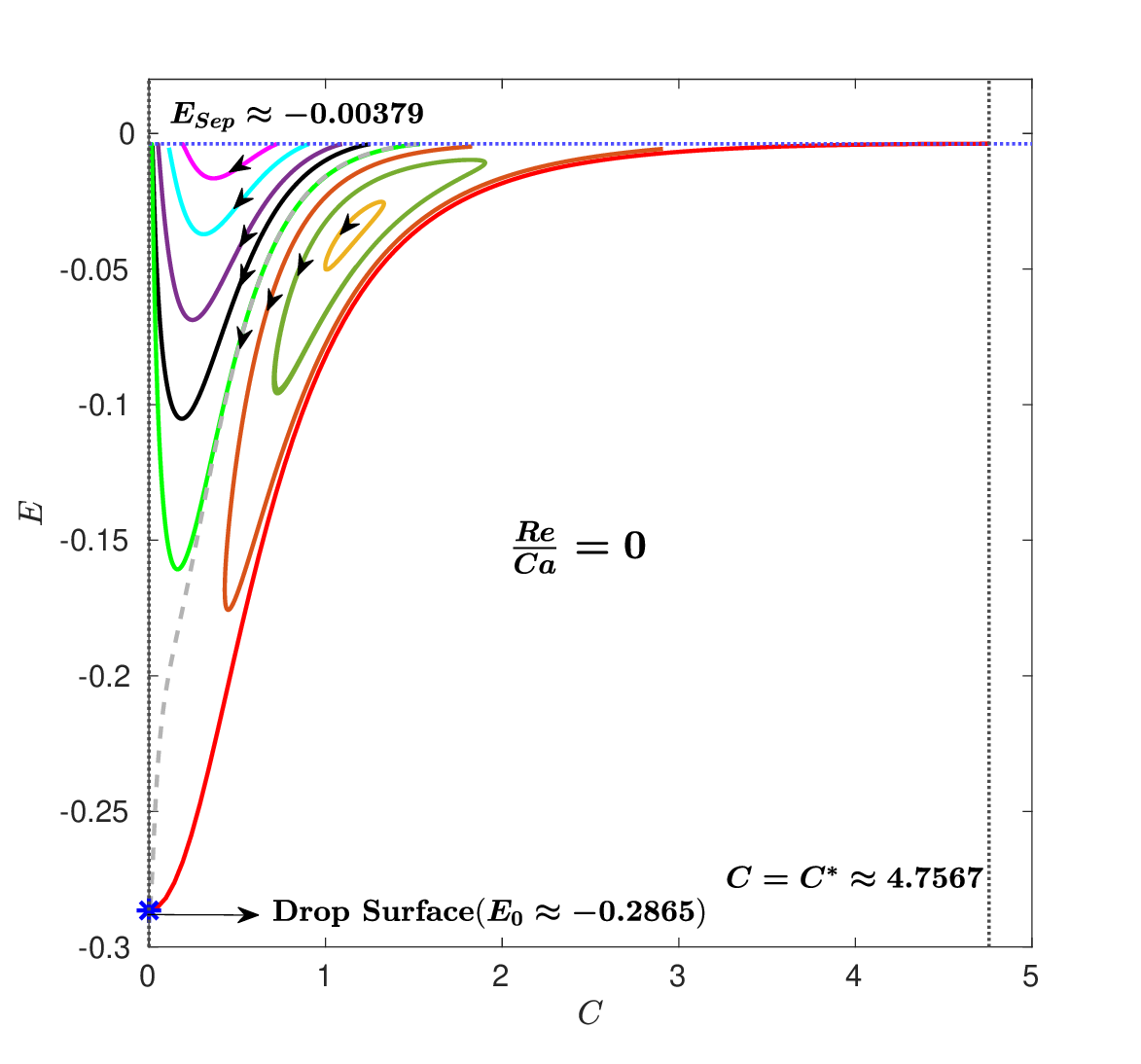}
\includegraphics[scale = 0.35]{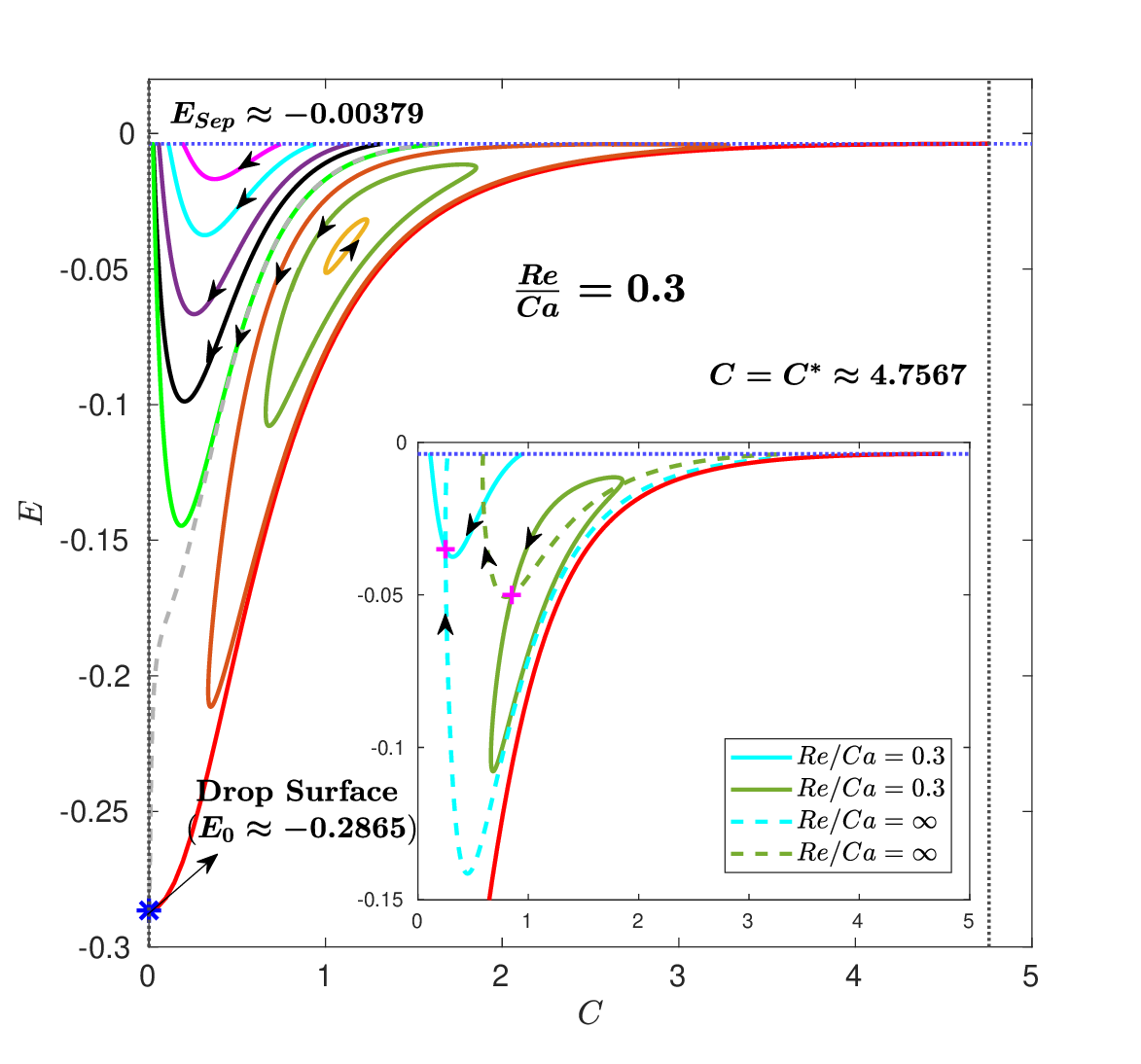}
\includegraphics[scale = 0.35]{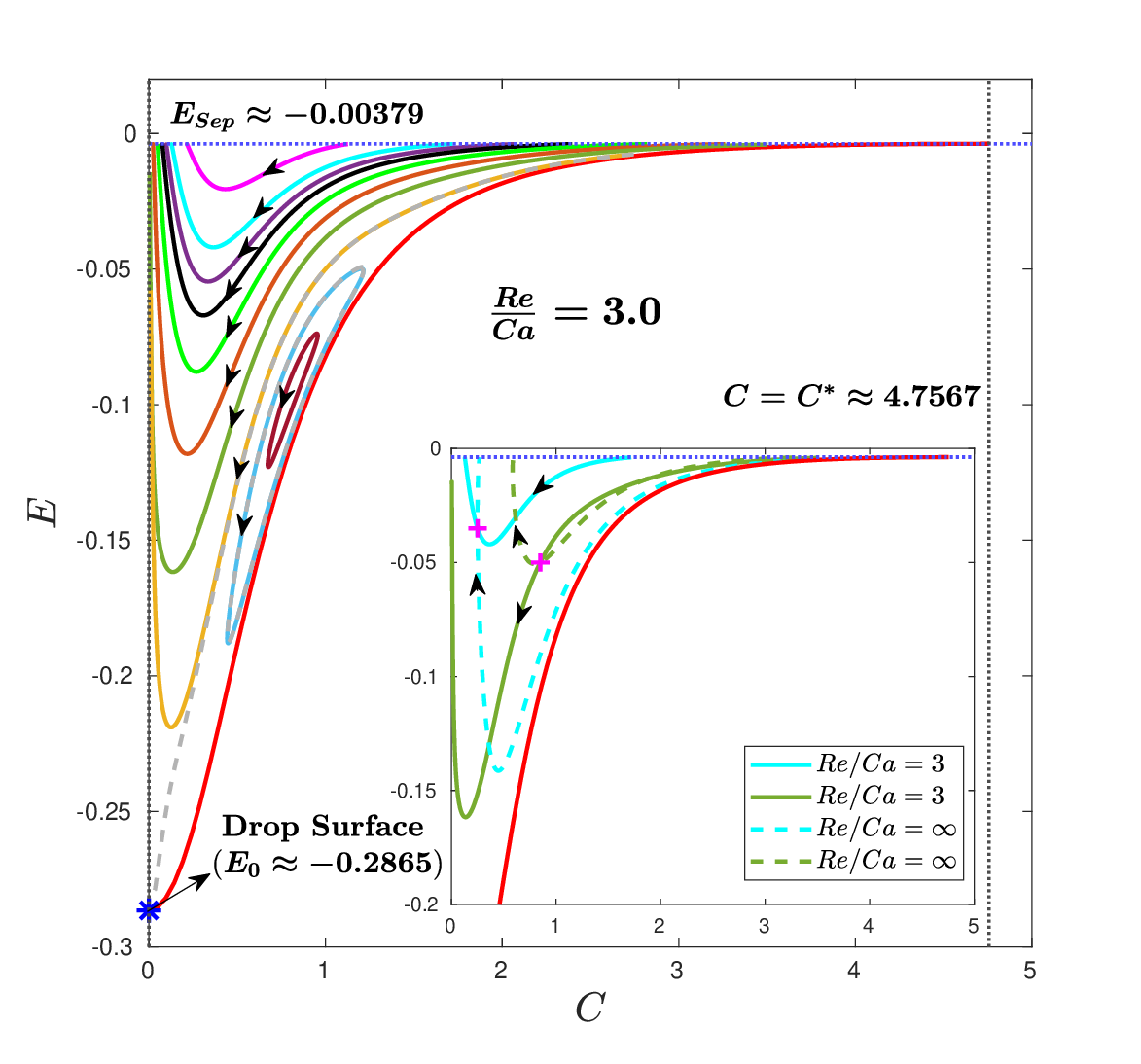}
\includegraphics[scale = 0.35]{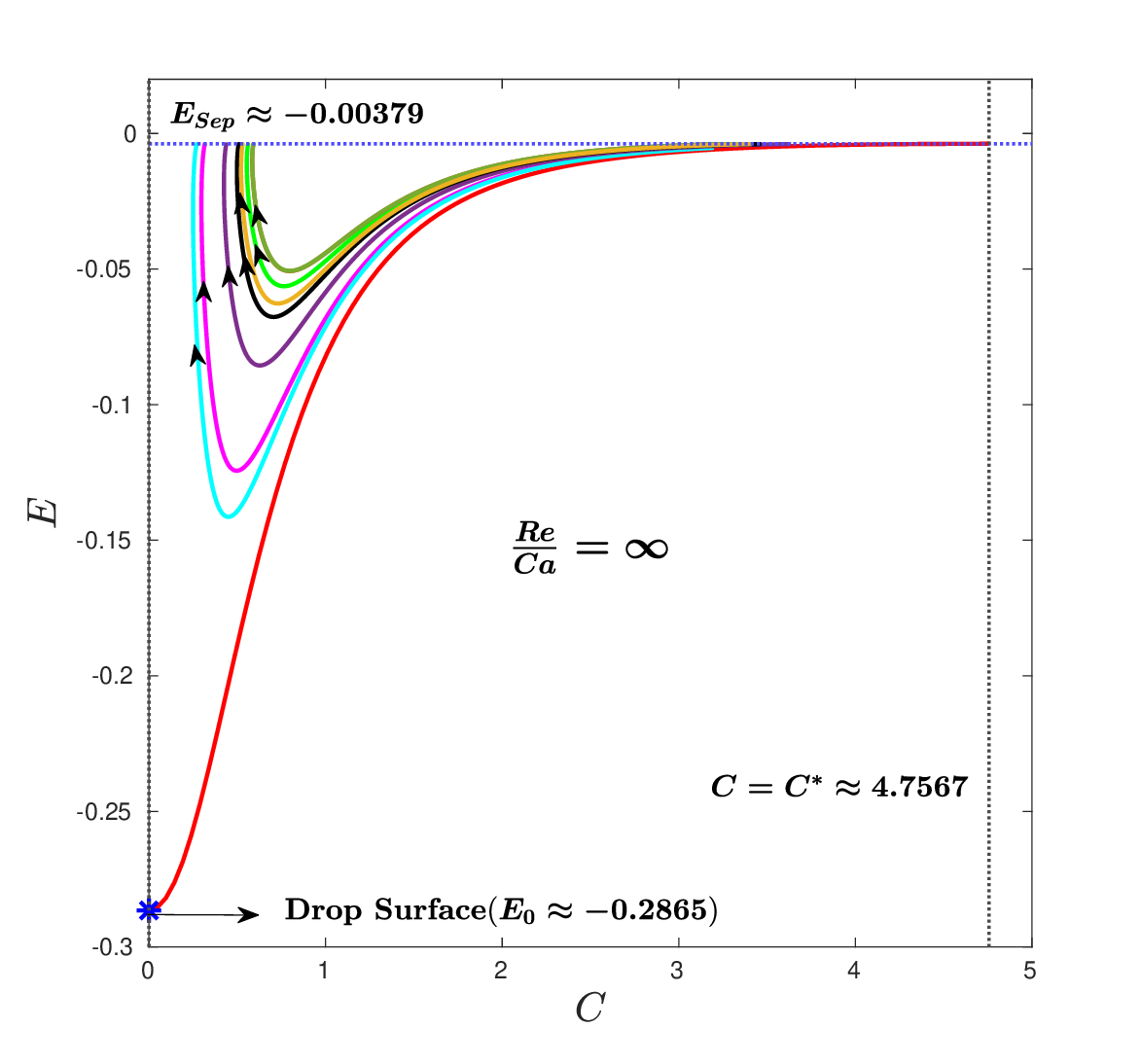}
\caption{The $C-E$ plane for $\lambda = 1$, $\hat{\alpha} = 10^{-4}$: $R\!e/C\!a =$ (a) $0.1$, (b) $0.3$, (c) $3$, and (d) $\infty$; $C\!a = 0.033$ in (a), (b) and (c), while $Re$ increases from $0.0033$ to $0.01$, and then to $0.1$. 
Trajectories of the same color, across the different subfigures, have the same initial point from which the forward-time and reverse-time integrations begin. The insets in (b) and (c) compare a pair of trajectories\,(solid curves) beginning from a given initial point\,(a magenta plus) for the particular $R\!e/C\!a$, to the corresponding pair for $R\!e/C\!a = \infty$\,(dashed curves). Finally, the grey dashed curves in (a), (b) and (c) correspond to lower-resolution trajectories\,($N_G=300$) that approach the drop; the corresponding higher resolution\,($N_G = 2000$) trajectories\,(the green curve in (a) and (b), and the yellow curve in (c) turn around and approach the separatrix instead.}
\label{fig:24}
\end{figure}

Herein, we briefly describe the $C-E$ plane representations, obtained within the averaging framework in a manner analogous to that in sections \ref{inertia:CE} and \ref{deformation:CE}, but now accounting for the combined effects of weak inertia and drop deformation. The limit is that of both $C\!a$ and $R\!e$ being small but finite, but the ratio $R\!e/C\!a$ being arbitrary, with the velocity field, used for obtaining the averaged equations, given by:
\begin{align}
\bm{u} = \bm{u}^{(0)} + Ca \left(\frac{Re}{Ca} \bm{u}^{(1)}_{O(Re)} + \bm{u}^{(1)}_{O(Ca)} \right) + O(Re^{3/2}) + O(Ca^2), \label{combined_velfield}
\end{align}
where $\bm{u}^{(1)}_{O(Re)}$ is given by (\ref{3.2}) and $\bm{u}^{(1)}_{O(Ca)}$ by (\ref{4.25main}). The averaged equations are again of the form (\ref{3.20}) and (\ref{3.21}), but with $\mathcal{F}_1$ and $\mathcal{G}_1$ being additionally dependent on $R\!e/C\!a$, as must be the case based on (\ref{combined_velfield}). The $C-E$ plane for the combined velocity field in (\ref{combined_velfield}) is shown for $\hat{\alpha} = 10^{-4}, \lambda = 1$, and for various values of $Re/Ca$, in Figs.\ref{fig:24}a-d. From Figs.\ref{fig:24}a and b, we see that, for small $Re/Ca$, the $C-E$ plane closely resembles that obtained with the deformation-induced correction alone\,(compare Fig.\ref{fig:18}). The region of nested closed curves persists in presence of weak
inertia, implying that the configuration of nested invariant tori in physical space is structurally stable to inertial perturbations. The region of closed curves, although smaller, is seen to exist even for $R\!e/C\!a = 3$ in Fig.\ref{fig:24}c. The insets in Figs.\ref{fig:24}b and c compare two of the trajectories to their purely inertial analogs\,($R\!e/C\!a = \infty$, depicted as dashed curves of the same color); the large disparity between the solid and dashed curves - the open green curve for $R\!e/C\!a = \infty$ is a closed one for $R\!e/C\!a = 0.3$ - shows that the effects of deformation remain dominant at least until $R\!e/C\!a = 3$. The final subfigure shows purely inertial trajectories which, for the aforementioned parameter values, correspond to the single-wake regime\,(see Fig.\ref{fig:10}a); thus, all trajectories start and end on the separatrix, proceeding towards smaller $C$ in between. Although the $R\!e/C\!a$ up to which trajectories deviate significantly from their limiting inertial forms depends on the particular choice of $\lambda$ and $\hat{\alpha}$, it is generally true that, for comparable $R\!e$ and $C\!a$, effects of drop deformation are dominant.

The streamline topology around a drop in simple shear flow, in presence of fluid inertia and deformation, has been examined by \citet{Singh11} using a front-tracking/finite-difference algorithm. The results reported were for non-zero $Re$ and $Ca$. Although the region of spiralling streamlines was found to depend on the magnitude of $R\!e$\,(rapidly decreasing in extent with increasing $R\!e$), the nature of spiralling was found to be largely insensitive to $R\!e/C\!a$ varying in the range $[0.5,10]$. Spiralling streamlines close to the flow-gradient plane always spiraled towards it, with those beyond a threshold distance spiralling away from it along the vorticity axis\,(Figs.9, 11 and 12 in the said paper). This is broadly consistent with the drop-deformation-induced change in the closed streamline topology described in section \ref{deformation:physical}. Moreover, the insensitivity of the spiralling to $R\!e/C\!a$ is consistent with the $C-E$ plane configurations in Fig.\ref{fig:24}, where the effect of drop deformation is dominant over fluid inertia. In attempting to relate their results to the inertial streamline topology around a rigid sphere known at that time\citep{Sub06a,Sub06b}, \citet{Singh11} also examined large $\lambda$ values\,(in their Fig.15). Although these results are likely outside the purview of a small-$Ca$ expansion\,(owing to $\lambda C\!a$ becoming of order unity), the authors continued to find streamlines that spiraled outward along the vorticity axis. Nevertheless, none of the numerical findings are not obviously suggestive of the presence of invariant tori. In this regard, it must be kept in mind that the computations are subject to periodic boundary conditions along the flow and gradient directions. In limiting the extent of the closed streamlines along the flow direction, periodicity indirectly acts to limit the size of the region of nested invariant tori, making numerical identification of such a configuration difficult.

\section{Conclusions} \label{sec:conclude}

\begin{figure}
\centering
\includegraphics[scale = 0.375]{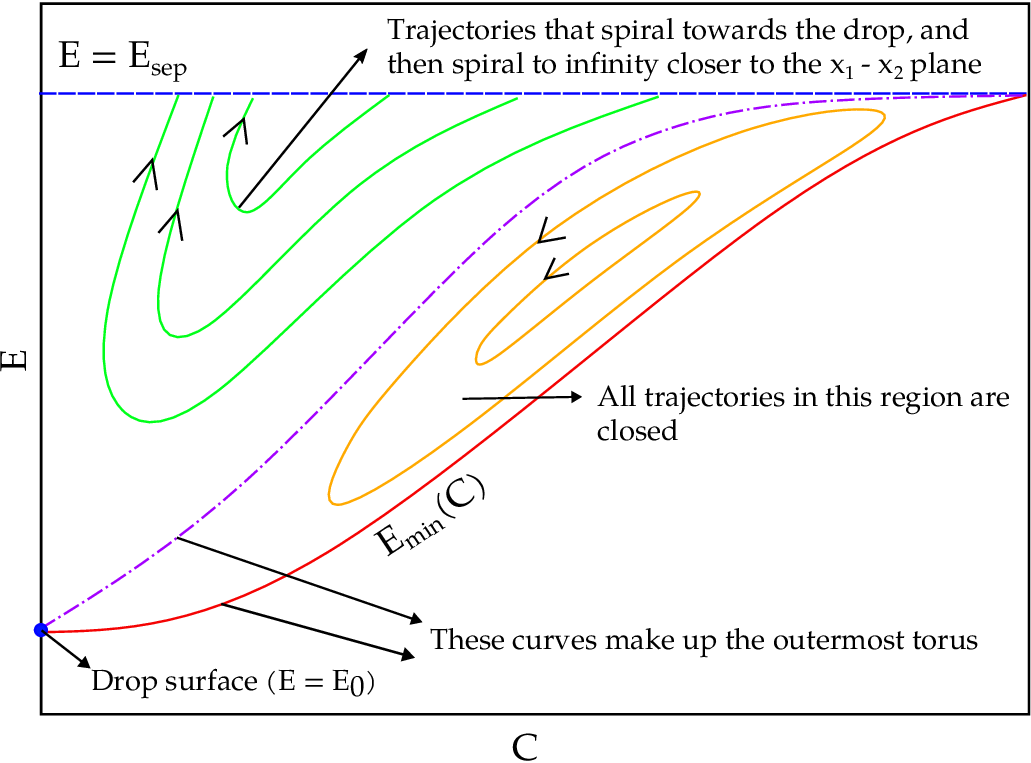}
\includegraphics[scale = 0.375]{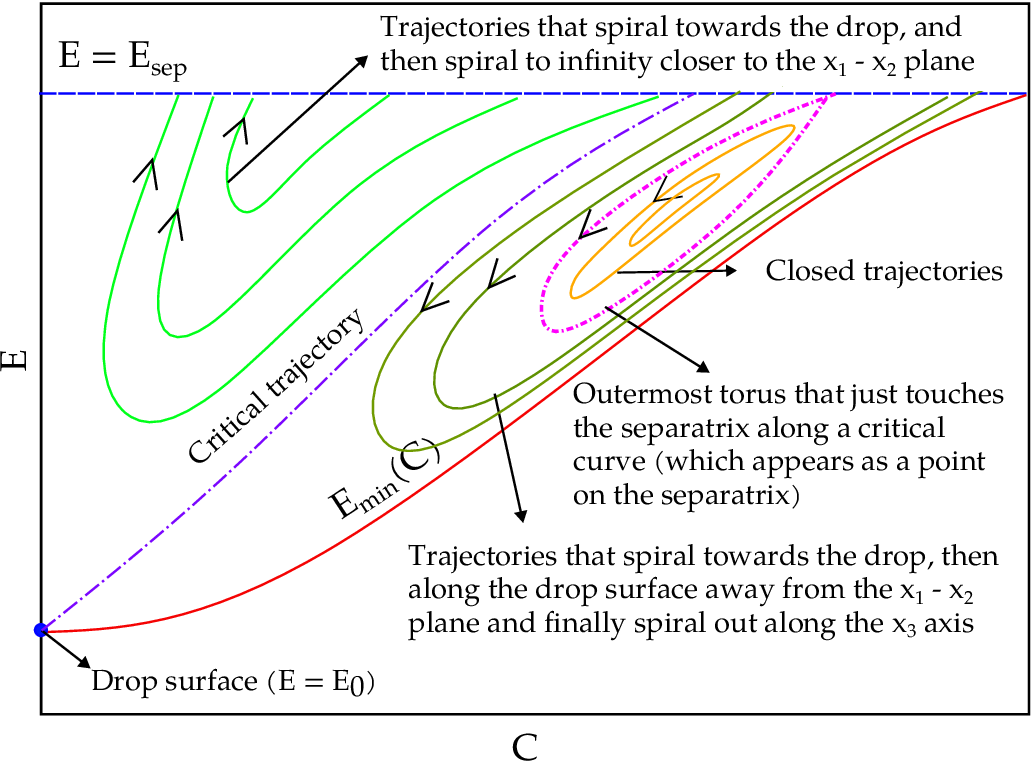}
\caption{Schematic of the $C-E$ plane representation of streamlines in exterior of a deformed drop for (a) $\hat{\alpha} = 0$ and (b) $0 < \hat{\alpha} < \hat{\alpha}_c$.}
\label{fig:25}
\end{figure}

The main contribution of this paper is the characterization of the exterior streamline topology around a deformed drop in an ambient hyperbolic linear flow, including the limiting case of simple shear flow. This has been done both analytically via the $O(C\!a)$ velocity field\,(sections \ref{deformation:physical} and \ref{deformation:CE}), and numerically via BEM computations\,(section \ref{BEM}). The schematics of the $C-E$ plane trajectories in Figs.\ref{fig:25}a and b summarize the streamline topology around the deformed drop\,(assuming infinite numerical resolution). 

The inertial streamline topology around a spherical drop is known from an earlier effort\citep{Kris18b}, and this has allowed us to briefly examine the exterior streamline topology under the combined effects of inertia and deformation\,(section \ref{ReCa:CE}). An unexpected finding is that of finite-$C\!a$ spiralling streamlines that wind densely around a set of nested invariant tori. The nested tori configuration appears to exist for all hyperbolic linear flows that have a closed streamlines region surrounding the originally spherical drop for $C\!a = 0$, and therefore, is present for all $\hat{\alpha} < \hat{\alpha}_c$; the nested tori also survive the inclusion of inertial perturbations. For simple shear flow alone, nested-tori configuration is infinite in extent within the flow-vorticity plane. In a recent effort, we have also characterized the effect of drop deformation on the interior streamline topology, and the resulting scalar transport rates\,\citep{Pavan_2023}. The interior streamlines, for small but finite $C\!a$, were again found to wind densely around nested invariant tori that foliated the drop volume. While such a dense-winding topology is expected on account of confinement\,(not obvious, however, owing to the possibility of chaotic wandering), its existence for the exterior streamlines is not.

Herein, we have only examined the streamline topology in the small-$C\!a$ limit such that $\lambda C\!a \ll 1$. It is of interest to extend this investigation to large $\lambda$, so as to cover the entire range of $\lambda C\!a$. As already mentioned in section \ref{ReCa:CE}, there exist numerical results\,\citep{Singh11} in the large-$\lambda$ limit suggestive of the spiralling streamline topology, outside of the invariant tori, remaining unaltered with increasing $\lambda C\!a$. It is worth noting that, while drop deformation changes from being $O(C\!a)$ to $O(1/\lambda)$ as $\lambda C\!a$ increases from small to large values, the destruction of the closed-streamline topology does not conform to this transition in scaling. While closed streamlines do open up at $O(C\!a)$ for $\lambda C\!a \ll 1$, as shown in section \ref{deformation:physical}, they remain closed at $O(1/\lambda)$ for $\lambda C\!a \gg 1$. The latter limit corresponds to a highly viscous miscible drop in the absence of surface tension, and therefore, the dynamics continues to be reversible in the absence of diffusion effects\,(which would tend to smooth out the jump in viscosity across the interface). Reversibility implies a destruction of the closed-streamline topology only at $O(1/\lambda^2 C\!a)$ for $\lambda C\!a \rightarrow \infty$. The analog of the $C-E$ plane representations, used in sections \ref{inertia:CE} and \ref{deformation:CE} to characterize the spiralling streamline topology, would now require the two families of invariant surfaces to account for the reversible drop deformation at $O(1/\lambda)$. This scaling has important implications for the scalar transport rate.
For the inertial case, \citet{Kris18b} showed that the scalar transport will be enhanced owing to the transition from a closed-streamline to a spiralling-streamline topology, and that the Nusselt number was 
$N\!u \sim O(R\!e P\!e)^{1/2}$ in the limit $R\!e \ll 1$, $R\!e P\!e \gg 1$. By analogy, one expects the deformation-induced transition to a spiralling-streamline topology to lead to $N\!u \sim O(C\!a P\!e)^{1/2}$ in the limit $C\!a, \lambda C\!a \ll 1$, $C\!a P\!e \gg 1$. In contrast, the aforementioned scaling arguments for the near-miscible limit imply $N\!u \sim (P\!e/(\lambda^2 C\!a))^{1/2}$ for $\lambda \gg 1$. 
In future, it would also be important to assess the relevance of these asymptotic scaling regimes by numerically evaluating the scalar transport rate over a range of finite $P\!e$. The role of the invariant tori configuration, in possibly inhibiting convective transport, might be of importance in the above context.


\appendix

\section{The corrections to the Stokesian velocity field around a spherical drop in an ambient linear flow} \label{AppA}

\subsection{The coefficients in the $O(Re)$ velocity field}

The $\hat{\alpha},\lambda$-dependent coefficients in the $O(Re)$ exterior velocity field, given by (\ref{3.2}) in Section \ref{Sec.3}, are:
\begin{equation}
u_1(r,\lambda) = \frac{1}{\lambda +1}\!\left[-\frac{\text{c}_1}{4 r^{11}}+\frac{\text{c}_2}{2 r^{10}}-\frac{7 \text{c}_3}{4 r^9}+\frac{\text{c}_4}{3 r^8}-\frac{\text{c}_5}{12 r^5} \right],
\end{equation}
\begin{equation}
u_2(r,\lambda) = \frac{1}{\lambda +1}\!\left[\frac{\text{c}_1}{18 r^9}-\frac{3 \text{c}_2}{32 r^8}-\frac{\text{c}_4}{36 r^6}+\frac{\text{c}_5}{18 r^3}+\frac{\text{c}_6}{r^7}-\frac{\text{c}_7}{2 r^5} \right], 
\end{equation}
\begin{equation}
u_3(r,\lambda) = \frac{1}{\lambda +1}\!\left[\frac{\text{c}_1}{18 r^9}-\frac{3 \text{c}_2}{32 r^8}-\frac{\text{c}_4}{36 r^6}-\frac{\text{c}_5}{9 r^3}+\frac{\text{c}_7}{2 r^5}+\frac{\text{c}_8}{r^7}\right], 
\end{equation}
\begin{equation}
u_4(r,\lambda) = \frac{1}{\lambda +1}\!\left[\frac{\text{c}_1}{36 r^9}-\frac{\text{c}_{10}}{r^5}-\frac{3 \text{c}_2}{32 r^8}-\frac{5 \text{c}_4}{36 r^6}+\frac{\text{c}_5}{36 r^3}+\frac{\text{c}_9}{r^7} \right], 
\end{equation}
\begin{equation}
u_5(r,\lambda) = \frac{1}{\lambda +1}\!\left[ \frac{\text{c}_1}{18 r^9}+\frac{\text{c}_{11}+\text{c}_9}{r^7}-\frac{3 \text{c}_2}{16 r^8}-\frac{5 \text{c}_4}{18 r^6}+\frac{\text{c}_5}{18 r^3} \right], 
\end{equation}
\begin{equation}
u_6(r,\lambda) = \frac{1}{\lambda +1}\!\left[\frac{\text{c}_1}{36 r^9}+\frac{\text{c}_{10}}{r^5}+\frac{\text{c}_{11}}{r^7}-\frac{3 \text{c}_2}{32 r^8}-\frac{5 \text{c}_4}{36 r^6}+\frac{\text{c}_5}{36 r^3} \right], 
\end{equation}
\begin{equation}
u_7(r,\lambda) = \frac{1}{\lambda +1}\!\left[-\frac{\text{c}_1}{126 r^7}-\frac{\text{c}_{17}}{r^5}+\frac{\text{c}_2}{32 r^6}+\frac{\text{c}_4}{36 r^4}+\frac{\text{c}_5}{9 r}-\frac{(\lambda +1)}{30 r^3} \right], 
\end{equation}
\begin{equation}
u_8(r,\lambda) = \frac{1}{\lambda +1}\!\left[-\frac{\text{c}_1}{126 r^7}-\frac{\text{c}_{18}}{r^5}+\frac{\text{c}_2}{32 r^6}+\frac{\text{c}_4}{36 r^4}-\frac{\text{c}_5}{18 r}+\frac{(\lambda +1)}{30 r^3} \right], 
\end{equation}
\begin{equation}
u_9(r,\lambda) = \frac{1}{\lambda +1}\!\left[-\frac{\text{c}_1}{126 r^7}-\frac{\text{c}_{12}}{r^5}+\frac{\text{c}_2}{32 r^6}+\frac{\text{c}_4}{36 r^4}+\frac{\text{c}_5}{9 r}-\frac{\text{c}_7}{6 r^3} \right], 
\end{equation}
\begin{equation}
u_{10}(r,\lambda) = \frac{1}{\lambda +1}\!\left[-\frac{\text{c}_1}{126 r^7}-\frac{\text{c}_{13}}{r^5}+\frac{\text{c}_2}{32 r^6}+\frac{\text{c}_4}{36 r^4}-\frac{\text{c}_5}{18 r}+\frac{\text{c}_7}{6 r^3} \right], 
\end{equation}
\begin{equation}
u_{11}(r,\lambda) = \frac{1}{\lambda +1}\!\left[-\frac{\text{c}_1}{252 r^7}-\frac{\text{c}_{12}+\text{c}_{13}}{2 r^5}+\frac{\text{c}_2}{48 r^6}+\frac{\text{c}_4}{18 r^4}-\frac{\text{c}_5}{36 r} \right],
\end{equation}
where the constants $c_i(\lambda)$ are given by:
\begin{align}
c_1(\lambda) = &\frac{5005 \lambda ^3+7722 \lambda ^2+2288 \lambda +112}{1144 (\lambda +1)^2} \\
c_2(\lambda) = &\frac{5 \lambda ^2+2 \lambda }{\lambda +1} \\
c_3(\lambda) = & \frac{19305 \lambda ^3+29172 \lambda ^2+11440 \lambda +1032}{10296 (\lambda +1)^2} \\
c_4(\lambda) = & \frac{(5 \lambda +2)^2}{4 (\lambda +1)}\\
c_5(\lambda) = & \frac{1}{2} (5 \lambda +2)\\
c_6(\lambda) = & \frac{42042 \lambda ^4+177177 \lambda ^3+204204 \lambda ^2+76996 \lambda +6304}{41184 (\lambda +1)^2 (2 \lambda +5)}\\
c_7(\lambda) = & \frac{\lambda }{2}\\
c_8(\lambda) = & \frac{35178 \lambda ^4+132561 \lambda ^3+133276 \lambda ^2+41532 \lambda +4016}{41184 (\lambda +1)^2 (2 \lambda +5)}\\
c_9(\lambda) = & \frac{426426 \lambda ^4+1728441 \lambda ^3+1942512 \lambda ^2+735778 \lambda +90412}{144144 (\lambda +1)^2 (2 \lambda +5)}\\
c_{10}(\lambda) = & \frac{3 \lambda ^2+3 \lambda +1}{9 (\lambda +1)}\\
c_{11}(\lambda) = & \frac{234234 \lambda ^4+959673 \lambda ^3+1077648 \lambda ^2+383426 \lambda +26348}{144144 (\lambda +1)^2 (2 \lambda +5)}\\
c_{12}(\lambda) = & \frac{282282 \lambda ^4+1134705 \lambda ^3+1268124 \lambda ^2+480828 \lambda +63504}{288288 (\lambda +1)^2 (2 \lambda +5)}\\
c_{13}(\lambda) = & \frac{138138 \lambda ^4+582153 \lambda ^3+675532 \lambda ^2+248596 \lambda +15456}{288288 (\lambda +1)^2 (2 \lambda +5)}\\
c_{14}(\lambda) = & \frac{\lambda ^2}{4 (\lambda +1)}\\
c_{15}(\lambda) = & \frac{3 \lambda ^2+3 \lambda +4}{18 (\lambda +1)}\\
c_{16}(\lambda) = & 0 \\
c_{17}(\lambda) = & \frac{1075074 \lambda ^4+4448301 \lambda ^3+5107388 \lambda ^2+1947684 \lambda +205408}{1441440 (\lambda +1)^2 (2 \lambda +5)} \\
c_{18}(\lambda) = & \frac{1027026 \lambda ^4+4135989 \lambda ^3+4610892 \lambda ^2+1699436 \lambda +189392}{1441440 (\lambda +1)^2 (2 \lambda +5)}
\end{align}
These above expressions for the constants are also given in \citet{Raja11}.

\subsection{The coefficients in the $O(Ca)$ velocity field}

The $\hat{\alpha},\lambda$-dependent coefficients in the $O(Ca)$ exterior velocity field, given by (\ref{4.25main}) in Section \ref{Sec.4}, are:
\begin{align}
 c_1(r,\lambda) &= -\frac{(19 \lambda +16) \left(7 (45 \lambda +22) r^2-9 (45 \lambda +4)\right)}{144 (\lambda +1)^2 r^{11}}, \\[5pt]
 \begin{split}
 c_2(r,\lambda) &= -\frac{(19 \lambda +16) \left(35 (\lambda +1) (45 \lambda +4)+18 \left(52 \lambda ^2+92 \lambda +31\right) r^4\right)}{1260 (\lambda +1)^3 r^9} \\
 &+\frac{(19 \lambda +16)\left(-15 (\lambda  (159 \lambda +233)+56) r^2\right)}{1260 (\lambda +1)^3 r^6}, 
 \end{split}\\[5pt]
 c_3(r,\lambda) &= \frac{(19 \lambda +16) \left(r^2-1\right) \left(9 \left(25 \lambda ^2+41 \lambda +4\right) r^2-5 \left(45 \lambda ^2+49 \lambda +4\right)\right)}{2520 (\lambda +1)^3 r^7}, \\[5pt]
 c_4(r) &= \frac{(19 \lambda +16) \left(-90 \lambda +(45 \lambda +22) r^2-8\right)}{72 (\lambda +1)^2 r^9}, \\[5pt]
 c_5(r,\lambda) &= -\frac{(19 \lambda +16) \left(\left(639 \lambda ^2+929 \lambda +182\right) r^2-10 \left(45 \lambda ^2+49 \lambda +4\right)\right)}{1260 (\lambda +1)^3 r^7}, \\[5pt]
 c_6(r,\lambda) &= \frac{(19 \lambda +16) \left((19 \lambda +16) r^2-5 (3 \lambda +2)\right)}{160 (\lambda +1)^2 r^7}, \\[5pt]
 c_7(r,\lambda) &= \frac{(3 \lambda +2) (19 \lambda +16)}{80 (\lambda +1)^2 r^5}.
\end{align}
An analytical expression for the interior field, to $O(Ca)$, is also available\citep{Greco02}, and the resulting altered streamline topology has recently been shown to lead to singularly enhanced transport rates\citep{Pavan_2023}.

\section{The divergence of the time period, of Stokesian closed streamlines, with approach towards the separatrix} \label{AppB}

\begin{figure}
\centering
 \includegraphics[scale = 0.4]{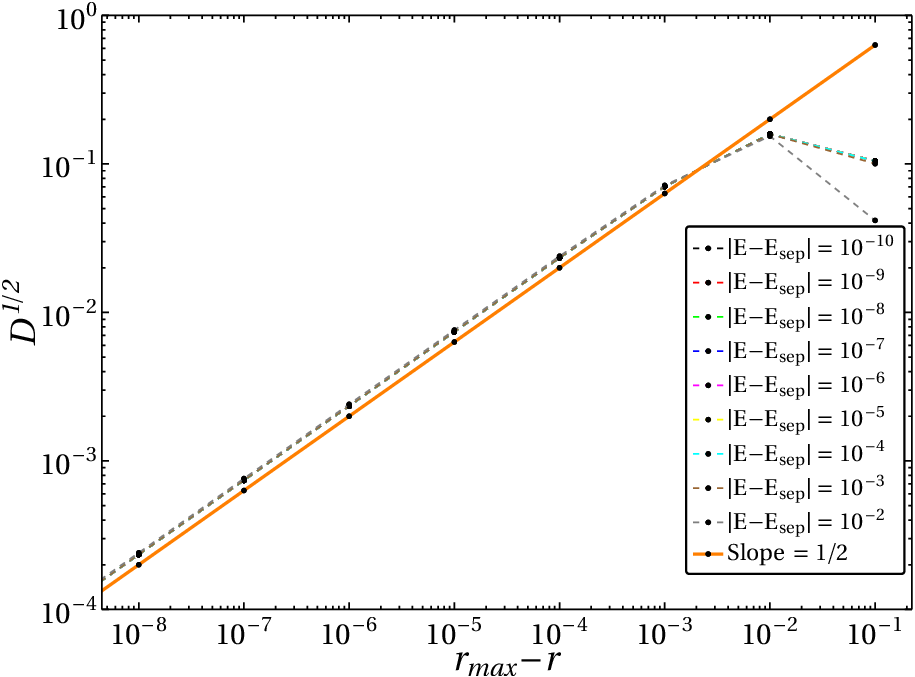}
 \hspace{0.15in}
 \includegraphics[scale = 0.4]{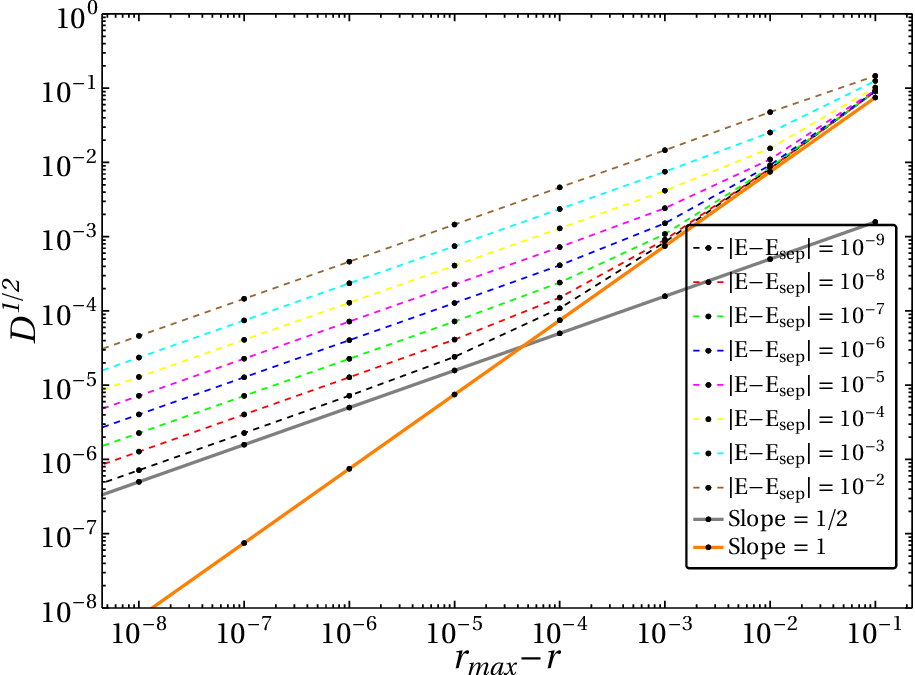}
\caption{The denominator of the integrand $D^{1/2}$ plotted against (a) $r - r_{min}$ and (b) $r_{max} - r$ for $\hat{\alpha} = 10^{-1},\,\lambda = 1$. As $E \rightarrow E_{sep}$, the denominator scales as $O(r - r_{min})^{1/2}$ in the former implying an integrable singularity and $O(r - r_{min})$, implying a logarithmic divergence of $t_{period}$. }
\label{fig:13}
\end{figure}

The nature of the time period divergence, on approach towards the separatrix, can be inferred from (\ref{3.25}). The denominator\,($D$) of the integrand\,($\equiv f(r)r^2/D^{\frac{1}{2}}$) is proportional to $u_r^{(0)}$, which must be zero at $r = r_{min}, r_{max}$, and the integrand therefore diverges for $r$ approaching these extremal locations. To infer the behavior of the integral, we expand $D$ about $r_{min}$ and $r_{max}$ as:
\begin{align}
 D =  \left( \frac{dD}{dr}\right)_{r_{max}} (r - r_{max}) + O(r - r_{max})^2, \label{3.26} \\
 D =  \left( \frac{dD}{dr}\right)_{r_{min}} (r - r_{min}) + O(r - r_{min})^2. \label{3.27}
\end{align}
One can easily show that:
\begin{align}
 \left( \frac{dD}{dr}\right)_{r_{max}} = -\left( \frac{dF_2}{dr} \right)_{r_{max}}, \label{3.28} \\
 \left( \frac{dD}{dr}\right)_{r_{min}} = -\left( \frac{dF_1}{dr} \right)_{r_{min}}, \label{3.29}
\end{align}
where $F_1$ and $F_2$, for $C= 0$, have been defined in (\ref{13a}); $F_2$ is independent of $C$ and $F_1(C) = F_1(0) -C^2$, as defined later in Section \ref{Sec.2}. Based on these expressions, one finds: 
\begin{align}
 \left(\frac{dF_2}{dr} \right)_{r_{max}} &= 
 \begin{cases}
      & 0 \; \text{as} \; E = E_{sep}\\
      & \text{finite} \; \forall \; E \neq E_{sep}     
 \end{cases} \label{3.31} \\
 \left(\frac{dF_1}{dr} \right)_{r_{min}} &= \text{finite} \: \forall \: E, \label{3.32} 
\end{align}
implying that:
\begin{align}
 \frac{f(r) r^2}{D^{1/2}} &\sim 
 \begin{cases}
  & O(r - r_{max})^{-1} \; \text{for} \; E \rightarrow E_{sep}, \\
  & O(r - r_{max})^{-1/2} \; \text{for} \; E \neq E_{sep},
 \end{cases} \label{3.33} \\
 \frac{f(r) r^2}{D^{1/2}} &\sim O(r - r_{min})^{-1/2} \; \forall \; E. \label{3.34}
\end{align}
The above asymptotic behavior is borne out from a numerical evaluation of the integral, as illustrated in Figs.\ref{fig:13}a and b. 
The persistence of the $(r_{max}-r)^{-1}$ scaling, in Fig.\ref{fig:13}b, over an increasing range of $r_{max}-r$, as $E \rightarrow E_{sep}$, is consistent with (\ref{3.33}), and points to a logarithmically divergent time period in this limit. 
The implication is that averaged trajectories never cross the separatrix, although actual streamlines\,(for non-zero $R\!e$ or $C\!a$) do.

\section{Validation of BEM Simulations}\label{Appc}

In order to ensure that the streamline topology around a deformed drop, obtained in section \ref{BEM} using BEM simulations, is reliable, we have performed a detailed validation exercise. We begin by comparing the time dependent deformation of an initially spherical drop in an impulsively started simple shear flow, obtained from simulations, to the known expression from theory. The shape of a weakly deformed drop, subject to an ambient linear flow, may be written as $F = 0$, with the scalar function $F$ given by:
\begin{equation}
F = r - \left[1 + \bm{A}(t;\lambda,Ca)\!:\!\bm{nn}  \right], \label{4.30}
\end{equation}
to linear order in the appropriate deformation parameter\,(which changes from $C\!a$ to $1/\lambda$ with increasing $\lambda$); $\bm{n}$ here is the unit outer normal which, to the said order, may be approximated as the unit radial vector. 
Without loss of generality, the shape tensor $\bm{A}$ in (\ref{4.30}) may be taken as symmetric, with $Tr{\bm A} = 0$ on account of incompressibility. For a simple shear flow imposed at $t = 0$, one has:
\begin{align}
 &A_{11} = \nonumber \\
 &\frac{5(16+19\lambda) \left( \left[ (16+19\lambda)(3+2\lambda) \cos t + 40(1+\lambda)Ca^{-1} \sin t\right]e^{-\frac{t}{\tau_{rel}}} -  (16+19\lambda)(3+2\lambda) \right)}{2 \left( (40(1+\lambda)Ca^{-1} )^2 + ((16+19\lambda)(3+2\lambda))^2 \right)}, 
 \label{4.32} \\
 &A_{12} = \nonumber \\
 &\frac{5(16+19\lambda) \left( \left[ 40(1+\lambda)Ca^{-1}\sin t - (16+19\lambda)(3+2\lambda) \cos t\right]e^{-\frac{t}{\tau_{rel}}} + 40(1+\lambda)Ca^{-1} \right)}{2 \left( (40(1+\lambda)Ca^{-1} )^2 + ((16+19\lambda)(3+2\lambda))^2 \right)},
 \label{4.33}
\end{align}
with $A_{11} = -A_{22}$ and $A_{12} = A_{21}$. The aforementioned expressions are corrected versions of those in \citet{Cox69}; specifically, the relaxation time, $\frac{19 \lambda Ca}{20}$, given therein, and that is only valid for $\lambda \gg 1, \lambda C\!a \sim O(1)$, has been replaced by $\frac{(3 + 2\lambda)(19 \lambda + 16)C\!a}{40(1+\lambda)}$, which remains valid for all $\lambda$\,\citep{Rall80}.

\begin{figure}
\includegraphics[scale = 0.4]{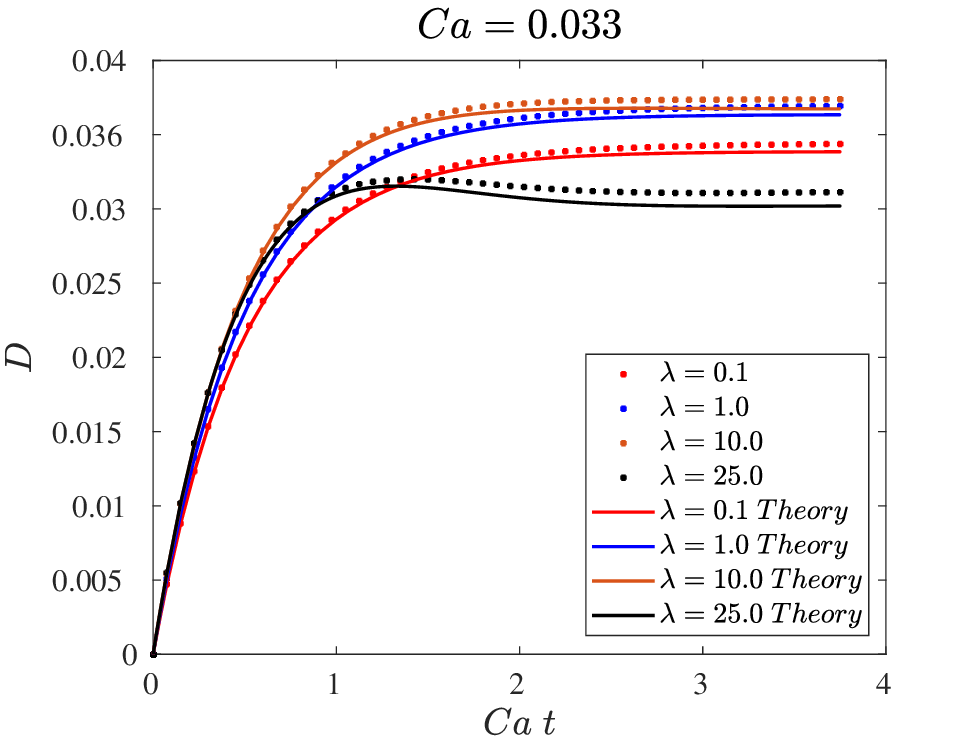}
\includegraphics[scale = 0.4]{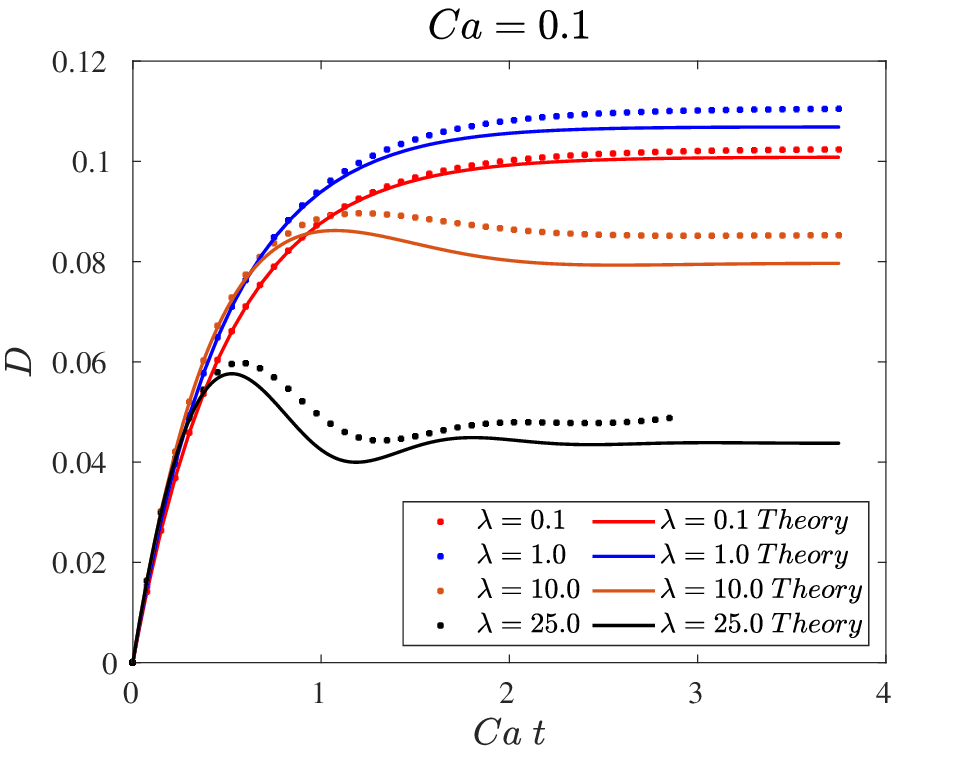}
\caption{Comparison of the analytical predictions and BEM results, for the Taylor deformation parameter, $D(t)$, as a function of time, for different $\lambda$; $C\!a\,t$ corresponds to time being made non-dimensional using the surface-tension-based scale\,($\sigma/\mu$). (a) $Ca = 0.033$; (b) $Ca = 0.1$. $N = 2048$ for all the BEM-curves.}
\label{fig:21B}
\end{figure}

Fig.\ref{fig:21B} compares the time dependent Taylor deformation parameter defined by $D(t) = (r_{max}-r_{min})/(r_{max}+r_{min})$, $r_{max}$ and $r_{min}$ being the semi-major and minor axes, obtained from the BEM simulations, to that derived from (\ref{4.32}) and (\ref{4.33}) as $D(t) = \sqrt{A_{11}^2 + A_{12}^2}$. Figs.\ref{fig:21B}a and b are for $C\!a = 0.033$ and $C\!a = 0.1$, respectively, with $\lambda$ in each figure ranging from $0.1$ to $25$. For both $C\!a$ values, there is a transition from an overdamped to an underdamped approach to the steady deformed shape, as $\lambda C\!a$ exceeds a value of order unity; the oscillations in the underdamped case scale with the inverse shear rate, and therefore, increase in frequency with increasing $\lambda$ - $C\!a\,t$ in Fig.\ref{fig:21B} is time made non-dimensional with $\sigma/\mu$. The BEM results satisfactorily match the theoretical predictions for all $\lambda$, for the lower $C\!a$. As expected, there is a larger discrepancy at the higher $C\!a$\,(although, there continues to be qualitative agreement). 

\begin{figure}
\includegraphics[scale = 0.45]{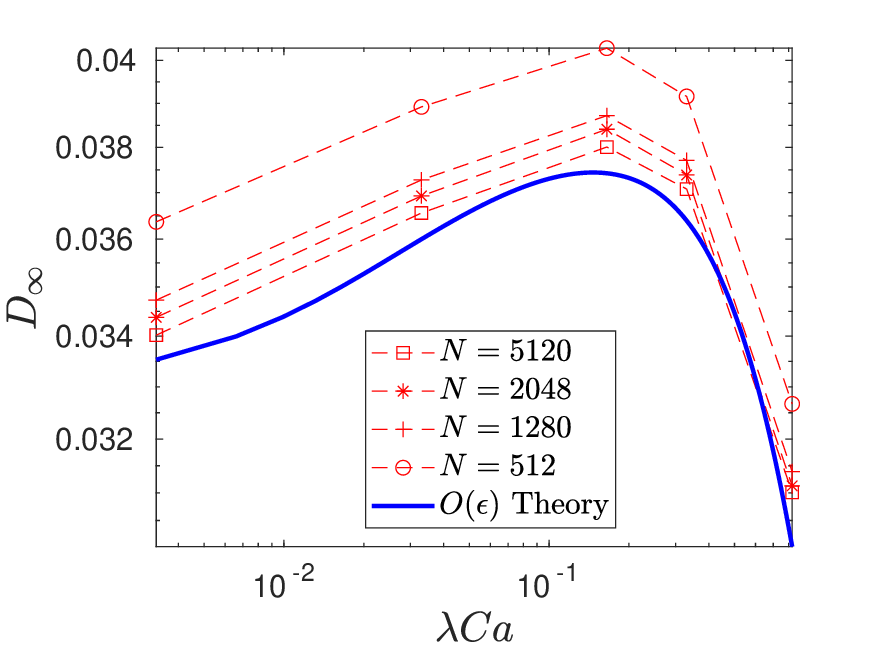}
\includegraphics[scale = 0.45]{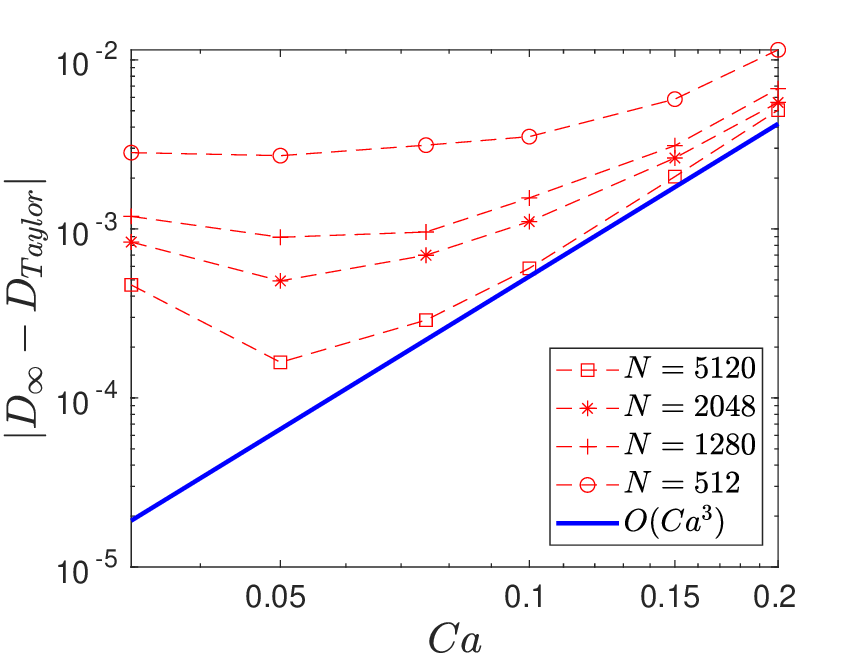}
\caption{Comparison of the analytical predictions and BEM results\,(with differing resolution $N$) for the steady state Taylor deformation parameter $D_{\infty}$: (a) compares the BEM results with the leading order theoretical prediction\,(termed the `$O(\epsilon)$ theory'), as a function of $\lambda C\!a$, for $Ca = 0.033$; note that $\epsilon \equiv C\!a$ for $\lambda C\!a \ll 1$, and $\epsilon \equiv 1/\lambda$ for $\lambda C\!a \gg 1$. (b) $|D^{\infty} - D^\infty_{\text{Taylor}}|$, for $\lambda = 1$, compared to the theoretically expected $O(Ca^3)$ scaling.}
\label{fig:21C}
\end{figure}

Next, in Figs.\ref{fig:21C}a and b, we compare the numerical and analytical values of the steady state deformation parameter, $D^\infty = \lim\limits_{C\!a t \rightarrow \infty} D(t)$ with $D(t)$ as defined above. Fig.\ref{fig:21C}a presents a comparison of $D^\infty$, as a function of $\lambda C\!a$\,($C\!a = 0.033$ with $\lambda$ varying), with the theoretical prediction $D^\infty = \frac{5 Ca (19 \lambda+16)}{2 \sqrt{1600(1+\lambda)^2 + [Ca((3+2\lambda)(19 \lambda+ 16)]^2}}$. Note that $\lim\limits_{\lambda C\!a \rightarrow 0} D^\infty = C\!a\frac{(19\lambda + 16)}{16(\lambda + 1)} = D_{\text{Taylor}}$, a result first given by \citet{Taylor32}. The theoretical curve has a small-$\lambda$ plateau\,($\lim\limits_{\lambda \to 0} D_{\text{Taylor}} = C\!a$), while decreasing as $\lambda^{-1}$ for $\lambda \gg 1$\,($\lim\limits_{\lambda C\!a \rightarrow \infty} D^\infty = \frac{5}{4\lambda}$). The BEM results compare well, and the comparison improves with increasing resolution\,($N$ increasing from $512$ to $5120$). In Fig.\ref{fig:21C}b, we plot $|D_{\infty}-D_{\text{Taylor}}|$ as a function of $C\!a$, with this difference expected to scale as $O(C\!a^3)$ for $C\!a \to 0$; 
although, its calculation requires expanding the velocity field to $O(C\!a^2)$, and an exact expression\,(for comparison purposes) does not appear to exist in the literature.  
The figure shows the BEM results for the lower resolutions\,($N = 512, 1280$) plateauing out for $C\!a \lesssim 0.15$. However, increasing $N$ does lead to the BEM curves conforming to the cubic scaling down to progressively lower $C\!a$. 


In addition to the low resolution, the poor comparison for smaller $C\!a$, in Fig.\ref{fig:21C}b, arises from $D^\infty$ being a pointwise measure, leading to a discrepancy between the true theoretical orientation corresponding to $r_{max}$\,(or $r_{min}$), and the numerical one closest to it\,(from the discrete set comprising the collocation points on the drop surface). To circumvent this error, we consider an integral (tensorial)\,measure of in-plane deformation, $F_{ij} = \int_{0}^{2\pi} (n_{i}n_{j}-\frac{1}{2}\delta_{ij}) r d\phi$, $r$ being given by (\ref{4.30}) with $\theta = \frac{\pi}{2}$, corresponding to the flow-gradient plane. The tensorial character of the measure has an additional advantage relative to the scalar measure $D^\infty$, in that deformation-induced contributions to the different scalar components of $\bm{F}$ occur at different orders in $C\!a$ in the limit $C\!a \rightarrow 0$, and can therefore be independently validated. 
This may be seen from the limiting analytical expressions for the two independent components $F_{11}$ and $F_{12}$, given by:
\begin{equation}
 Ca \ll 1, \: \lambda \sim O(1): \begin{cases}
F_{11} = Ca^2 \pi \left(\frac{16 +19\lambda}{8(1+\lambda)} \right)^2 \left(\frac{3 + 2\lambda}{20} \right), \\[5pt]
                                  F_{12} = Ca \; \pi \left(\frac{16+19\lambda}{32(1+\lambda)}\right),
                                 \end{cases} \label{4.35}
\end{equation}
\begin{equation}
 \lambda \gg 1, \: \lambda\,Ca \sim O(1): \begin{cases}
                                  F_{11} = \frac{\pi}{2\lambda} \left(\frac{1805 \lambda^2 Ca^2}{1444 \lambda^2 Ca^2 + 1600}  \right), \\[5pt]
                                  F_{12} =\frac{\pi}{2\lambda} \left(\frac{1900 \lambda Ca}{1444 \lambda^2 Ca^2 + 1600} \right).
                                 \end{cases} \label{4.36}
\end{equation}
\begin{figure}
\centering
 \includegraphics[scale = 0.45]{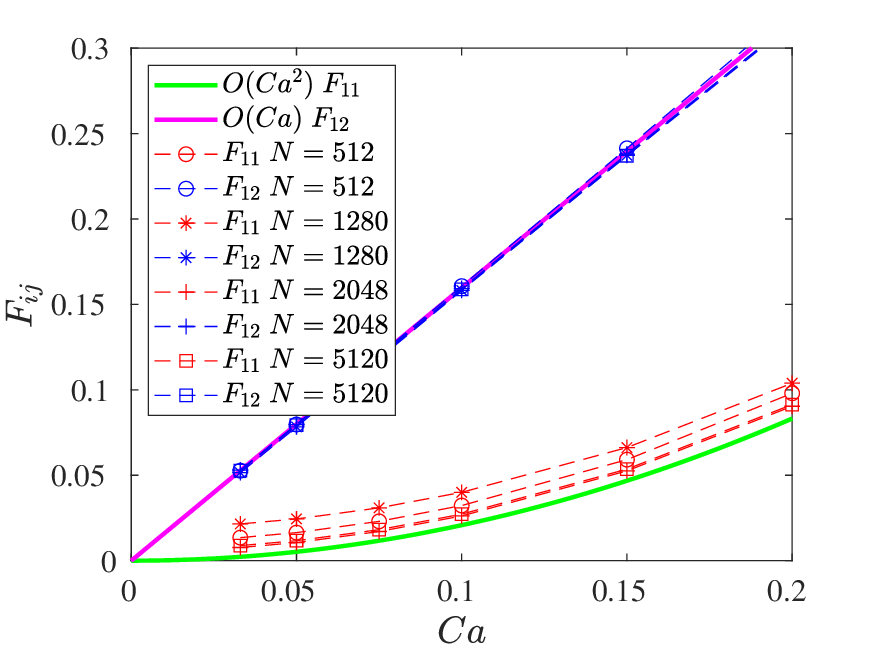}
 \includegraphics[scale = 0.45]{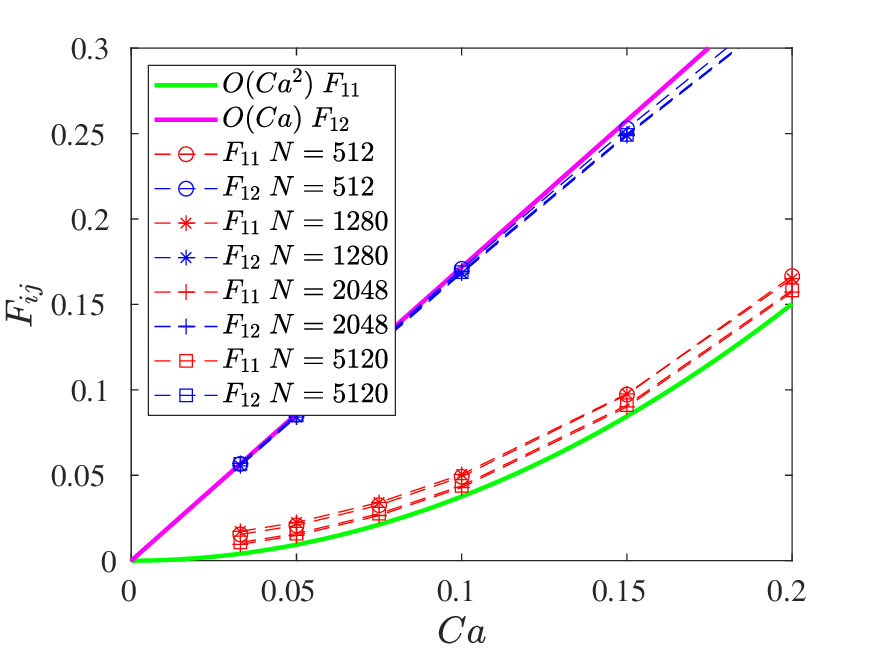}
 \includegraphics[scale = 0.45]{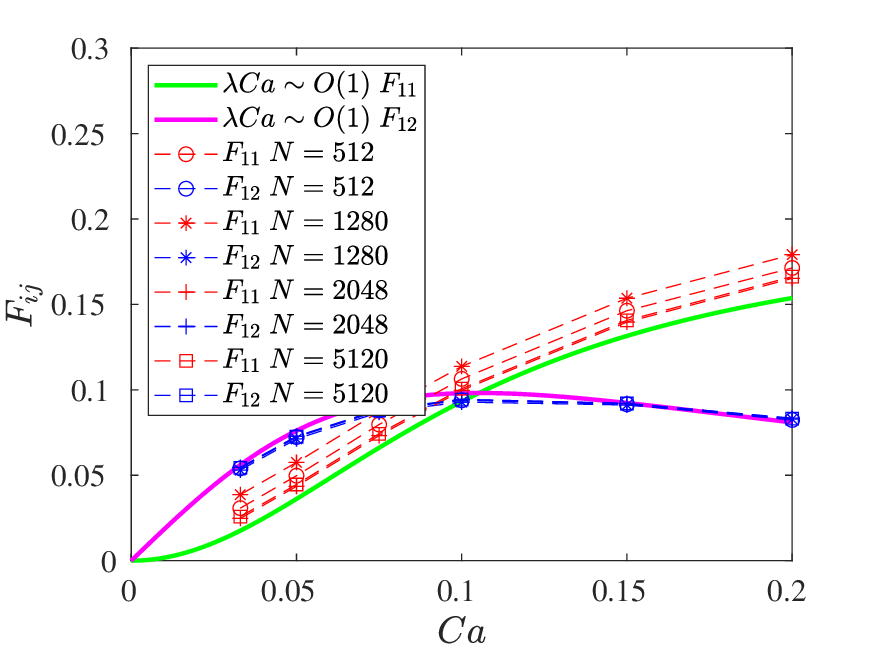}
\caption{Comparison between the BEM results and theoretical predictions for the components $F_{11}$ and $F_{12}$, of the tensorial deformation measure $\bm{F}$, as a function of $Ca$: (a) $\lambda = 0.1$, (b) $\lambda = 1$ and (c) $\lambda = 10$.} 
\label{appc2}
\end{figure}
The BEM results for $F_{11}$ and $F_{12}$, for different $N$, are plotted as a function of $C\!a$ in Figs.\ref{appc2}a, b and c. The comparison for $F_{12}$ exhibits excellent agreement for all $N$; this includes the case of the highest $\lambda$ in Fig.\ref{appc2}c where, as predicted by theory, there is a transition from an $O(C\!a)$ increase to an eventual $O(C\!a^{-1})$ decrease with increasing $C\!a$. Owing to the $O(C\!a^2)$ scaling behavior, $F_{11}$ is evidently a more sensitive measure for small $C\!a$, but the analysis and BEM numerics nevertheless exhibit good agreement down to the smallest $C\!a\,(=0.033)$ for the highest resolution\,($N= 5120$).

\begin{figure}
\centerline{\includegraphics[scale = 0.4]{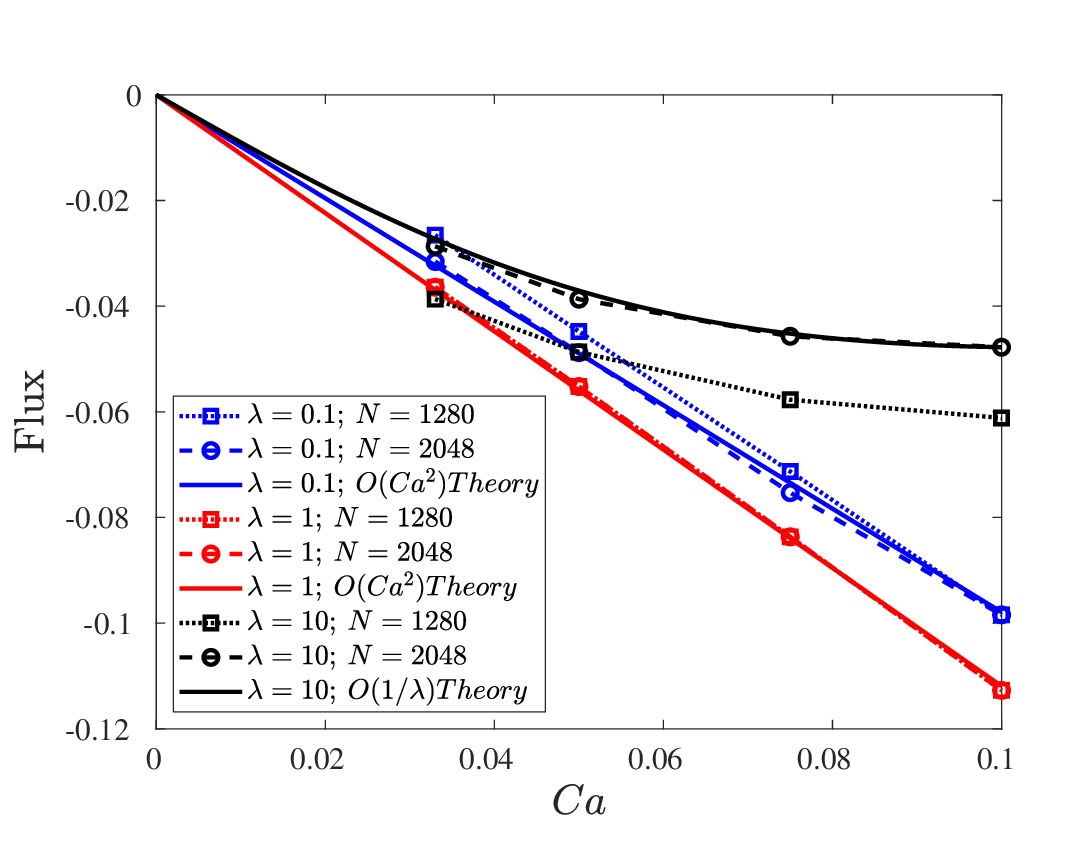}}
\caption{Comparison of the analytical predictions\,(see (\ref{4.37}) and (\ref{4.38})) and BEM results for the inplane flux, as a function of $C\!a$, for different $\lambda$; (a) $N = 1280$ and (b) $N = 2048$. For $\lambda = 1$, simulations converge to theoretical predictions even for the smaller $N$ owing to the absence of the double layer potential.}
\label{fig:21}
\end{figure}

As a final check, we plot the in-plane flux across a specific closed contour\,(a circle of radius $r_0$) surrounding the projection of the deformed drop in the flow-gradient plane. The flux, defined as $\int_0^{2\pi} \bm{u.n} \; r_0 \; d\phi$, must equal zero for any choice of contour, for $C\!a = 0$, and is thereby a sensitive indicator of the deformation-induced alteration of the streamline topology. Using (\ref{smallCa:expansion}), one has:
\begin{equation}
\begin{split}
 &\int_0^{2\pi} \bm{u} \cdot \bm{n} \; r_0 \; d\phi = C\!a \int_0^{2\pi} \bm{u}^{(1)} \cdot \bm{n} \; r_0 \; d\phi = \\
 &-\frac{\pi  Ca (19 \lambda +16) \left(-5 (\lambda +1) (45 \lambda +4)+12 (\lambda  (25 \lambda +41)+4) r_0^4+(182-\lambda  (75 \lambda +37)) r_0^2\right)}{6720 (\lambda + 1)^3 r_0^5},
 \end{split}\label{4.37}
\end{equation}
for $C\!a \ll 1, \lambda \sim O(1)$. In contrast, 
\begin{align}
   &\int_0^{2\pi} \bm{u} \cdot \bm{n} \; r_0 \; d\phi = -\frac{2375 \pi  \text{Ca} [4 r_0^2-1)r_0^2-3)}{28 r_0^6 \left(361 \text{Ca}^2 \lambda ^2+400\right)}, \label{4.38}
\end{align}
for $\lambda \gg 1, \lambda C\!a \sim O(1)$, where, in contrast to (\ref{smallCa:expansion}), we have used an expression for $\bm{u}$ that accounts for drop deformation in the large-$\lambda$ limit. In principle, this velocity field may be used for an examination of the streamline topology in this complementary limit; although, as stated in the introduction, the focus here is only on $C\!a \ll 1$. From (\ref{4.37}) and (\ref{4.38}), the in-plane flux is seen to be sensitively dependent on $C\!a$ regardless of $\lambda$. It is $O(C\!a)$ for $\lambda C\!a \ll 1$, while being $O(\lambda^{-2}C\!a^{-1})$ for $\lambda C\!a \gg 1$, with the latter order being the one at which one expects the closed-streamline topology to be destroyed for $\lambda \gg 1$ - see section \ref{sec:conclude}. 
Fig.\ref{fig:21} shows the comparison between analytical predictions above and the corresponding BEM results, for $r_0 = 1.25$, as a function of $Ca$ and for different $\lambda$; (\ref{4.37}) is used for $\lambda = 0.1$ and $1$, while (\ref{4.38}) is used for $\lambda = 10$. The agreement is seen to be best for $\lambda = 1$. Even for the non-unity $\lambda$'s\,(in which case one has to solve an integral equation for the velocity field), however, the agreement is good for the higher resolution\,($N = 2048$). 

\section{Numerical integration of $C-E$ plane trajectories} \label{AppD}
As stated in section \ref{inertia:CE}, the system of averaged equations is:
\begin{align}
 \frac{dC}{d \hat{t}} &= -\frac{\int_{r_{min}}^{r_{max}} \left( \sin \theta f(r) u_{\theta}^{(1)} + \cos \theta f'(r) u_{r}^{(1)} \right)f(r)^{-2} (u_r^{(0)})^{-1} dr}{\int_{r_{min}}^{r_{max}} (u_r^{(0)})^{-1} dr}, \label{D1} \\
 \frac{dE}{d \hat{t}} &= \frac{\int_{r_{min}}^{r_{max}} \left( u_\theta^{(1)} I_1 + u_r^{(1)} I_2 + u_\phi^{(1)} I_3 \right) (u_r^{(0)})^{-1} dr}{\int_{r_{min}}^{r_{max}} (u_r^{(0)})^{-1} dr}, \label{D2}
\end{align}
where $\hat{t} = C\!a\,t$ or $R\!e\,t$ depending on whether deformation or inertia perturbs the Stokesian closed-streamline topology around a spherical drop. Here, $\bm{u}^{(0)}$ is defined by (\ref{2}-\ref{4}), $\bm{u}^{(1)}$ is given by (\ref{3.2})\,(inertia) or (\ref{4.25main})\,(drop deformation), with the $I_i$'s being defined below (\ref{3.12}). We use the adaptive Runge-Kutta method\,(the in-built RK4(5) tool in Mathematica) for numerical integration, which requires evaluations of the RHS of (\ref{D1}) and (\ref{D2}) at each time step. To do this, one calculates $\theta$ and $\phi$ as functions of $r$\,(varying in the range $[r_{min},r_{max}]$), using (\ref{3.4}-\ref{3.5}), as:
\begin{align}
\theta &= \cos^{-1} \left( C \; f(r) \right) \\
\phi &= \sin^{-1}\left( \frac{1}{\sin \theta} \left( \frac{\hat{\alpha}}{1+\hat{\alpha}} + E f(r)^2 + \frac{\beta \lambda f(r)^2 g(r)}{1+\lambda} \right)^{1/2} \right).
\end{align}
Next, one determines the limits, $r_{max}$ and $r_{min}$, 
of the definite integrals in (\ref{D1}-\ref{D2}). These quantities are calculated as functions of $C$ and $E$ by numerically solving (\ref{3.22}-\ref{3.23}), 
with the aforementioned integrals evaluated using Gaussian quadrature. 
Once all these quantities are calculated, the RHS is used to advance to the next time step, a tolerance of $10^{-13}$ being used to ensure sufficient accuracy. 

\begin{figure}
\includegraphics[scale = 0.35]{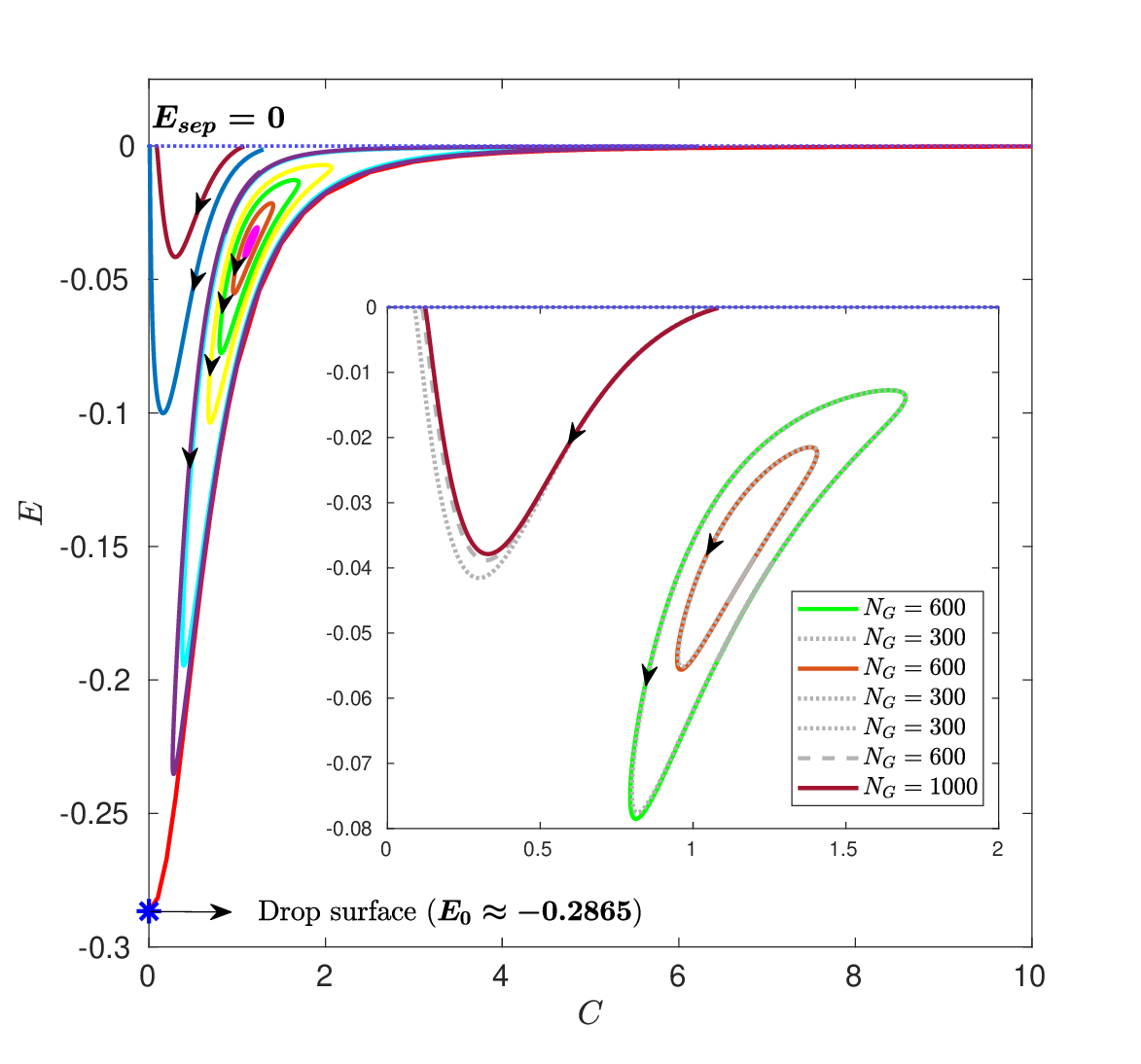}
\includegraphics[scale = 0.35]{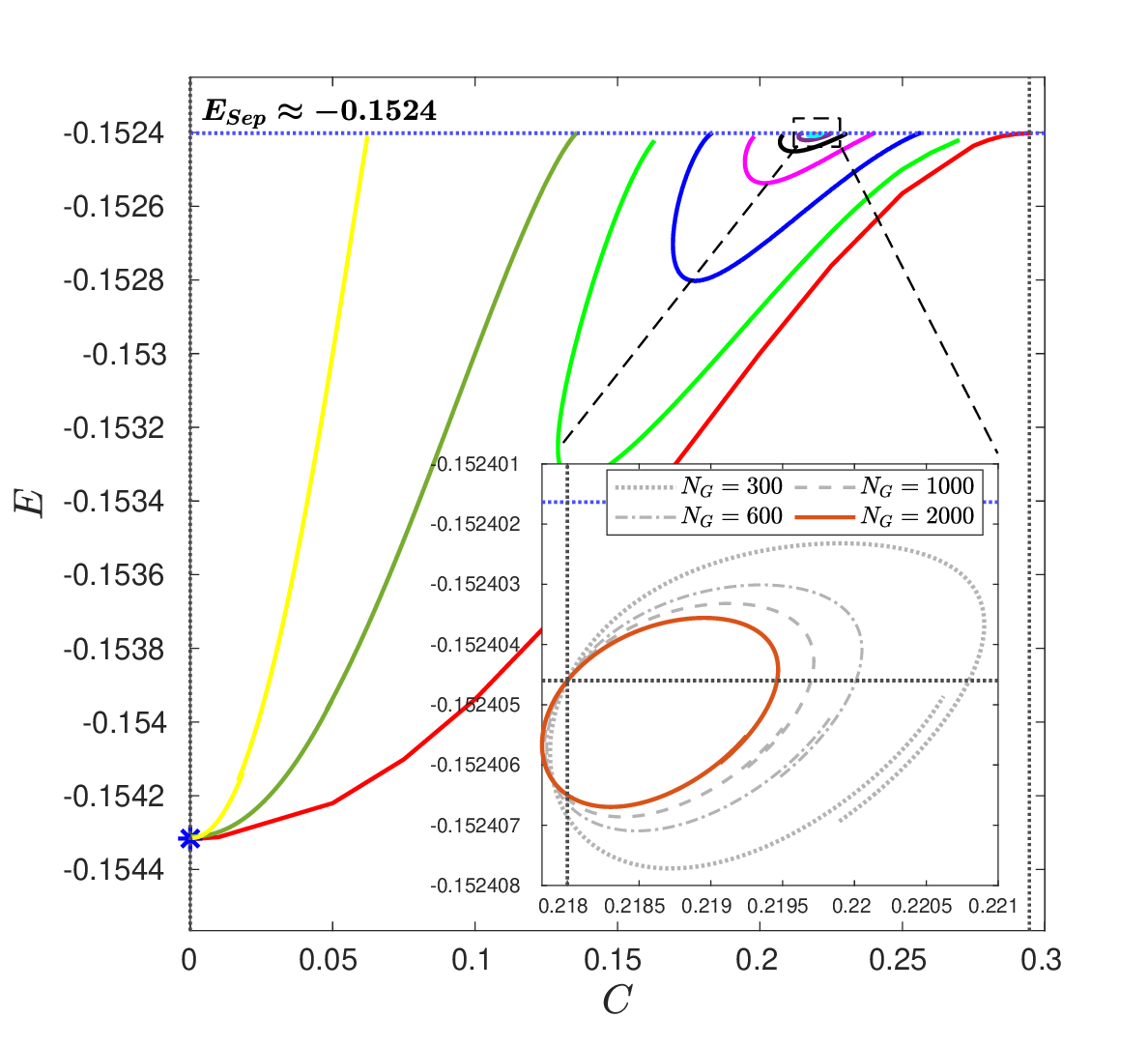}
\caption{Plot of the $CE$ trajectories for $\lambda = 1$ and (a) $\hat{\alpha} = 0$, and (b)  $\hat{\alpha} = 0.3$. The inset in each figure shows select trajectories plotted for different values of Gaussian quadrature points $N_G = 300, 600$ etc. and shows that the trajectories, specifically the closed ones, are converged with respect to number of quadrature points used in the evaluating the integrals in (\ref{D1}-\ref{D2}). In (b), the inset shows a trajectory close to separatrix plotted for multiple values of Gaussian quadrature points $N_G$ and shows that the trajectories don't converge with respect to $N_G$. The trajectories in the main figures are plotted with an $N_G$ for which the trajectories are converged.}
\label{fig:22N}
\end{figure}

The most important factor controlling the accuracy of the numerics is the number of quadrature points, $N_G$, used for the integral evaluations. The critical role of $N_G$ is illustrated in Figs.\ref{fig:22N}a and b. For simple shear flow, where the region of closed curves is the largest in extent, $N_G = 600$ is found to be sufficient for convergence. In contrast, for $\hat{\alpha}= 0.3$, where the region of closed curves sits very close to the separatrix, the inset in Fig.\ref{fig:22N}b illustrates the lack of convergence with increasing $N_G$, even for $N_G$ as large as $2000$. This lack of convergence also underlies the difficulty in isolating the singular curve representative of the outermost invariant torus, and that is expected to touch the separatrix at a single point\,(a corner). 


\bibliographystyle{jfm}

\end{thebibliography}

\end{comment}


\end{document}